\documentclass[aps, prd, reprint,10pt, notitlepage, a4paper,floats, floatfix,amsmath, amssymb, amsfonts,superscriptaddress,showpacs, showkeys,nofootinbib,longbibliography,tightenlines,nobibnotes,aps]{revtex4-2}

\usepackage{graphicx} 
\usepackage{amsmath,amsfonts,amssymb,amsthm,bm}
\usepackage{amsfonts}
\usepackage{hyperref} 
\usepackage{color}
\usepackage{xcolor}
\usepackage{graphics}
\usepackage[font=small,justification=justified,format=plain]{caption}
\usepackage{array} 
\usepackage{booktabs} 
\usepackage{subcaption}
\usepackage{multirow}
\usepackage[normalem]{ulem}
\usepackage{ragged2e}

\usepackage[markup=underlined,authormarkup=none,highlightmarkup=background]{changes}
\definechangesauthor[name={HE}, color=blue]{HE}

\def\mr{\mathrm}

\def\LNhat{\bm{l}_N}
\def\dotLNhat{\dot{\bm{l}}_N}

\def\J2P{J\rightarrow P}
\def\I2J{I\rightarrow J}
\def\I2P{I\rightarrow P}

\newcommand{\SEOBASYM}{\texttt{SEOBNRv5PHM}$_{\mr{w/asym}}$ }

\newcommand{\AEI}{\affiliation{Max Planck Institute for Gravitational Physics (Albert Einstein Institute), Am M\"uhlenberg 1, Potsdam 14476, Germany}}
\newcommand{\Maryland}{\affiliation{Department of Physics, University of Maryland, College Park, MD 20742, USA}}

\definecolor{dodgerblue}{HTML}{1E90FF}
\hypersetup{citecolor=dodgerblue}

\begin{document}

\makeatletter
\long\def\@makecaption#1#2{%
  \vskip\abovecaptionskip
  \small
  \sbox\@tempboxa{#1: #2}%
  \ifdim \wd\@tempboxa >\hsize
    \parbox{\hsize}{\justifying\small #1: #2}%
  \else
    \global \@minipagefalse
    \hb@xt@\hsize{\hfil\box\@tempboxa\hfil}%
  \fi
  \vskip\belowcaptionskip}
\makeatother

\title{Adding equatorial-asymmetric effects for spin-precessing binaries into the SEOBNRv5PHM waveform model}

\author{H\'ector Estell\'es}\AEI
\author{Alessandra Buonanno}\AEI\Maryland
\author{Raffi Enficiaud}\AEI
\author{Cheng Foo}\AEI
\author{Lorenzo Pompili}\AEI

\date{\today}

\begin{abstract}

    Gravitational waves from spin-precessing binaries exhibit equatorial asymmetries absent in non-precessing systems, leading to net linear momentum emission and contributing to the remnant's recoil. This effect, recently incorporated into only a few waveform models, is crucial for accurate recoil predictions and improved parameter estimation. We present an upgrade to the \texttt{SEOBNRv5PHM} model --- \SEOBASYM --- which includes equatorial asymmetric contributions to the $\ell=m\leq 4$ waveform modes in the co-precessing frame. The model combines post-Newtonian inputs with calibrated amplitude and phase corrections and a phenomenological merger-ringdown description, tuned against 1523 quasi-circular spin-precessing numerical relativity waveforms and single-spin precessing test-body plunging-geodesic waveforms. We find that \SEOBASYM improves the agreement with NR waveforms across inclinations, with median unfaithfulness reduced by up to 50\% compared to \texttt{SEOBNRv5PHM}, and achieves 30--60\% lower unfaithfulness than \texttt{IMRPhenomXPNR} and 76--80\% lower than \texttt{TEOBResumS\_Dali}. The model significantly improves the prediction of the recoil velocity, reducing the median relative error with numerical relativity from 70\% to 1\%. Bayesian inference on synthetic injections demonstrates improved recovery of spin orientations and mass parameters, and a reanalysis of GW200129 shows a threefold increase in the spin-precessing Bayes factor, highlighting the importance of these effects for interpreting spin-precessing events.

\end{abstract}

\maketitle


\section{Introduction}

Since the first direct observation of gravitational waves (GWs) from a compact binary coalescence by the LIGO detectors in 2015 \cite{LIGOScientific:2016aoc}, GW astronomy has rapidly developed into a powerful observational tool. To date, the LIGO-Virgo-KAGRA (LVK) Collaboration --- comprising the two LIGO interferometers in Livingston and Hanford (USA) \cite{LIGOScientific:2014pky}, Virgo in Cascina (Italy) \cite{VIRGO:2014yos}, and KAGRA in Toyama (Japan) \cite{KAGRA:2020tym} --- has reported over 90 GW detections consistent with compact binary signals. Most of these events appear in the successive releases of the Gravitational-Wave Transient Catalog (GWTC) \cite{LIGOScientific:2018mvr,LIGOScientific:2020ibl,LIGOScientific:2021usb,KAGRA:2021vkt}, with additional detections reported by independent groups \cite{Nitz:2021zwj,Wadekar:2023gea}. These observations have enabled a wide range of scientific studies, including population inference of binary systems \cite{KAGRA:2021duu}, constraints on the neutron star equation of state \cite{LIGOScientific:2018cki}, measurements of the Hubble constant \cite{LIGOScientific:2017adf}, and tests of general relativity \cite{LIGOScientific:2021sio}.

A key factor behind the scientific success of GW observations has been the development of accurate and efficient waveform models for compact binary systems. These models combine analytical and semi-analytical results --- derived from post-Newtonian (PN) theory, post-Minkowskian theory, and black hole perturbation theory --- with phenomenological methods and high-fidelity numerical relativity (NR) simulations. The main modeling programs --- the \texttt{IMRPhenom} family (in both frequency \cite{Pan:2007nw,Ajith:2007qp,Ajith:2009bn,Santamaria:2010yb,Hannam:2013oca,Husa:2015iqa,Khan:2015jqa,London:2017bcn,Khan:2018fmp,Khan:2019kot,Dietrich:2019kaq,Pratten:2020fqn,Garcia-Quiros:2020qpx,Pratten:2020ceb,Hamilton:2021pkf,Thompson:2023ase,Colleoni:2024knd,xpnr_inprep} and time domains \cite{Estelles:2020osj,Estelles:2020twz,Estelles:2021gvs,Rossello-Sastre:2024zlr,Planas:2025feq}), the effective-one-body (EOB) \cite{Buonanno:1998gg,Buonanno:2000ef,Damour:2000we,Damour:2001tu,Buonanno:2005xu} formalism with its two main branches \texttt{SEOBNR} \cite{Buonanno:2006ui,Buonanno:2007pf,Buonanno:2009qa,Pan:2009wj,Pan:2011gk,Taracchini:2012ig,Taracchini:2013rva,Bohe:2016gbl,Cotesta:2018fcv,Ossokine:2020kjp,Pompili:2023tna,Ramos-Buades:2023ehm,Gamboa:2024hli} and \texttt{TEOBResumS} \cite{Damour:2014sva,Nagar:2015xqa,Nagar:2018zoe,Nagar:2019wds,Nagar:2020pcj,Riemenschneider:2021ppj,Nagar:2020pcj,Chiaramello:2020ehz,Gamba:2021ydi,Gamba:2024cvy,Albanesi:2025txj}, and the NR surrogate models \cite{Blackman:2015pia,Blackman:2017dfb,Blackman:2017pcm,Varma:2018mmi,Varma:2019csw,Williams:2019vub,Rifat:2019ltp,Islam:2021mha,Islam:2022laz,Islam:2024zqo} --- differ in the amount and type of analytical and numerical input, the approximations employed, and their coverage of the parameter space.

These models are continuously being refined to incorporate additional physical effects, reduce systematic uncertainties, and improve computational efficiency. In the Fourier-domain \texttt{IMRPhenom} approach, recent advances have focused on improving the widely used quasi-circular multipolar spin-precessing model \texttt{IMRPhenomXPHM} \cite{Pratten:2020fqn,Garcia-Quiros:2020qpx,Pratten:2020ceb}. Improvements include a refined treatment of spin dynamics during inspiral (\texttt{IMRPhenomXPHM\_SpinTaylor} \cite{Colleoni:2024knd}), and calibration to NR waveforms in the late inspiral, merger, and ringdown (\texttt{IMRPhenomXO4a} \cite{Hamilton:2021pkf,Thompson:2023ase}), leveraging a new set of single-spin precessing simulations generated with the BAM code \cite{Hamilton:2023qkv}. This calibration includes ringdown frequency corrections for spin-precessing systems \cite{Hamilton:2023znn} and the modeling of the dominant spin-precessing antisymmetric mode \cite{Ghosh:2023mhc}. These developments have recently been unified into a new model, \texttt{IMRPhenomXPNR} \cite{xpnr_inprep}. In the time domain, the \texttt{IMRPhenomT} waveform family has been extended to eccentric systems (\texttt{IMRPhenomTEHM} \cite{Planas:2025feq}, currently limited to aligned-spin binaries) and to include the real $(2,0)$ harmonic with non-linear memory effects \cite{Rossello-Sastre:2024zlr,Valencia:2024zhi}.

In the EOB approach, the \texttt{TEOBResumS} program has advanced beyond its quasi-circular spin-precessing model \texttt{TEOBResumS\_Giotto} \cite{Akcay:2020qrj,Gamba:2021ydi} to produce \texttt{TEOBResumS\_Dali}, capable of modeling binaries in eccentric and unbound orbits. This model was recently extended to generic precessing-spin systems \cite{Gamba:2024cvy,Albanesi:2025txj} and now includes real $m=0$ modes with non-linear memory \cite{Albanesi:2024fts}. The \texttt{SEOBNR} program has released the fifth generation of quasi-circular multipolar waveform models: \texttt{SEOBNRv5HM} for aligned-spin systems \cite{Pompili:2023tna} and \texttt{SEOBNRv5PHM} for precessing systems \cite{Ramos-Buades:2023ehm}. These models are implemented in the modular Python infrastructure \texttt{pySEOBNR} \cite{Mihaylov:2023bkc}, which incorporates state-of-the-art PN resummations \cite{Khalil:2023kep} and second-order gravitational self-force results \cite{vandeMeent:2023ols}. Additional developments in this framework include eccentric extensions (\texttt{SEOBNRv5EHM} \cite{Gamba:2024cvy}), models for binary neutron stars (\texttt{SEOBNRv5THM} \cite{Haberland:2025luz}), tests of beyond-GR theories (\texttt{SEOBNRv5PHM\_ESGB} \cite{Julie:2024fwy}), and the incorporation of post-Minkowskian and self-force information into the conservative dynamics \cite{Buonanno:2024byg,Leather:2025nhu}.

NR surrogates have emerged as a highly successful modeling strategy. The Simulating eXtreme Spacetimes (SXS) Collaboration has released three major public waveform catalogs, the latest being the third SXS catalog \cite{Scheel:2025jct}, which includes over 3700 waveforms for quasi-circular and eccentric binaries with aligned and misaligned spins. These datasets have enabled the construction of accurate surrogate models. Notably, the \texttt{NRSur7dq4} model \cite{Varma:2019csw}, built from more than 1500 simulations in the second SXS catalog \cite{Boyle:2019kee}, achieves accuracy comparable to intrinsic NR precision, although its domain is limited to the catalog's coverage. Recent developments include the first surrogate models for eccentric binaries \cite{Islam:2021mha}, for large mass-ratio systems using test-body waveforms \cite{Islam:2022laz, Rink:2024swg}, and including non-linear memory effects \cite{Yoo:2023spi}.

While current waveform models are sufficiently accurate for most detected signals --- where statistical uncertainties dominate --- future detectors will observe events with higher signal-to-noise ratio (SNR), where waveform systematics could lead to biased inference \cite{Purrer:2019jcp,Hu:2022rjq,Nagar:2023zxh}. Several works have identified systematic discrepancies in existing models \cite{Foo:2024exr,MacUilliam:2024oif,Kapil:2024zdn,Dhani:2024jja} that can impact both astrophysical and fundamental-physics conclusions \cite{Hu:2022bji,Toubiana:2023cwr,Gupta:2024gun}. One mitigation strategy includes marginalizing over waveform uncertainties during inference \cite{Moore:2014pda,Moore:2015sza,Jan:2020bdz,Read:2023hkv,Hoy:2024vpc,Khan:2024whs,Pompili:2024yec}, which helps address inaccuracies in the physical modeling but does not capture the impact of missing physics.

An effect which is not yet fully incorporated into all major waveform families is equatorial asymmetry. These asymmetries arise from in-plane spin effects and break the emission equatorial symmetry \cite{Boyle:2014ioa} of non-precessing systems, leading to net emission of linear momentum perpendicular to the orbital plane and enhancing the recoil kick of the remnant black hole. Recoil velocities can become especially large in “superkick” configurations \cite{Bruegmann:2007bri,Campanelli:2007cga,Gonzalez:2007hi,Zlochower:2010sn}, and even larger when combined with “hang-up” effects \cite{Lousto:2011kp,Lousto:2012su,Lousto:2019lyf}. In extreme cases, such kicks may eject the remnant from its host environment, with implications for black hole populations \cite{Gerosa:2021mno,Mahapatra:2021hme,Alvarez:2024dpd,Borchers:2025sid}. The connection between waveform asymmetries and gravitational recoil is well understood \cite{Boyle:2014ioa,Ma:2021znq,Mielke:2024kya,Leong:2025raf} --- in general, accurate modeling of the waveform modes during merger and ringdown is crucial for correct recoil predictions \cite{CalderonBustillo:2018zuq,Borchers:2022pah,CalderonBustillo:2022ldv} ---, and recent work suggests that modeling asymmetric effects improves parameter inference in spin-precessing binaries \cite{Kolitsidou:2024vub,Borchers:2024tdi}, particularly for spin misalignment angles that are sensitive to binary formation channels \cite{Gerosa:2013laa,Vitale:2015tea,Farr:2017gtv,Mapelli:2021taw}. Currently, these effects are included in the \texttt{NRSur7dq4} model for all modes, and more recently in the dominant harmonic of \texttt{IMRPhenomXO4a} and \texttt{IMRPhenomXPNR} \cite{Ghosh:2023mhc}.

In this work, we present \SEOBASYM, an extension of \texttt{SEOBNRv5PHM} that includes equatorial-asymmetric contributions in the $(2,2)$, $(3,3)$, and $(4,4)$ waveform modes in the co-precessing frame. During the inspiral, we incorporate linear-in-spin PN contributions projected onto the orbital plane \cite{Arun:2008kb,Boyle:2014ioa}, including next-to-leading order (NLO) terms for the dominant mode and leading-order terms for the subdominant ones. These are further enhanced by amplitude and non-quasicircular phase corrections. For the merger-ringdown, we adopt the existing \texttt{SEOBNRv5PHM} phenomenology and recalibrate the dominant-mode amplitude. The model is calibrated using 1523 precessing NR simulations from the SXS catalog (including simulations from Ref.~\cite{Ossokine:2020kjp}) and over 6000 test-body Teukolsky waveforms from plunging geodesics in the large mass-ratio limit \cite{Apte:2019txp,Lim:2019xrb}.

We assess the accuracy of \SEOBASYM by computing the unfaithfulness with respect to NR waveforms across different source inclinations, using two independent datasets: the second SXS catalog and a set of precessing simulations from the BAM code. In both cases, we find a significant improvement over \texttt{SEOBNRv5PHM}, with a reduction of the median unfaithfulness by up to 50\% for systems viewed face-on, and a notable decrease in the number of cases exceeding the 1\% unfaithfulness threshold, particularly at low inclinations. Compared to other state-of-the-art models, \texttt{IMRPhenomXPNR} and \texttt{TEOBResumS\_Dali}, \SEOBASYM exhibits consistently lower median unfaithfulness --- by 30--50\% and 77--81\%, respectively, depending on inclination.

We also evaluate the performance of \SEOBASYM in predicting the recoil (kick) velocity of the remnant, by computing the linear momentum flux for the NR simulations and comparing it to the predictions from \texttt{SEOBNRv5PHM} and \SEOBASYM evaluated on the same configurations. The inclusion of antisymmetric mode contributions in \SEOBASYM allows it to closely reproduce the distribution of NR kick velocities, including an accurate prediction of the maximum recoil magnitude present in the catalog.

We validate the model's performance through Bayesian inference on three challenging NR injections in zero-noise realizations, observing improved recovery of spin and, in some cases, mass and inclination parameters. Finally, we reanalyze GW200129\_065458 (which is denoted as GW200129 in the rest of the text) --- a loud signal potentially sourced by a spin-precessing system \cite{Hannam:2021pit}, with known data quality issues \cite{Payne:2022spz} --- to assess the impact of equatorial asymmetries in the modes. Our analysis confirms the findings of \cite{Kolitsidou:2024vub} regarding the importance of the asymmetric contributions: \SEOBASYM shows closer agreement with \texttt{NRSur7dq4} and yields significantly higher Bayes factors in favor of the spin-precessing hypothesis.

This paper is organized as follows. In Sec.~\ref{sec:model}, we introduce equatorial-asymmetric effects and describe their modeling during inspiral and merger-ringdown. Sec.~\ref{sec:fitting} details the calibration of amplitude and phase corrections using NR and test-particle datasets. In Sec.~\ref{sec:unfaith}, we assess model accuracy through unfaithfulness comparisons with state-of-the-art models. Sec.~\ref{subsec:recoil} presents kick velocity predictions. In Sec.~\ref{subsec:nrinj}, we analyze synthetic NR injections. Sec.~\ref{sec:gw200129} revisits GW200129 and quantifies the impact of antisymmetric modes. We conclude in Sec.~\ref{sec:conclusions} with a summary and outlook. We include two appendices, showing an unfaithfulness study against the \texttt{NRSur7dq4} model in Appendix~\ref{sec:nrsur} and a comparison of the unfaithfulness results against NR between \texttt{IMRPhenomXPHM\_ST}, \texttt{IMRPhenomXO4a} and \texttt{IMRPhenomXPNR} in Appendix~\ref{sec:xphm}.

\subsection*{Notation}

Throughout this work, we adopt geometric units, setting $G = c = 1$ unless otherwise specified. We consider binary black hole (BBH) systems with component masses $m_1$ and $m_2$, labeled such that $m_1 \geq m_2$, and associated spin vectors $\bm{S}_1$ and $\bm{S}_2$. Several combinations of these quantities are useful:
\begin{equation}
    \begin{gathered}
    M = m_1 + m_2, \quad \mu = \frac{m_1 m_2}{M}, \quad q = \frac{m_1}{m_2},\\
    \nu = \frac{\mu}{M}, \quad \delta = \frac{m_1 - m_2}{M},
    \end{gathered}
\end{equation}
corresponding to the total mass, reduced mass, mass ratio, symmetric mass ratio, and normalized mass difference, respectively. A particularly relevant combination for GW data analysis is the \textit{chirp mass} \cite{Sathyaprakash:2009xs}, given by
\begin{equation}
\mathcal{M} = \nu^{3/5} M.
\label{eq:chirpMass}
\end{equation}

The dimensionless spin vectors are defined as
\begin{equation}
\bm{\chi}_{\mr i} = \frac{\bm{S}_{\mr i}}{m_{\mr i}^2}.
\end{equation}
We also use the total symmetric and rescaled-antisymmetric spin combinations,
\begin{subequations}
    \begin{align}
        \bm{S} &= \bm{S}_1 + \bm{S}_2,\\
        \bm{\Sigma} &= M\left(\frac{\bm{S}_2}{m_2} - \frac{\bm{S}_1}{m_1}\right).
    \end{align}
\end{subequations}

The relative position and momentum vectors in the binary's center-of-mass frame are $\bm{r}$ and $\bm{p}$, with
\begin{equation}
\bm{p}^2 = p_r^2 + \frac{L^2}{r^2}, \quad
p_r = \bm{n} \cdot \bm{p}, \quad
\bm{L} = \bm{r} \times \bm{p},
\end{equation}
where $\bm{n} = \bm{r} / r$ is the radial unit vector, and $\bm{L}$ is the orbital angular momentum, with magnitude $L$. The total angular momentum is $\bm{J} = \bm{L} + \bm{S}_1 + \bm{S}_2$. The Newtonian angular momentum vector is given by
\begin{equation}
\bm{L}_{\rm N} = \mu\, \bm{r} \times \dot{\bm{r}},
\end{equation}
with normalized direction $\LNhat = \bm{L}_{\rm N} / |\bm{L}_{\rm N}|$.

To describe the orbital geometry and precession, we use a triad of basis vectors $\{ \bm{n}, \bm{\lambda}, \LNhat \}$, where
\begin{subequations}
    \begin{align}
        \bm{n} &= \frac{\bm{r}}{|\bm{r}|},\\
        \bm{\lambda} &= \LNhat \times \bm{n}.
    \end{align}
\end{subequations}
The orbital phase is defined as
\begin{equation}
\label{eq:orbphase}
\phi(t) = \arctan\left[ \frac{\bm{n}(t)}{\bm{n}(t_{\rm ref})} \right],
\end{equation}
where $t_{\rm ref}$ is an arbitrary reference time such that $\phi(t_{\rm ref}) = 0$. The orbital frequency is $\Omega(t) = \dot{\phi}(t)$, and we define the PN velocity as $v(t) = (M \Omega)^{1/3}$.

We denote projections of spin vectors onto the precessing basis as
\begin{equation}
\begin{gathered}
S_{i,n} = \bm{S}_i \cdot \bm{n}, \quad
S_{i,\lambda} = \bm{S}_i \cdot \bm{\lambda}, \quad
S_{i,l} = \bm{S}_i \cdot \LNhat,
\end{gathered}
\end{equation}
with analogous definitions for $\bm{S}$ and $\bm{\Sigma}$. The magnitudes of the in-plane projections are
\begin{equation}
\begin{gathered}
S_\perp = \sqrt{S_n^2 + S_\lambda^2}, \quad
\Sigma_\perp = \sqrt{\Sigma_n^2 + \Sigma_\lambda^2},
\end{gathered}
\end{equation}
and the azimuthal directions of the spin vectors are
\begin{equation}
\begin{gathered}
\phi_S = \arctan\left(\frac{S_\lambda}{S_n}\right), \quad
\phi_\Sigma = \arctan\left(\frac{\Sigma_\lambda}{\Sigma_n}\right).
\end{gathered}
\end{equation}

Two effective spin parameters are employed in this work: the \textit{effective aligned spin} $\chi_{\rm eff}$ \cite{Damour:2001tu,Racine:2008qv,Santamaria:2010yb},
\begin{equation}
\chi_{\rm eff} = \frac{1}{M} \left( m_1 \bm{\chi}_1 + m_2 \bm{\chi}_2 \right) \cdot \LNhat,
\label{eq:chi_eff}
\end{equation}
and the \textit{effective precession spin} $\chi_{\rm p}$ \cite{Schmidt:2014iyl},
\begin{equation}
\chi_{\rm p} = \frac{1}{B_1 m_1^2} \max \left( B_1 m_1^2 \chi_{1,\perp},\; B_2 m_2^2 \chi_{2,\perp} \right),
\label{eq:chi_p}
\end{equation}
where $B_1 = 2 + \tfrac{3}{2} m_2 / m_1$, $B_2 = 2 + \tfrac{3}{2} m_1 / m_2$, and $\chi_{i,\perp}$ denotes the magnitude of the projection of $\bm{\chi}_i$ onto the orbital plane.

It is useful to define the \textit{PN-weighted effective aligned spin} $\hat{\chi}$, which enters the GW phase at leading order in the PN expansion \cite{Cutler:1994ys,Khan:2015jqa,Ajith:2011ec,Poisson:1995ef}
\begin{equation}
    \hat{\chi}=\frac{\chi_{\rm eff}-(38\nu/113)\left(\bm{\chi}_1 +\bm{\chi}_2 \right) \cdot \LNhat}{1-76\nu/113},
\end{equation}
and the \textit{antisymmetric aligned spin}
\begin{equation}
    \chi_a=\frac{\left(\bm{\chi}_1 - \bm{\chi}_2 \right) \cdot \LNhat}{2}.
\end{equation}

We define two non-inertial frames associated with the binary: the \textit{co-rotating} frame and the \textit{co-precessing} frame. The co-rotating frame (consistent with the definition in Ref.~\cite{Boyle:2013nka} employed in Ref.~\cite{Boyle:2014ioa}) is given by
\begin{subequations}
    \begin{align}
    \hat{\bm{x}}_{\rm corot} &= \bm{n},\\
    \hat{\bm{y}}_{\rm corot} &= \bm{\lambda},\\
    \hat{\bm{z}}_{\rm corot} &= \LNhat,
    \end{align}
\end{subequations}
while in the co-precessing frame, the basis vectors satisfy
\begin{subequations}
\label{eq:coprec_frame}
    \begin{align}
    \bm{n} &= (\cos\phi, \sin\phi, 0)_{\rm coprec},\\
    \bm{\lambda} &= (-\sin\phi, \cos\phi, 0)_{\rm coprec},\\
    \LNhat &= (0, 0, 1)_{\rm coprec},
    \end{align}
\end{subequations}
where $\phi$ is the orbital phase (defined in Eq.\eqref{eq:orbphase}). At the reference time, the co-rotating and co-precessing frames coincide.

\section{Modeling equatorial asymmetric effects in the waveform modes}\label{sec:model}

The GW signal emitted by a coalescing binary can be described in terms of two real polarization states, $h_+$ and $h_{\times}$, typically combined into a complex strain:
\begin{equation}
\begin{aligned}
    h(t,\boldsymbol{\Lambda},d_{\rm L},\iota,\varphi_{\rm ref}) &= h_+(t,\boldsymbol{\Lambda},d_{\rm L},\iota,\varphi_{\rm ref}) \\
    &- i h_{\times}(t,\boldsymbol{\Lambda},d_{\rm L},\iota,\varphi_{\rm ref}),
\end{aligned}
\end{equation}
which depends on the intrinsic parameters of the binary $\bm{\Lambda} = \{m_1, m_2, \bm{\chi}_1(t), \bm{\chi}_2(t)\}$ (for a quasi-circular BBH), the orientation of the source with respect to the observer, commonly parameterized by the inclination angle $\iota$ and azimuthal angle $\varphi_{\rm ref}$, and the luminosity distance of the source $d_{\rm L}$. To separate intrinsic parameters from orientation, the complex strain is typically decomposed into spin-weighted spherical harmonics (SWSHs):
\begin{equation}
    h(t,\boldsymbol{\Lambda},d_{\rm L},\phi(t),\iota,\varphi_{\rm ref}) = \sum_{\ell,m} \frac{h_{\ell m}(t,\boldsymbol{\Lambda},\phi(t))}{d_{\rm L}}\;{}_{-2}Y_{\ell m}(\iota,\varphi_{\rm ref}),
\end{equation}
where the SWSHs encode directional dependence, while the modes $h_{\ell m}$ depend only on the intrinsic binary properties. We explicitly indicate the dependence on the orbital phase $\phi(t)$ (we omit in the following the time dependence in $\phi(t)$, and we remind that $\phi(t=t_{\rm ref})=0$).

The behavior of the waveform modes under parity transformations is important for understanding how spin-precession affects the waveform structure. The full parity operation is a spatial inversion:
\begin{equation}
    (x, y, z) \to (-x, -y, -z),
\end{equation}
which corresponds in spherical coordinates to:
\begin{equation}
    \iota \to \pi - \iota, \quad \varphi_{\rm ref} \to \varphi_{\rm ref} + \pi,\quad \phi \to \phi + \pi.
\end{equation}
Under this transformation, the complex strain (we drop from now on the dependency on the intrinsic parameters, for brevity) transforms as \cite{Arun:2008kb}
\begin{equation}
h(t, \phi,\iota,\varphi_{\rm ref}) = h^*(t, \phi+\pi,\pi - \iota, \varphi_{\rm ref} + \pi). 
\end{equation}
The SWSHs obey the identity
\begin{equation}
    {}_{-2}Y_{\ell m}(\pi - \iota, \varphi_{\rm ref} + \pi) = (-1)^{\ell} {}_{-2}Y_{\ell, -m}(\iota, \varphi_{\rm ref}),
\end{equation}
which implies that the modes transform as \cite{Arun:2008kb}
\begin{equation}
    \label{eq:parity_symmetry}
    h_{\ell m}(t, \phi) = (-1)^{\ell + m} h^*_{\ell, -m}(t, \phi + \pi).
\end{equation}
The $\phi\rightarrow\phi + \pi$ shift in the orbital phase is a consequence of the azimuthal coordinate transformation in spherical coordinates under full parity.

It is particularly useful to consider this symmetry in the co-precessing frame (defined in Eq.~\eqref{eq:coprec_frame}), where the explicit orbital phase dependence can be factored out of the waveform modes:
\begin{equation}
    \label{eq:coprec_modes}
    h_{\ell m}^{\mathrm{coprec}}(t,\phi) = \hat{h}_{\ell m}(t,\phi) e^{-i m \phi},
\end{equation}
while the reduced modes $\hat h_{\ell m}$ still retain an implicit
dependence on the orbital phase through projections with the basis vectors $\bm{n}$ and $\bm{\lambda}$.

Applying the symmetry relation of Eq.~\eqref{eq:parity_symmetry} in this frame yields:
\begin{equation}
\begin{aligned}
    h_{\ell m}^{\mathrm{coprec}}(t,\phi) 
    &= (-1)^{\ell + m} \hat{h}^*_{\ell,-m}(t,\phi + \pi) e^{i m (\phi + \pi)} \\
    &= (-1)^{\ell} \hat{h}^*_{\ell,-m}(t,\phi + \pi) e^{i m \phi}.
\end{aligned}
\end{equation}

To proceed, we examine how the amplitudes $\hat{h}_{\ell m}(\phi)$ transform under $\phi \rightarrow \phi + \pi$. The orbital configuration of the binary is parameterized by the basis vectors $\{\bm{n}, \bm{\lambda}, \LNhat\}$, which transform under this shift as:
\begin{subequations}
\begin{align}
    \bm{n} &\rightarrow -\bm{n}, \\
    \bm{\lambda} &\rightarrow -\bm{\lambda}, \\
    \LNhat &\rightarrow \LNhat.
\end{align}
\end{subequations}
Hence, if $\hat{h}_{\ell m}$ depends only on mass parameters and the spin projections onto $\LNhat$, and is independent of in-plane projections onto $\bm{n}$ and $\bm{\lambda}$, then it is invariant under $\phi \rightarrow \phi + \pi$:
\begin{equation}
    \hat{h}_{\ell m}(t,m_i, S_{i,\ell}, \phi + \pi) = \hat{h}_{\ell m}(t,m_i, S_{i,\ell}, \phi).
\end{equation}
Eq.~\eqref{eq:parity_symmetry} then reduces to the \textit{equatorial symmetry} relation:
\begin{equation}
    \label{eq:equatorial_symmetry}
    h_{\ell m}^{\mathrm{coprec}}(t, m_i, S_{i,\ell}, \phi) = (-1)^{\ell} h_{\ell, -m}^{\mathrm{coprec} *}(t, m_i, S_{i,\ell}, \phi).
\end{equation}

The equatorial symmetry relation is closely related to the \textit{$z$-parity} operation, which corresponds to a reflection through the $x-y$ plane (which corresponds to the orbital plane in the co-precessing frame):
\begin{equation}
    (x, y, z) \to (x, y, -z),
\end{equation}
or in spherical coordinates:
\begin{equation}
    \iota \to \pi - \iota, \quad \varphi_{\rm ref} \to \varphi_{\rm ref}, \quad \phi \to \phi.
\end{equation}
Under $z$-parity, the SWSHs transform as:
\begin{equation}
    {}_{-2}Y_{\ell m}(\pi - \iota, \varphi_{\rm ref}) = (-1)^{\ell} {}_{-2}Y_{\ell, -m}(\iota, \varphi_{\rm ref}),
\end{equation}
and invariance of the full strain requires the same relation:
\begin{equation}
    h^{\mathrm{coprec}}_{\ell m}(t, \phi) = (-1)^{\ell} h^{\mathrm{coprec}*}_{\ell, -m}(t, \phi).
\end{equation}
The key is that this relation arises from requiring the invariance of the strain under $z$-parity transformations. This is not a general symmetry of the strain, as Eq.~\eqref{eq:parity_symmetry} is, and only holds for non-precessing systems \cite{Boyle:2014ioa}.

Based on whether Eq.~\eqref{eq:equatorial_symmetry} is satisfied, it is convenient to decompose the waveform modes in the co-precessing frame into a part that satisfies this symmetry (referred to as the \textit{symmetric} part), and a part that transforms with the opposite sign (referred to as the \textit{antisymmetric} part). These are defined as:
\begin{subequations}
\label{eq:formula_sym_asym}
\begin{align}
    h^{\mathrm{sym}}_{\ell m} &= \frac{h^{\mathrm{coprec}}_{\ell m} + (-1)^{\ell} h^{\mathrm{coprec}*}_{\ell, -m}}{2}, \\
    h^{\mathrm{asym}}_{\ell m} &= \frac{h^{\mathrm{coprec}}_{\ell m} - (-1)^{\ell} h^{\mathrm{coprec}*}_{\ell, -m}}{2}.
\end{align}
\end{subequations}
This terminology follows the naming convention in the literature, where ``symmetric'' and ``antisymmetric'' refer to the co-precessing frame modes satisfying or not the equatorial symmetry relation coming from $z-$parity invariance. We emphasize that the difference between the $z-$parity transformation and the full parity transformation is precisely the shift $\phi\rightarrow\phi + \pi$, which ensures that the full symmetry of Eq.~\eqref{eq:parity_symmetry} is always satisfied.

The decomposition formally defined through Eq.~\eqref{eq:formula_sym_asym} has direct physical relevance. The antisymmetric contributions correspond to an imbalance in GW emission between the northern and southern hemispheres of the source. This asymmetry leads to a net flux of linear momentum perpendicular to the orbital plane \cite{Boyle:2014ioa,Ma:2021znq,Mielke:2024kya}, which in turn imparts a recoil (kick) to the remnant black hole. The dominant contribution to the out-of-plane linear momentum flux can be shown to depend on the $(\ell, m) = (2,2)$ symmetric and antisymmetric components as follows \cite{Mielke:2024kya}:
\begin{equation}
\begin{aligned}
    \dot{P}_{l} &\approx \lim_{r \to \infty} \frac{r^2}{6\pi} \, \dot{\phi}_{22}^{\rm sym} \, \dot{\phi}_{22}^{\rm asym} \, |h_{22}^{\rm sym}|\,|h_{22}^{\rm asym}| \\
    &\quad \times \cos\left(\phi_{22}^{\rm asym} - \phi_{22}^{\rm sym}\right),
\end{aligned}
\end{equation}
where $\phi_{22}^{\rm sym}$ and $\phi_{22}^{\rm asym}$ denote the phases of the symmetric and antisymmetric components. Other modes also contribute to the kick, but the dominant $(2,2)$ antisymmetric part typically accounts for the leading effect. When antisymmetric components are absent, the net linear momentum flux in this direction vanishes.

Traditionally, waveform models for spin-precessing binary systems have been constructed by exploiting the approximate separation between precessional and orbital timescales \cite{Buonanno:2002fy}. In this approach, the individual modes of the GW signal are modeled in the co-precessing frame (Eq.~\eqref{eq:coprec_frame}), where the dynamics of the system closely resemble that of a non-precessing (aligned-spin) binary. These co-precessing frame modes are typically built by adapting aligned-spin waveform models --- often calibrated to NR simulations --- to include the time-dependent evolution of the spin components. The full signal is then recovered in the inertial frame through a time-dependent rotation of the co-precessing waveform using the so-called “twisting-up” prescription \cite{Buonanno:2002fy,Boyle:2007ft,Schmidt:2010it,Schmidt:2012rh,OShaughnessy:2012iol}:
\begin{equation}
    \label{eq:rotation_coprec_I}
    h_{\ell m}^{\rm I}(t) = \mathcal{D}^{\ell}_{m', m}[\boldsymbol{R}(t)]\, h^{\rm coprec}_{\ell m'}(t),
\end{equation}
where $\mathcal{D}^{\ell}_{m', m}$ are Wigner D-matrix elements and $\boldsymbol{R}(t)$ denotes the time-dependent rotation from the co-precessing to the inertial frame, determined by the precession of the orbital angular momentum and the individual spin vectors. Waveform models constructed in this way inherit the equatorial symmetry from the underlying aligned-spin models. Therefore, all antisymmetric contributions (as defined in Eq.~\eqref{eq:formula_sym_asym}) to the waveform are neglected \cite{Ramos-Buades:2020noq}.

Only a few waveform models currently go beyond this approximation by explicitly incorporating the antisymmetric contributions to the co-precessing frame modes. Notably, the \texttt{NRSur7dq4} model includes these effects in all the co-precessing modes. Similarly, the \texttt{IMRPhenomXO4a} model introduces an analytic description of the dominant antisymmetric multipoles $(\ell = 2, m = \pm 2)$, calibrated to a set of single-spin precessing NR simulations from the BAM code. More recently, this model has been extended to \texttt{IMRPhenomXPNR} \cite{xpnr_inprep}, which improves the inspiral treatment of spin-precession \cite{Colleoni:2024knd} while retaining the same treatment of antisymmetries and merger-ringdown description as \texttt{IMRPhenomXO4a}.

In this work, we develop a complementary approach by incorporating antisymmetric contributions into the \texttt{SEOBNRv5PHM} model \cite{Ramos-Buades:2023ehm}, which is based on the EOB formalism. Throughout the paper, we refer to these contributions interchangeably as antisymmetric modes, asymmetric contributions, or equatorial-asymmetric effects --- terms which all refer to the same definition given in Eq.~\eqref{eq:formula_sym_asym}. 

\subsection{Modeling spin-precessing binaries with \texttt{SEOBNRv5PHM}}
\label{sec:SEOBNRv5PHM}

The EOB formalism \cite{Buonanno:1998gg,Buonanno:2000ef} provides a Hamiltonian framework for modeling the two-body dynamics of compact binaries. In this approach, the full EOB Hamiltonian $H_{\rm EOB}$ is constructed from an effective Hamiltonian $H_{\rm eff}$ describing a test particle of mass $\mu$ moving in a deformed Kerr spacetime of total mass $M$. The deformation, expressed in terms of the symmetric mass ratio $\nu$, is derived such that the EOB Hamiltonian matches the PN Hamiltonian order by order through a canonical transformation. An energy map relates the effective and EOB Hamiltonians:
\begin{equation}
H_{\rm EOB} = M \sqrt{1+2 \nu \left(\frac{H_{\rm eff}}{\mu}-1 \right)}.
\label{eq:energy_map}
\end{equation}

The dynamical variables in the EOB formalism are the relative position vector $\bm{r}$, its conjugate momentum $\bm{p}$, and the individual spin vectors $\bm{S}_{1,2}$. Their evolution is governed by the Hamilton equations:
\begin{equation}
    \label{eq:fullSpin}
    \begin{gathered}
    \dot{\bm{r}} = \frac{\partial H^{\rm prec}_{\rm EOB}}{\partial \bm{p}}, \qquad
    \dot{\bm{p}} = -\frac{\partial H^{\rm prec}_{\rm EOB}}{\partial \bm{r}} + \bm{\mathcal{F}}, \\
    \dot{\bm{S}}_{1,2} = \frac{\partial H^{\rm prec}_{\rm EOB}}{\partial \bm{S}_{1,2}} \times \bm{S}_{1,2},
    \end{gathered}
\end{equation}
where $H^{\rm prec}_{\rm EOB}$ includes spin-precession effects and $\bm{\mathcal{F}}$ is the radiation-reaction force. For the explicit construction and PN content (including all generic-spin information up to 4PN, and the non-spinning dynamics incorporated up to 4PN with partial 5PN results) of $H^{\rm prec}_{\rm EOB}$, see Ref.~\cite{Khalil:2023kep}.

In the quasi-circular, spin-precessing \texttt{SEOBNRv5PHM} model \cite{Ramos-Buades:2023ehm}, the full set of equations of motion of Eq.~\eqref{eq:fullSpin} is approximated by a set of planar equations solved in the co-precessing frame:
\begin{subequations}
    \label{eq:EOBEOMs}
    \begin{align}
    \dot{r}&=\xi(r) \frac{\partial H_{\rm EOB}^\text{pprec}}{\partial p_{r_*}},\\
    \dot{\phi}&=\frac{\partial H_{\rm EOB}^\text{pprec}}{\partial p_{\phi}},\\
    \dot{p}_{r_*}&=-\xi(r)\frac{\partial H_{\rm EOB}^\text{pprec}}{\partial r} +\mathcal{F}_{r},\\
    \dot{p}_{\phi}&=\mathcal{F}_{\phi},
    \end{align}
\end{subequations}
where $H_{\rm EOB}^\text{pprec}$ is a partially precessing Hamiltonian including orbit-averaged in-plane spin effects for circular orbits, and $\mathcal{F}_{\phi}$, $\mathcal{F}_{r}$ are the azimuthal and radial components for the radiation-reaction force. The evolution of the radial momentum is performed using the tortoise-coordinate $p_{r_*} = p_r\xi(r)$ with $\xi(r) = dr/dr_*$, which improves the stability during the plunge and merger \cite{Damour:2007xr,Pan:2009wj}. In the aligned-spin limit, $H_{\rm EOB}^\text{pprec}$ reduces to the Hamiltonian of \texttt{SEOBNRv5HM} \cite{Pompili:2023tna}. For full details, see Ref.~\cite{Ramos-Buades:2023ehm}.

These orbital equations are coupled with PN-expanded spin-precession equations:
\begin{subequations}
\label{eq:SLeqns}
\begin{align}
\dot{v}_{\rm PN} &= \left [\frac{\dot{E}(v_{\rm PN})}{dE(v_{\rm PN})/dv_{\rm PN}} \right ]_{\rm PN-expanded},\\
\dot{\bm S}_{\mr i} &= \bm{\Omega}_{S_{\mr i}}\times \bm{S}_{\mr i}, \\
\bm{L} &= \bm{L}(\LNhat,v_{\rm PN},\bm{S}_{\mr i}),\\
\dotLNhat &= \dotLNhat(\LNhat,v_{\rm PN},\bm{S}_{\mr i}),
\end{align}
\end{subequations}
where $v_{\rm PN}=\Omega_{\rm PN}^{1/3}$ is the PN-expansion velocity parameter of the system, $E(v_{\rm PN})$ is the PN binding energy of the system, $\dot{E}(v)$ is the circular-orbit PN-expanded energy flux, and $\bm{\Omega}_{S_{\mr i}}$ are the spin-precessing frequencies. These equations are obtained by PN-expanding and orbit-averaging the full Hamilton equations for the spin evolution up to 4PN order, including spin-orbit (SO) and spin-spin (SS) contributions to next-to-next-to-leading order (NNLO) (see Ref.~\cite{Khalil:2023kep} for the explicit expressions of these equations). In order to obtain the complete solution for all the dynamical variables, \texttt{SEOBNRv5PHM} first solves the PN-expanded spin-evolution equations Eq.~\eqref{eq:SLeqns}, before using $\dot{\phi}$ at each time-step to evaluate the spin vectors $\bm{S}_{\mr i}(\Omega_{\rm PN})$ within $H_{\rm EOB}^\text{pprec}$ and the flux components $\mathcal{F}_{\phi}$ and $\mathcal{F}_{r}$. 

With the full dynamical solution at hand, given by Eq.~\eqref{eq:EOBEOMs} and Eq.~\eqref{eq:SLeqns}, the co-precessing waveform modes $h_{\ell m}^{\rm coprec}(t)$ are constructed as a combination of inspiral-plunge and merger-ringdown contributions:
\begin{equation}
\label{eq:h_match}
h_{\ell m}^{\rm coprec}(t)= \begin{cases}h_{\ell m}^{\text {insp-plunge }}(t), & t<t_{\text {match }}^{\ell m} \\
h_{\ell m}^{\text {merger-RD }}(t), & t>t_{\text {match }}^ {\ell m }\end{cases},
\end{equation}
where $t_{\text{match}}^{\ell m}$ is typically the amplitude peak of the $(2,2)$ mode, except for $(5,5)$ where $t_{\text{match}}^{55} = t_{\text{peak}}^{22} - 10M$.

The inspiral-plunge modes are built using factorized-resumed PN expressions \cite{Damour:2007xr,Damour:2008gu,Pan:2010hz, Taracchini:2012ig,Taracchini:2013rva}:
\begin{equation}
    \label{eq:hlmFactorized}
    h^{\rm insp-plunge}_{\ell m}(t) = h_{\ell m}^{\mathrm{N}}\hat{S}_{\rm eff}f_{\ell m}T_{\ell m}e^{i\delta_{\ell m}}N_{\ell m},
\end{equation}
where $h_{\ell m}^{\mathrm{N}}$ is the Newtonian waveform, $\hat{S}_{\rm eff}$ is an effective source term, $f_{\ell m}$ and $e^{i\delta_{\ell m}}$ are a resummation of the PN amplitude and phase corrections, $T_{\ell m}$ is a resummation of the tail terms (see Ref.~\cite{Pompili:2023tna} for detailed expressions of these quantities), and $N_{\ell m}$ are non-quasi-circular (NQC) corrections used to improve the accuracy in the last cycles by matching with NR and test-body-limit waveforms, given in terms of the EOB dynamical variables as 
\begin{equation}
    \label{eq:nqcsym}
    \begin{gathered}
        N_{\ell m} =\left[1+\frac{\hat{p}_{r_*}^2}{(r \Omega)^2}\left(a_1^{h_{\ell m}}+\frac{a_2^{h_{\ell m}}}{\hat{r}}+\frac{a_3^{h_{\ell m}}}{\hat{r}^{3 / 2}}\right)\right] \\
        \quad \times \exp \left[i\left(b_1^{h_{\ell m}} \frac{\hat{p}_{r_*}}{r \Omega}+b_2^{h_{\ell m}} \frac{\hat{p}_{r_*}^3}{r \Omega}\right)\right].
     \end{gathered}
\end{equation}

The five coefficients $a_i^{h_{\ell m}}, b_i^{h_{\ell m}}$ are fixed by matching to NR and BH-perturbation waveforms at $t^{\ell m}_{\rm match}$ using the conditions~\cite{Taracchini:2013rva,Bohe:2016gbl,Cotesta:2018fcv}:
\begin{subequations}
    \label{eq:nqcivmatching}
    \begin{align}
        \left| h_{\ell m}^{\textrm{insp-plunge}} \right| &= \left|h_{\ell m}^\textrm{NR}\right|,\\
        \frac{d}{dt}\left| h_{\ell m}^{\textrm{insp-plunge}} \right| &= \frac{d}{dt}\left| h_{\ell m}^{\textrm{NR}} \right|,\\
        \frac{d^2}{dt^2}\left| h_{\ell m}^{\textrm{insp-plunge}} \right| &= \frac{d^2}{dt^2}\left| h_{\ell m}^{\textrm{NR}} \right|,\\
        \omega_{\ell m}^{\textrm{insp-plunge}} &= \omega_{\ell m}^{\textrm{NR}},\\
        \dot{\omega}_{\ell m}^{\textrm{insp-plunge}} &= \dot{\omega}_{\ell m}^{\textrm{NR}},
    \end{align}
\end{subequations}
where the right-hand-side of Eqs.~\eqref{eq:nqcivmatching} are called \emph{input values} and are fitted across the aligned-spin parameter space using SXS and perturbation-theory waveforms \cite{Pompili:2023tna}.

The merger-ringdown waveform is modeled using a phenomenological model (inspired by the work of \cite{Damour:2014yha})
\begin{equation}
    \label{eq:mrd_waveform}
    h_{\ell m}^{\text {merger-RD }}(t)=\nu \tilde{A}_{\ell m}(t) e^{i \tilde{\phi}_{\ell m}(t)} e^{-i \sigma_{\ell m 0}\left(t-t_{\text {match }}^{\ell m}\right)},
\end{equation}
with quasi-normal mode (QNM) frequency $\sigma_{\ell m 0} = \sigma_{\ell m}^\mathrm{R} - i \sigma_{\ell m}^\mathrm{I}$ and phenomenological models for amplitude and phase:
\begin{subequations}
\label{eq:ansatz_hlm_mrd}
\begin{align}
    \tilde{A}_{\ell m}(t)&=c_{1, c}^{\ell m} \tanh \left[c_{1, f}^{\ell m}\left(t-t_{\text {match }}^{\ell m}\right)+c_{2, f}^{\ell m}\right]+c_{2, c}^{\ell m},\\
    \tilde{\phi}_{\ell m}(t)&=\phi_{\text {match }}^{\ell m}-d_{1, c}^{\ell m} \log \left[\frac{1+d_{2, f}^{\ell m} e^{-d_{1, f}^{\ell  m}(t-t_{\text {match}}^{\ell_{m}})}}{1+d_{2, f}^{\ell m}}\right].
\end{align}
\end{subequations}
where the coefficients $c_{1, c}^{\ell m}$, $c_{2, c}^{\ell m}$ and $d_{1, c}^{\ell m}$ are fixed by imposing continuity and differentiability of the full waveform at the matching time, and the coefficients $c_{1, f}^{\ell m}$, $c_{2, f}^{\ell m}$, $d_{2, f}^{\ell m}$ and $d_{2, f}^{\ell m}$ are calibrated to NR and test-particle waveforms (see Ref.~\cite{Pompili:2023tna} for a detailed description of the fitting procedure and the dataset employed). Following Ref.~\cite{Hamilton:2023znn}, the QNM frequency $\sigma_{\ell m 0}$ is shifted to account from the frame transformation between the remnant frame and the co-precessing frame used for modeling.

Finally, the inertial-frame polarizations are constructed by rotating the co-precessing modes $h_{\ell m}^{\rm coprec}(t)$ to a frame consistent with LVK conventions \cite{Schmidt:2017btt}. First, having obtained $\LNhat(t)$ from Eq.~\eqref{eq:SLeqns}, a time-dependent rotation from the co-precessing frame (P-frame) to an inertial frame aligned with the total angular momentum $J$ (J-frame) is determined, using Euler angles:
\begin{subequations}
    \label{eq:euler_angles}
    \begin{align}
    \alpha^{\rm \J2P}(t)&=\arctan(l_{N,y}(t),l_{N,x}(t)),\\
    \cos\beta^{\rm \J2P}(t)&=l_{N,z},\\
    \dot{\gamma}^{\rm \J2P}(t)&=\dot{\alpha}^{\rm \J2P}(t)\cos\beta^{\rm \J2P}(t),
    \end{align}
\end{subequations}
where the $\LNhat$ components are given in the $J$-frame. A global rotation then maps from the $J$-aligned frame to the LVK inertial frame (I-frame), and the resulting composite rotation Euler angles are employed to rotate the spin-weighted spherical harmonic basis and construct the polarizations
\begin{equation}
\begin{gathered}
    \label{eq:direct_hpc}
    h^{\rm I}(t)=e^{2i\alpha^{\rm \I2P}(t)}\sum_{\ell, m} {}_{-2} Y_{\ell m}(\gamma^{\rm \I2P}(t),\beta^{\rm \I2P}(t))\,h^{\rm coprec}_{\ell m}(t).
\end{gathered}
\end{equation}
Equivalently, one can also rotate the waveform modes to the LVK frame using Eq.~\eqref{eq:rotation_coprec_I} and construct the polarizations from them.

\begin{figure*}[htb!]
    \includegraphics[width=\textwidth]{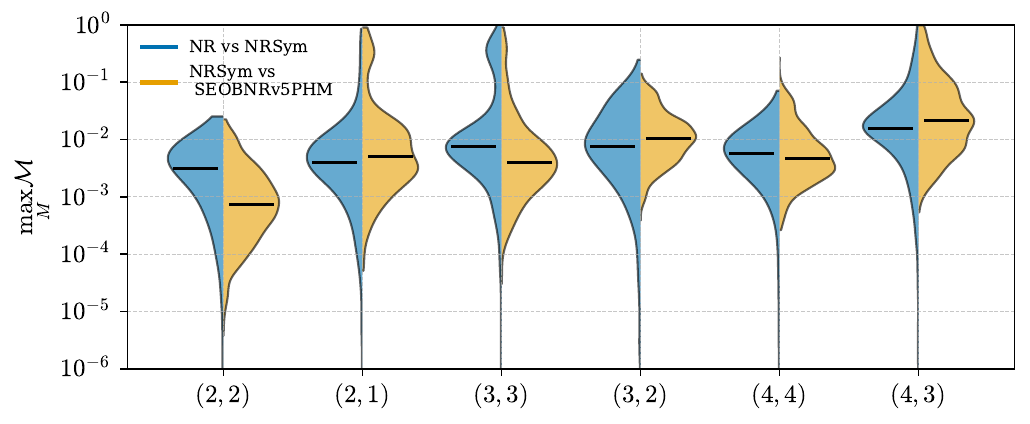}
    \caption{Distribution of the unfaithfulness between symmetrized spin-precessing NR waveforms (\texttt{NRSym}) from the SXS catalog and two alternatives: the original NR waveforms (with antisymmetric components) (NR, blue), and the \texttt{SEOBNRv5PHM} model without antisymmetric modes (orange). The unfaithfulness is computed mode by mode. 
    }
    \label{fig:unfaith_asym_nr_sym}
\end{figure*}

\subsection{Assessing the impact of neglecting equatorial asymmetry}

As previously mentioned, the co-precessing waveform modes $h^{\rm coprec}_{\ell m}$ in \texttt{SEOBNRv5PHM} are constructed from the aligned-spin model \texttt{SEOBNRv5HM}. As a result, these modes depend only on the spin projections along the Newtonian orbital angular momentum direction, $\chi_{i,l_N}$, and they are symmetric under the transformation defined in Eq.~\eqref{eq:equatorial_symmetry}. Consequently, the spin-precessing waveforms constructed from \texttt{SEOBNRv5PHM} do not include equatorial asymmetric contributions by construction. Before presenting our modeling of these contributions, we assess the potential accuracy loss that results from neglecting them.

To this end, we compute the unfaithfulness (also referred to as the mismatch; see Sec.~\ref{sec:unfaith} for a detailed definition) between two sets of co-precessing modes: (i) the full NR modes and their symmetrized counterparts (\texttt{NRSym}), obtained via Eq.~\eqref{eq:formula_sym_asym}; and (ii) the symmetrized NR modes and the co-precessing modes of \texttt{SEOBNRv5PHM}. In both cases, we optimize only over global time and phase shifts. The goal is to determine whether the omission of antisymmetric contributions constitutes the dominant source of modeling error in the current version of \texttt{SEOBNRv5PHM}.

The unfaithfulness is computed using the \texttt{aLIGOHighPowerZeroDetuned} power spectral density (PSD)~\cite{Barsotti:2018hvm}, varying the total binary mass between $10\,M_{\odot}$ and $300\,M_{\odot}$ to assess accuracy across the detector's sensitive band.

Figure~\ref{fig:unfaith_asym_nr_sym} presents the cumulative unfaithfulness distributions for each multipole mode. For the $(2,2)$ and $(3,3)$ modes, we find that the median unfaithfulness is lower in the comparison between \texttt{SEOBNRv5PHM} and \texttt{NRSym} than in the comparison between full NR and \texttt{NRSym}, indicating that the missing antisymmetric contribution constitutes a significant part of the residual error. This trend is even more pronounced in the maximum unfaithfulness, particularly for the $(2,2)$ mode. For the $(4,4)$ mode, the median unfaithfulness is comparable between both comparisons, but the maximum value is substantially higher in the NR vs \texttt{NRSym} case.

These results suggest that the absence of antisymmetric contributions is currently a limiting factor in the accuracy of \texttt{SEOBNRv5PHM}, especially for the $(2,2)$ mode, and to a lesser extent the $(3,3)$ mode. We note that the tail of high unfaithfulness in the NR vs \texttt{NRSym} comparison for the $(3,3)$ mode predominantly corresponds to systems with near-equal masses. In such cases, the symmetric contribution to the $(3,3)$ mode is strongly suppressed, and the antisymmetric component dominates the signal.

In contrast, for the $\ell \neq m$ modes, the unfaithfulness between \texttt{SEOBNRv5PHM} and \texttt{NRSym} is higher than that between NR and \texttt{NRSym}, implying that other modeling inaccuracies dominate for those modes. We conclude that, for these subdominant multipoles, the omission of antisymmetric contributions is not the main source of modeling error.

These findings motivate the development of an explicit model for the antisymmetric modes, which we present in the next sections.

\subsection{Inspiral-plunge antisymmetric modes}
\label{sec:insp_plunge_asym}

The antisymmetric contributions to the waveform modes arise from spin components in the orbital plane (in-plane spins). In PN theory, such contributions have been computed at 1.5PN order in Ref.~\cite{Arun:2008kb} and extended to 2PN in Ref.~\cite{Boyle:2014ioa}, where the waveform modes are explicitly decomposed into symmetric and antisymmetric components as defined by Eq.~\eqref{eq:formula_sym_asym}.

Writing the co-precessing PN waveform modes as the sum of symmetric and antisymmetric components,
\begin{equation}
    h_{\ell m}^{\rm PN}(t)= h_{\ell m}^{\rm PN, sym}(t) + h_{\ell m}^{\rm PN, asym}(t),
\end{equation}
we can express the antisymmetric contribution in the form:
\begin{equation}
    \label{eq:asym_mode_pn}
    h_{\ell m}^{\rm asym,PN}=\frac{2\nu M v^2}{R}\sqrt{\frac{16\pi}{5}}\hat{h}_{\ell m}^{\rm asym}e^{-im\phi},
\end{equation}
where $\hat{h}_{\ell m}^{\rm asym}$ are complex amplitudes listed in Ref.~\cite{Boyle:2014ioa} for the relevant modes in \texttt{SEOBNRv5PHM}. The dominant $(2,2)$ and $(2,1)$ modes are known to NLO:
\begin{subequations}
\begin{align}
\hat{h}_{22}^{\rm asym} &= -\frac{v^2(\Sigma_{\lambda}+i\Sigma_{n})}{2M^2} + \frac{v^4}{84M^2}[182i\delta S_n + 19\delta S_{\lambda} \nonumber \\
&\quad + 14i(7-20\nu)\Sigma_n + (5-43\nu)\Sigma_{\lambda}],\\
\hat{h}_{21}^{\rm asym} &= -\frac{v^3(25S_n + 4iS_{\lambda} + 13\delta\Sigma_{n}+4i\Sigma_{\lambda})}{6M^2} \nonumber \\
&\quad +\frac{3v^4}{2M^4\nu}(S_{1,l}S_{2,n}+S_{1,n}S_{2,l}).
\end{align}
\end{subequations}
For the other modes, only leading-order expressions are available:
\begin{subequations}
\begin{align}
\hat{h}_{33}^{\rm asym} &= -\sqrt{\frac{10}{21}}\frac{v^3}{M^2}[S_n-iS_{\lambda}+\delta(\Sigma_n-i\Sigma_{\lambda})],\\
\hat{h}_{44}^{\rm asym} &= \sqrt{\frac{5}{7}}\frac{9v^4}{8M^2}[\delta(S_{\lambda}+iS_n)+(1-3\nu)(\Sigma_{\lambda}+i\Sigma_n)],\\
\hat{h}_{32}^{\rm asym} &= \sqrt{\frac{5}{7}}\frac{v^4}{24M^2}[25\delta(S_{\lambda}-4iS_n) - 4i(13-55\nu)\Sigma_n \nonumber \\
&\quad +(17-83\nu)\Sigma_{\lambda}],
\end{align}
\end{subequations}
with no antisymmetric contributions arising at 2 PN order for the $(4,3)$ and $(5,5)$ modes. 

\begin{figure*}[htpb!]
    \includegraphics[width=\textwidth]{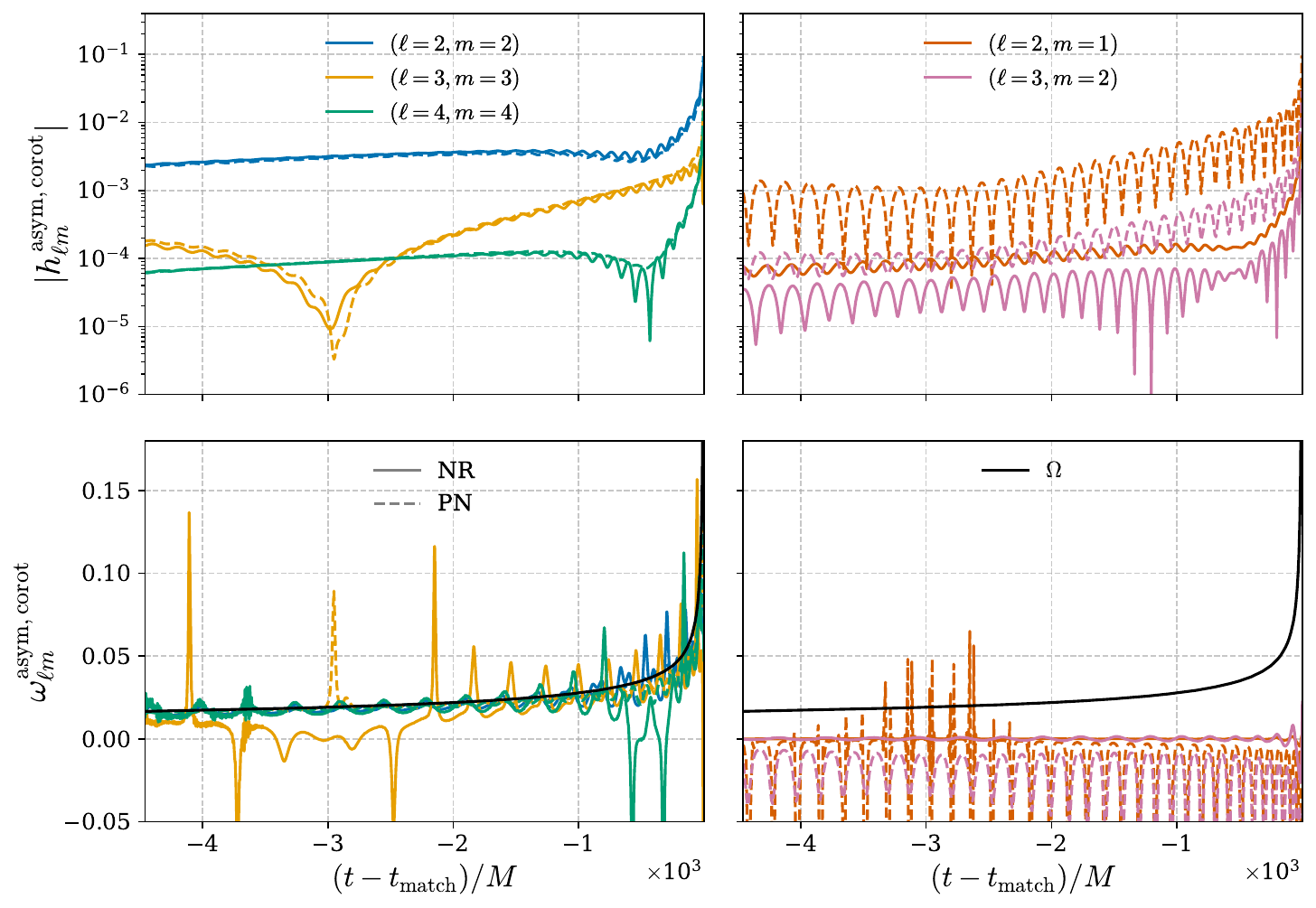}
    \caption{Comparison of PN and NR antisymmetric modes for an example case with parameters $q=2$, $\chi_{1,\perp}=\chi_{2,\perp}=0.85$ (\texttt{SXS:BBH:1207}).
    Top row: amplitude of the antisymmetric modes. Bottom row: wave-frequency of the antisymmetric modes in the co-rotating frame. Solid colored lines denoted NR results, and dashed lines PN results. Solid black line in the bottom panel denotes the PN orbital frequency.}
    \label{fig:inspiral_nr_pn}
\end{figure*}

At leading PN order, the $\ell=m$ antisymmetric contributions depend on the combinations $S_n-iS_\lambda$ and $\Sigma_n-i\Sigma_\lambda$, and thus grow with the in-plane spin magnitude. These amplitudes oscillate approximately with the orbital frequency (due to the evolution of the triad vectors $\boldsymbol{n}$ and $\boldsymbol{\lambda}$) plus the spin-precession frequency (due to the evolution of the orientation of the spins in the precessing time-scale),
\begin{equation}
    \label{eq:omega_asym_inspiral}
    \omega^{\rm asym,corot}_{\ell ,m=\ell}\approx\Omega+\dot{\alpha},
\end{equation}
where $\alpha$ denotes the precession angle (related to the Euler angles of Eq.~\eqref{eq:euler_angles}). In the co-precessing frame, this yields:
\begin{equation}
    \label{eq:omega_asym_inspiral_coprec}
    \omega^{\rm asym,coprec}_{\ell ,m=\ell}\approx\dot{\alpha} - (m-1)\Omega_{\rm orb},
\end{equation}
which is in agreement with the findings of Ref.~\cite{Ghosh:2023mhc} for the dominant $\ell=m=2$ antisymmetric contribution. Nevertheless, for $\ell\neq m$ modes, the explicit PN expressions show a more complicated structure at leading-order, with different coefficient values for the $\boldsymbol{n}$-projection and $\boldsymbol{\lambda}$-projection of the spin combinations $\boldsymbol{S}$ and $\boldsymbol{\Sigma}$. This translates into high frequency oscillations both in amplitude and frequency.

We confirm this behavior in SXS NR simulations, as shown in Fig.~\ref{fig:inspiral_nr_pn} for an example case (\texttt{SXS:BBH:1207}). For $\ell=m$ modes, PN and NR results qualitatively agree, especially in amplitude modulation and frequency behavior. In contrast, for $\ell\neq m$ modes like $(2,1)$ and $(3,2)$, PN predictions fail to capture the structure observed in NR, and the morphology of the amplitude and frequency in the NR modes is not reproduced by the PN expressions. The NR $(2,1)$ and $(3,2)$ modes seem to be actually non-oscillatory in the co-rotating frame, while PN predicts a negative frequency secular trend. For the amplitudes of these two modes, both NR and PN show high frequency oscillations, but the average value of the amplitudes and the precessing time-scale morphology are different.

Given this discrepancy and the lesser impact of antisymmetric contributions in $\ell\neq m$ modes (see Fig.~\ref{fig:unfaith_asym_nr_sym}), we restrict antisymmetric contributions to \SEOBASYM{} to the $(2,2)$, $(3,3)$, and $(4,4)$ modes. We evaluate their PN complex amplitudes $\hat{h}_{\ell m}^{\rm asym}$ using the spin dynamics solutions of Eq.~\eqref{eq:SLeqns} until the matching time $t_{\rm match}=t_{\rm peak}^{22}$.

We apply a phase-only non-quasi-circular (NQC) correction of the form
\begin{equation}
N_{\ell m}^{\rm asym} = \exp \left[i\left(b_{1,\ell m} \frac{\hat{p}_{r_*}}{r \Omega}\right)\right],
\end{equation}
and fix $b_1^{h_{\ell m}^{\rm asym}}$ to enforce frequency agreement with a calibrated input value at $t_{\rm match}$:
\begin{equation}
    b_{1,\ell m} = \dfrac{\dot{\phi}_{\ell m}^{\rm asym,NR}(t_{\rm match})-\dot{\phi}_{\ell m}^{\rm asym,PN}(t_{\rm match})}{-\frac{d}{dt}\Big(\frac{\hat{p}_{r_*}}{r\Omega}\Big)|_{t=t_{\rm match}}}.
\end{equation}

To correct the amplitude of $h_{22}^{\rm asym}$ near merger, we apply a phenomenological correction, inspired by the findings of Ref.~\cite{Ghosh:2023mhc}
\begin{equation}
A_{22}^{\rm asym,corr} \equiv (1 + b_6 v^6),
\end{equation}
where $b_6$ is calibrated by matching the NR amplitude at $t_{\rm match}$:
\begin{equation}
    b_{6} = \frac{|h_{22}^{\rm asym,NR}(t_{\rm match})|-|h_{22}^{\rm asym,PN}(t_{\rm match})|}{v^6|h_{22}^{\rm asym,PN}(t_{\rm match})|}.
\end{equation}
For the subdominant modes, we find that imposing an amplitude correction leads, in general, to waveform instabilities, while not affecting significantly the accuracy of the reconstruction, and therefore we decide to not apply any correction.

The final inspiral-plunge expressions read:
\begin{align}
    h_{22}^{\rm asym,insp-plunge} &= A_{22}^{\rm asym,corr} N_{22}^{\rm asym} h_{22}^{\rm asym,PN}, \\
    h_{\ell m}^{\rm asym,insp-plunge} &= N_{\ell m}^{\rm asym} h_{\ell m}^{\rm asym,PN}, \quad \ell=m=3,4.
\end{align}
and, as mentioned, depends on two free parameters for the dominant mode ($b_{1,22}$ and $b_6$), and one extra free phase parameter for each of the subdominant modes ($b_{1,33}$, $b_{1,44}$).

\begin{figure}[h]
    \includegraphics[width=\columnwidth]{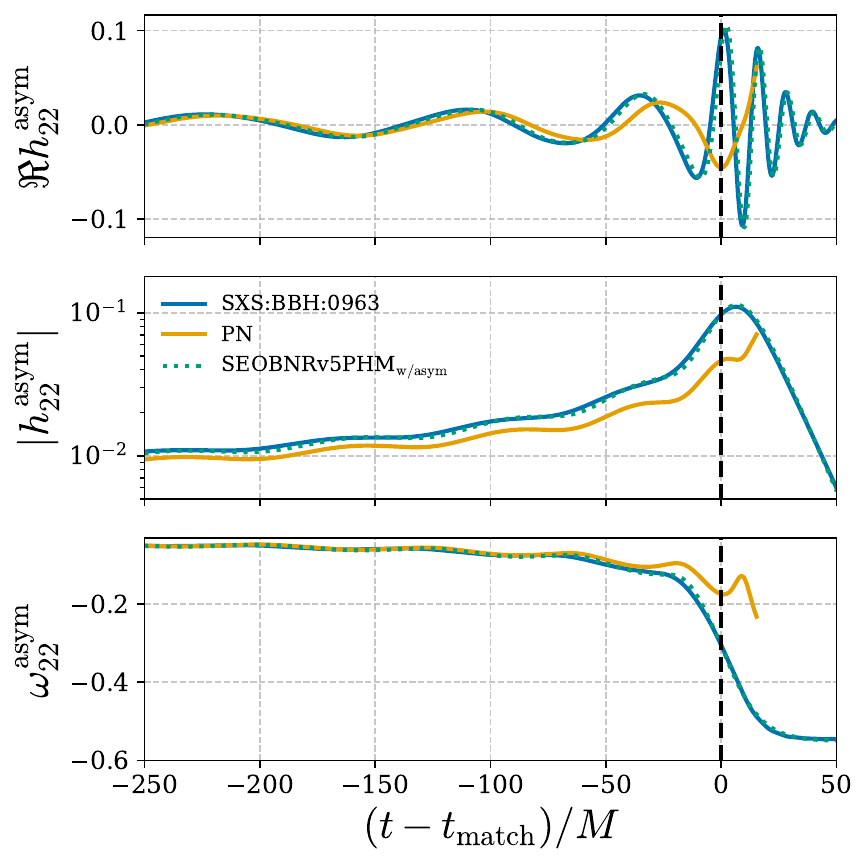}
    \caption{Waveform reconstruction comparison for an example NR simulation \texttt{SXS:BBH:0963} (with parameters $q=1$,$\chi_{1,\perp}=\chi_{2,\perp}=0.8$). Top: real part of the co-precessing antisymmetric (2,2) mode; middle: absolute value of the co-precessing antisymmetric (2,2) mode; bottom: phase derivative of the co-precessing antisymmetric (2,2) mode. The NR simulation is shown in blue. The orange curve corresponds to the evaluation of $h_{22}^{\rm asym,PN}$, while the dotted green curve has an additional amplitude correction factor and non-quasi-circular correction for the phase. The attachment time is shown as a vertical dashed black line, and after that time the green curve corresponds to the merger-ringdown model described in Sec.\ref{sec:mrd_antisym}.}
    \label{fig:wv_recons}
\end{figure}

Figure~\ref{fig:wv_recons} illustrates these corrections for an example NR waveform (\texttt{SXS:BBH:0963}), demonstrating the improved agreement in both amplitude and frequency upon applying the NQC and amplitude correction factors.

\subsection{Merger-ringdown antisymmetric modes}
\label{sec:mrd_antisym}

The antisymmetric contributions are generally stronger in the merger-ringdown regime, with the emission in this region determining most of the gravitational recoil velocity. Although we lack an analytical description of these contributions in this regime, we expect the late ringdown emission to be dominated by corresponding complex QNM frequencies $\sigma_{n=0,\ell m}$, as in the case of the symmetric contributions.

\begin{figure}[h]
    \includegraphics[width=\columnwidth]{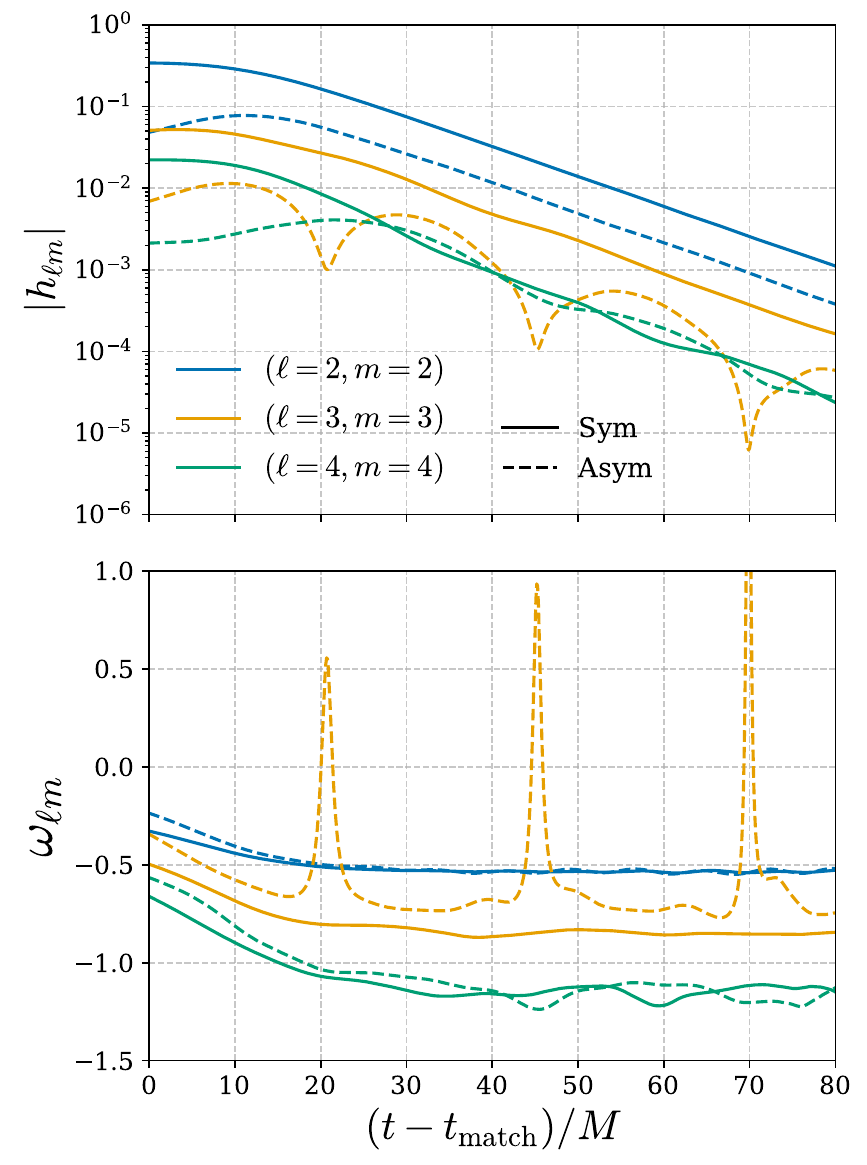}
    \caption{Comparison of symmetric and antisymmetric $\ell=m$ (with $\ell\leq 4$) modes for the NR simulation \texttt{SXS:BBH:1216} (with parameters $q=2$, $\chi_{1,\perp}=0.85$, $\chi_{\rm{eff}}=-0.25$). Top panel: amplitude of the modes. Bottom panel: wave-frequency of the modes. Solid lines denote symmetric modes, and dashed lines antisymmetric modes.}
    \label{fig:mrd_sym_asym}
\end{figure}

In Fig.~\ref{fig:mrd_sym_asym} we compare the amplitude and mode frequency for the symmetric and antisymmetric parts of the $(2,2)$, $(3,3)$ and $(4,4)$ for an example NR simulation, although we observe in general a similar behavior across parameter space. Typically, the antisymmetric contributions have a maximum amplitude peak after the peak of the symmetric dominant mode $(2,2)$, and the antisymmetric contributions for $(2,2)$ and $(4,4)$ show the same late time behavior as the corresponding symmetric contributions -- they reach the same asymptotic frequency and the decay time is comparable -- validating the expectation of being described by the same set of complex QNM frequencies. 

Nevertheless, we observe a different phenomenology in the antisymmetric $(3,3)$ mode, which exhibits post-merger oscillations reminiscent of mode-mixing \cite{Buonanno:2006ui, Kelly:2012nd}. This behavior is related to the mismatch between the \textit{spherical} harmonic basis used for extraction in NR simulations, and the \textit{spheroidal} harmonics adapted to the perturbation theory of Kerr BHs. We have inspected the frequency content in the merger-ringdown of the $(3,3)$ antisymmetric mode but were unable to relate it to frequencies of other $m=3$ modes, and we leave for future work a better understanding of this effect. 

We decide to apply the same phenomenological ansatz as for the symmetric co-precessing modes in \texttt{SEOBNRv5PHM} given by Eq.~\eqref{eq:mrd_waveform} and Eq.~\eqref{eq:ansatz_hlm_mrd}. This is motivated by the similar phenomenology observed for the $(2,2)$ and $(4,4)$ modes and the aim that the $(3,3)$ mode, despite showing a more complicated structure, could be reasonably modeled using a simpler description.

For the dominant $(2,2)$ mode, one difference between the symmetric and antisymmetric parts is that the former peaks at the matching time, by construction, and the latter peaks later. Therefore, in order to reconstruct the peak structure of the antisymmetric mode with the merger-ringdown ansatz, the amplitude free coefficients of Eq.~\eqref{eq:ansatz_hlm_mrd} have to be recalibrated. Nevertheless, the reconstruction of the phase is qualitatively similar, transitioning from the phase at the matching time (which is different for symmetric and antisymmetric parts) to the same asymptotic frequency. Therefore, we decide to fix the phase free coefficients with the same values as for the symmetric part. We have checked that this does not impact the accuracy of the waveform construction significantly, and it simplifies the behavior of the model across parameter space. 

The resulting model for the $(2,2)$ merger-ringdown antisymmetric waveform is 
\begin{equation}
    \label{eq:mrd_waveform_asym}
    h_{22}^{\text {asym,merger-RD }}(t)=\nu \tilde{A}^{\rm asym}_{22}(t) e^{i \tilde{\phi}_{22}^{\rm asym}(t)} e^{-i \sigma_{220}\left(t-t_{\text {match }}^{22}\right)},
\end{equation}
with
\begin{subequations}
    \label{eq:ansatz_hlm_mrd_asym}
    \begin{align}
        \tilde{A}^{\rm asym}_{22}(t)&=c_{1, c}^{22,\rm asym} \tanh \left[c_{1, f}^{22,\rm asym}\left(t-t_{\text {match }}\right)+c_{2, f}^{22,\rm asym}\right] \nonumber \\&+c_{2, c}^{22,\rm asym},\\
        \tilde{\phi}^{\rm asym}_{22}(t)&=\phi_{\text {match }}^{22,\rm asym} \nonumber \\
        &-d_{1, c}^{22,\rm asym} \log \left[\frac{1+d_{2, f}^{22,\rm asym} e^{-d_{1, f}^{22,\rm asym}\left(t-t_{\text {match}}\right)}}{1+d_{2, f}^{22,\rm asym}}\right].
    \end{align}
\end{subequations}
Here, the constrained coefficients $c_{i,c}^{22,\rm asym}$ and $d_{i,c}^{22,\rm asym}$ take different values for the antisymmetric case, since the inspiral-plunge waveform that is connected is different. The amplitude free coefficients $c_{i, f}^{22,\rm asym}$ have to be recalibrated for the antisymmetric case, and the phase free coefficients $d_{i, f}^{22,\rm asym}$ are fixed with the value of the symmetric case.

For the subdominant $(3,3)$ and $(4,4)$ modes, the amplitude free coefficients of the symmetric case already account for the amplitude peak in the merger-ringdown region, and therefore we decide to not recalibrate these free parameters. We emphasize that the constrained parameters are still different, and therefore the shape of the early ringdown region, including the position and value of the amplitude peak, is different for the antisymmetric modes despite having the same value for the free coefficients. This simplifies the behavior of the model, avoiding construction pathologies, while not affecting significantly the accuracy of the reconstructed waveforms.
\section{Calibrating antisymmetric modes with Numerical Relativity and test-body simulations}\label{sec:fitting}

The antisymmetric modes model in \SEOBASYM requires six quantities to be obtained from numerical data (NR and test-body-limit simulations)
\begin{equation}\nonumber
    \left\{
    \begin{aligned}
        &|h_{22}^{\rm asym,NR}(t_{\rm match})|, \\
        &\dot{\phi}_{\ell m}^{\rm asym,NR}(t_{\rm match}),\quad 2\leq \ell=m\leq 4 \\
        &c_{1, f}^{22,\rm asym},\\
        & c_{2, f}^{22,\rm asym}.
    \end{aligned}
    \right.
\end{equation}

This is a classical regression problem in which estimators $\hat{\bm{y}}$ for the target quantities $\bm{y}$ have to be obtained as functions of some input quantities $\bm{\Theta}$
\begin{equation}
    y_i\approx\hat{y}_i(\bm{\Theta}),
\end{equation}
in this case $\bm{\Theta}$ being some parameterization of the seven-dimensional parameter space of spin-precessing binaries.

Fitting model quantities across the 7-dimensional parameter space of spin-precessing BBH is substantially more complex than the traditional fitting in the aligned-spin parameter space, which most models nowadays routinely incorporate \cite{Jimenez-Forteza:2016oae,Bohe:2016gbl,Pompili:2023tna}. Nevertheless, there are examples in the literature of spin-precessing calibration in waveform models. In \texttt{IMRPhenomXO4a} \cite{Hamilton:2021pkf,Thompson:2023ase}, the symmetric co-precessing modes, the dominant antisymmetric mode \cite{Ghosh:2023mhc}, and the rotation angles from co-precessing to inertial frame are calibrated using single-spin precessing simulations, and also neglecting the global spin orientation. The number of degrees of freedom is therefore reduced from seven to three, and the traditional methods of aligned-spin calibration can be employed. 

The NR surrogate model \texttt{NRSur7dq4} \cite{Varma:2019csw} deals with the full 7-dimensional problem, with co-precessing waveform modes (symmetric and antisymmetric) and spin-dynamics being modeled using the \textit{empirical time interpolation} method \cite{Field:2013cfa, Yvon2009, Chaturantabut2010}. The data pieces are decomposed in a set of time nodes and these quantities are fitted for each node across parameter space using a \textit{forward-stepwise greedy fit} algorithm \cite{Blackman:2017dfb}, a class of sparse polynomial regression with feature selection. Additionally, the remnant quantities and ringdown amplitudes of spin-precessing binaries have also been fitted using Gaussian process regression (GPR) \cite{Varma:2018aht,Nobili:2025ydt}.

We decide to a employ a sparse polynomial regressor based on orthogonal matching pursuit (OMP) \cite{Mallat1993,Rubinstein2008}, a method similar to the \textit{forward-stepwise greedy fit} employed for \texttt{NRSur7dq4}. This section details the waveform datasets, the extraction of training targets, the choice of input features, and the OMP procedure.

\subsection{Waveform datasets}
\label{sec:datasets}

We combine $1523$ precessing SXS simulations ($1405$ from the second SXS catalog~\cite{Boyle:2019kee}\footnote{The SXS waveforms considered here do not contain gravitational-wave memory effects, recently added to the third release of the SXS catalog~\cite{Scheel:2025jct}} and $118$ from Ref.~\cite{Ossokine:2020kjp}) with $5872$ test-body waveforms computed with a time-domain Teukolsky solver along geodesic plunges (Refs.~\cite{Apte:2019txp,Lim:2019xrb}) for single-spin precessing systems. 

The SXS sample has most of its coverage in the region $q\leq 4$ with spin magnitudes up to $0.8$, with a limited number of simulations outside this region; the test--particle set covers $0\le\Sigma_\perp\le1$ and a dense grid in the in--plane spin angle $\phi_\Sigma$.

The inclusion of the test-body body dataset helps to inform the fits in the $\nu\rightarrow 0$ limit, improving therefore the extrapolation behavior. Additionally, it contains a good coverage of the in-plane spin direction in the single-spin limit, which helps to inform the fits on this degree of freedom.

To reduce numerical noise, we discard SXS cases whose highest-- and second--highest--resolution results disagree by more than the $85^{\mathrm{th}}$ percentile in any target quantity. For the test-body set we remove trajectories with orbital inclination $\pi/2\pm0.3$\,rad, where we observe waveform pathologies in the co-precessing frame, complicating the extraction of input values and the fit of the merger-ringdown coefficients. We additionally perform outlier removal based on inspection of the distributions for each target quantities.

\subsection{Target quantities}
\label{sec:targets}

We transform the target quantities to improve the quality of the fits. For the amplitude input value, we factor out the dependence on $\Sigma_{\rm in-plane}$ that comes from the leading-order PN expression, since this ensures that the quantities reconstructed from the fit smoothly go to zero in the limit $\Sigma_{\rm in-plane}\rightarrow 0$. We also factor out a linear dependence on the symmetric mass-ratio coming from the common Newtonian prefactor. For the wave frequency input value, we subtract the orbital frequency, approximated as half the wave-frequency of the symmetric $(2,2)$ mode, since we expect the frequency of the antisymmetric mode to be close to this frequency with some correction.

The final set of target quantities we will fit for the dominant antisymmetric mode is then
\begin{subequations}
    \label{eq:target_y_22}
\begin{align}
    y_1 &=\frac{1}{\nu \Sigma_{\rm in-plane}}|h^{\rm asym,NR}_{22}|(t_{\rm match}),\\
    y_2 &=\dot{\phi}^{\rm asym,NR}_{22}(t_{\rm match})-\Omega^{\rm NR}(t_{\rm match}),\\
    y_3 &=\Omega^{\rm NR}(t_{\rm match}),\\
    y_4 &=c^{\rm asym,22}_{1,f},\\
    y_5 &=c^{\rm asym,22}_{2,f},
\end{align}
\end{subequations}
where $\Omega^{\rm NR}(t_{\rm attach})$ is the orbital frequency from the numerical data at the matching time (that we approximate as $\Omega^{\rm NR}\equiv\tfrac12\dot{\phi}^{\mathrm{sym,NR}}_{22}$). From these quantities, we can reconstruct the needed input values for the model (and the merger-ringdown coefficients which do not require any additional transformation).

For the subdominant $(3,3)$ and $(4,4)$ modes we just fit the corresponding frequency input value, without additional transformations
\begin{subequations}
    \label{eq:target_y_lm}
\begin{align}
    y_6 &=\dot{\phi}^{\rm asym,NR}_{33}(t_{\rm match}),\\
    y_7 &=\dot{\phi}^{\rm asym,NR}_{44}(t_{\rm match}).
\end{align}
\end{subequations}

We process the waveforms in the combined dataset in order to obtain the co-precessing frame waveforms modes, before applying Eq.~\eqref{eq:formula_sym_asym} to obtain the symmetric and antisymmetric parts for the co-precessing $(\ell,m)$ modes. In practice, we employ the python package \texttt{scri} \cite{mike_boyle_2020_4041972, Boyle:2013nka, Boyle:2015nqa} to perform these operations. From this, we read the input values needed for constructing the first three target quantities of the dominant mode from Eq.~\eqref{eq:target_y_22} and the two target quantities for the subdominant modes from Eq.~\eqref{eq:target_y_lm}.

The target quantities for the merger-ringdown coefficients $y_4$ and $y_5$ are extracted for each case using a similar procedure as in \cite{Pompili:2023tna}, where the amplitude and phase ansatz in Eq.~\eqref{eq:ansatz_hlm_mrd_asym} are fitted independently using the co-precessing waveform modes from the matching time to around $80M$ after the matching time. As discussed in Sec.~\ref{sec:SEOBNRv5PHM}, since in this case we deal with spin-precessing systems, we apply a frequency shift to the QNM frequencies for evaluating the ansatz, in order to account for the frame transformation between the inertial frame where the perturbation theory QNM frequencies are computed, and the co-precessing frame that is employed for modeling (see Ref.~\cite{Hamilton:2023qkv} for a detailed discussion on this).

\subsection{Input features}
\label{sec:spin-param}

We approximate the target quantities $y_i$ with functions $\hat{y}(\bm{\Theta})$ of some parameterization $\bm{\Theta}$ of the seven-dimensional spin-precessing parameter space. For this, we first extract the individual masses $m_i$ and individual spins $\bm{S}_i$ from the combined dataset described in Sec.~\ref{sec:datasets}, as primary quantities to build a set of input features $\bm{\Theta}$.

For spin-precessing systems, the individual spins $\bm{S}_i$ evolve and therefore an extraction time or frequency has to be selected, with the selection potentially impacting the ability to model the target quantities \cite{Varma:2018mmi,Hamilton:2021pkf}. We select this extraction time as $t_{\rm match}$ since we find that with this choice the dependency of the input values on the composite spin directions is simplified. 

Nevertheless, at this time, the spin values predicted by \texttt{SEOBNRv5PHM} and the numerical waveforms are expected to disagree due to intrinsic inaccuracies in the spin evolution equations from Eq.~\eqref{eq:SLeqns} near merger. To reduce systematic biases in the evaluation of the target quantities $y_i$, we therefore extract the spin values predicted by SEOBNRv5PHM for an equivalent system, rather than using the values directly from the simulations.

Once we have extracted the primary quantities, we construct input features $\bm{\Theta}_{y_i}$ for each target quantity $y_i$, with the selection motivated by the main correlations observed in the data with a large set of composite quantities. For the target quantities associated to input values, we define the following parameterization
\begin{equation}
    \label{eq:feature-vector-ivs}
    \begin{gathered}
  \bm{\Theta}_{y_{1,2,6,7}}=\Bigl(
    \nu,S_{\rm in-plane},\Sigma_{\rm in-plane},\\
    \chi_{\rm eff},\chi_a,\cos2\phi_{\Sigma},\sin2\phi_{\Sigma},
  \Bigr),
    \end{gathered}
\end{equation}
while for $y_3$ (the value of the orbital frequency at the matching time) we define:
\begin{equation}
    \label{eq:feature-vector-omegaorb}
    \begin{gathered}
  \bm{\Theta}_{y_{3}}=\Bigl(
    \nu,S_{n},S_{\lambda},\hat{\chi},\Sigma_{n},\Sigma_{\lambda},\chi_a
  \Bigr).
    \end{gathered}
\end{equation}

For $y_4$ ($c^{\rm asym,22}_{1,f}$), we employ the following parameterization
\begin{equation} 
    \bm{\Theta}_{y_{4}}=\Bigl(\nu,\chi_{\rm eff},\chi_a, \cos2\phi_{S},\sin2\phi_{S}\Bigr),
\end{equation}
while for $y_5$ ($c^{\rm asym,22}_{2,f}$) we additionally include a dependence on $y_4$, since we observe a correlation of both quantities in the data
\begin{equation} 
    \bm{\Theta}_{y_{5}}=\Bigl(\nu,c^{\rm asym,22}_{1,f},\chi_{\rm eff},\chi_a, \cos2\phi_{S},\sin2\phi_{S}\Bigr).
\end{equation}

\subsection{Fitting procedure}
\label{sec:omp}

We employ polynomial regression with feature selection to fit each target quantity $y_i$ as a function of relevant input parameters~$\bm{\Theta}_{y_i}$, using the Orthogonal Matching Pursuit (OMP) algorithm implemented in \texttt{scikit-learn} \cite{scikit-learn}. This approach builds a sparse but accurate approximation to each fit while avoiding overfitting.

We assume a polynomial structure for each target:
\begin{equation}
  y_i = \mathcal{P}(\bm{\Theta}_{y_i}) = \sum_M c_j \psi_j(\bm{\Theta}_{y_i}),
  \label{eq:poly-model}
\end{equation}
where the functions $\psi_j(\bm{\Theta}_{y_i})$ are multivariate monomials formed from the input features,
\[
  \psi_j(\bm{\Theta}_{y_i}) = \prod_{k=1}^{K} \Theta_k^{\,p_{jk}}, \quad \text{with} \quad \sum_{k=1}^{K} p_{jk} \le d_{\max}.
\]
This generates a complete basis of polynomial terms of total degree $\le d_{\max}$ in the $K$ features, resulting in
\begin{equation}
  M = \binom{d_{\max}+K}{d_{\max}}
\end{equation}
monomials. To ensure numerical stability, all monomials are evaluated on the training points and then $Z$-score normalized. The total degree $\le d_{\max}$ employed for each target quantity is specified in Table~\ref{tab:omp_summary}.

Rather than fitting all $M$ terms, which would be computationally expensive and prone to overfitting, OMP builds a sparse model of the form
\begin{equation}
  \hat{y}(\bm{\Theta}_{y_i}) = \sum_{j \in \mathcal{S}} c_j\, \psi_j(\bm{\Theta}_{y_i}), \qquad |\mathcal{S}| \ll M,
  \label{eq:omp-exp}
\end{equation}
where only the most informative terms are included. The OMP algorithm constructs this sparse fit in a step-by-step, greedy fashion. It begins with a constant baseline and iteratively adds one polynomial term at a time. At each iteration, it selects the term that reduces the current error on the training data the most. After each addition, the model is refitted using all selected terms to ensure they jointly contribute useful and non-redundant information. 

\begin{table}[htbp]
    \centering
    \renewcommand{\arraystretch}{1.2}
    \setlength{\tabcolsep}{8pt}
    \begin{tabular}{lcccc}
        \hline
        $y_i$ & $d_{\max}$ & \# terms & RMS (train) & RMS (test) \\
        \hline
        
        $y_1$ &   3    &    49     &      0.0184       &      0.0189        \\
        $y_2$ &    4   &    50     &       0.0151      &      0.0164        \\
        $y_3$ &    5   &    50     &        0.0024     &       0.0026       \\
        $y_4$ &   4    &    13     &      0.0061       &      0.0070        \\        
        $y_5$ &   3    &    10     &      0.0894       &     0.0945         \\       
        $y_6$ &    3   &    50     &      0.0504       &      0.0515        \\       
        $y_7$ &    3   &     42    &       0.0422      &       0.0564       \\        \hline
    \end{tabular}
    \caption{Summary of the regression fits for each target quantity $y_i$. We list the maximum polynomial degree used ($d_{\max}$), the number of polynomial terms selected by the OMP algorithm, and the RMS error on the training and test datasets.}
    \label{tab:omp_summary}
\end{table}

The procedure stops when either a desired accuracy is reached or a maximum number of terms has been added, which we select to be $50$, keeping the evaluation cost low while maintaining accuracy and generalization. In practice, some of the target quantities require the maximum number of $50$ terms, but we do not observe a significant reduction on the fitting error by increasing further the number of terms.  

For training, we randomly split the each dataset described in Sec.~\ref{sec:datasets} into 70\% for training and 30\% for testing, and combine them into a final training and test sets. We find that the root-mean-square (RMS) errors on the test set are comparable to those on the training set, indicating that the fits generalize well and overfitting is avoided. We additionally compare the RMS error in each dataset to prevent that the test-body-limit dataset (since it contains more waveforms) dominates the fit. 

In Table~\ref{tab:omp_summary}, we display for each target quantity the maximum polynomial degree used ($d_{\max}$), the number of polynomial terms selected by the OMP algorithm, and the RMS error on the training and test datasets. The polynomial coefficients for all fitted quantities are made available in the \texttt{pySEOBNR} package, along with routines to evaluate the fits. The computational cost of evaluating the polynomial regressions is negligible compared to the rest of the waveform generation pipeline.

\section{Unfaithfulness with Numerical Relativity}
\label{sec:unfaith}
In this section, we assess the accuracy improvement of \SEOBASYM{}, the upgraded version of the multipolar spin-precessing \texttt{SEOBNRv5PHM} model, by comparing both versions to NR simulations of quasi-circular precessing-spin BBH systems. We also include comparisons with other state-of-the-art waveform models: \texttt{IMRPhenomXPNR}~\cite{xpnr_inprep}, which is calibrated to a set of single-spin precessing simulations and includes the dominant-mode asymmetry, and \texttt{TEOBResumS\_Dali}~\cite{Albanesi:2025txj},\footnote{We use \texttt{TEOBResumS\_Dali} from the public Bitbucket repository \url{https://bitbucket.org/teobresums} with git hash \texttt{6558331f8986a7d710fcf810bdaabbbe6a309112} and tag version \texttt{v1.1.0-Dali}} which is a model for generic orbits, evaluated here in its quasi-circular limit.

We compare against two NR datasets. The first is the SXS dataset described in Sec.~\ref{sec:datasets}, which includes simulations from the second SXS catalog~\cite{Boyle:2019kee} and 118 simulations from Ref.~\cite{Ossokine:2020kjp}, used in the calibration of the antisymmetric modes of \SEOBASYM{}. The second is the BAM catalog released in 2023~\cite{Hamilton:2023qkv}, containing 80 single-spin precessing BBH simulations gridded over mass ratio, primary spin magnitude, and tilt angle. This dataset was used in the calibration of both \texttt{IMRPhenomXO4a}~\cite{Hamilton:2021pkf,Thompson:2023ase} and \texttt{IMRPhenomXPNR}~\cite{xpnr_inprep}.

\subsection{Faithfulness and mismatch metrics}
\label{sec:mismatch}

We compute the averaged SNR-weighted unfaithfulness between detector responses using the procedure of Refs.~\cite{Cotesta:2018fcv,Ossokine:2020kjp,Ramos-Buades:2023ehm}, equivalent to the unfaithfulness metrics introduced in Refs.~\cite{Harry:2016ijz,Harry:2017weg}. For completeness, we summarize the main definitions.

The strain measured by a detector for a quasi-circular, spin-precessing BBH system is characterized by fifteen parameters: the intrinsic parameters $\bm{\Lambda} = \{m_1, m_2, \bm{\chi}_1(t), \bm{\chi}_2(t)\}$, and the extrinsic parameters $\{d_{\rm L}, \iota, \varphi_{\rm ref}, \psi, \alpha_{\rm sky}, \delta_{\rm sky}, t_c\}$, where $d_{\rm L}$ is the luminosity distance, $\psi$ is the polarization angle, $\alpha_{\rm sky}$ (right ascension) and $\delta_{\rm sky}$ (declination) parameterize the sky position of the source, and $t_c$ is the coalescence time.
The expression for the detector strain is
\begin{equation}
h(t) = F_+(\alpha_{\rm sky}, \delta_{\rm sky}, \psi)\, h_+(t) + F_\times(\alpha_{\rm sky}, \delta_{\rm sky}, \psi)\, h_\times(t),
\label{eq:detector_strain}
\end{equation}
where $F_{+,\times}$ are the antenna pattern functions~\cite{Sathyaprakash:1991mt,Finn:1992xs}. This expression can be rewritten in terms of an effective polarization angle $\kappa$ and amplitude $\mathcal{A}$ as~\cite{Cotesta:2018fcv,Ossokine:2020kjp}
\begin{equation}
h(t) = \mathcal{A}(\alpha_{\rm sky}, \delta_{\rm sky})\left[h_+(t)\cos \kappa + h_\times(t)\sin \kappa\right].
\label{eq:strain_kappa}
\end{equation}

The noise-weighted inner product between two waveforms is defined in the frequency domain as
\begin{equation}
\langle h_1, h_2 \rangle = 4\, \mathrm{Re} \int_{f_{\mathrm{in}}}^{f_{\mathrm{max}}} \frac{\tilde{h}_1(f)\, \tilde{h}_2^*(f)}{S_n(f)}\, \mathrm{d}f,
\label{eq:inner_prod}
\end{equation}
where tildes denote Fourier transforms, a star denotes complex conjugation, and $S_n(f)$ is the one-sided noise power spectral density. We use the zero-detuned high-power LIGO design curve~\cite{Barsotti:2018hvm}. For full-band signals, we set $(f_{\mathrm{in}}, f_{\mathrm{max}}) = (10, 2048)\,$Hz. For short NR signals, we use $f_{\mathrm{in}} = 1.35 f_{\mathrm{peak}}$, where $f_{\mathrm{peak}}$ is the peak frequency of the signal's spectrum. The factor 1.35 mitigates Fourier artifacts~\cite{Hinder:2017sxy,Ossokine:2020kjp}.

\begin{figure*}[ht!]
    \centering
    \includegraphics[width=\textwidth]{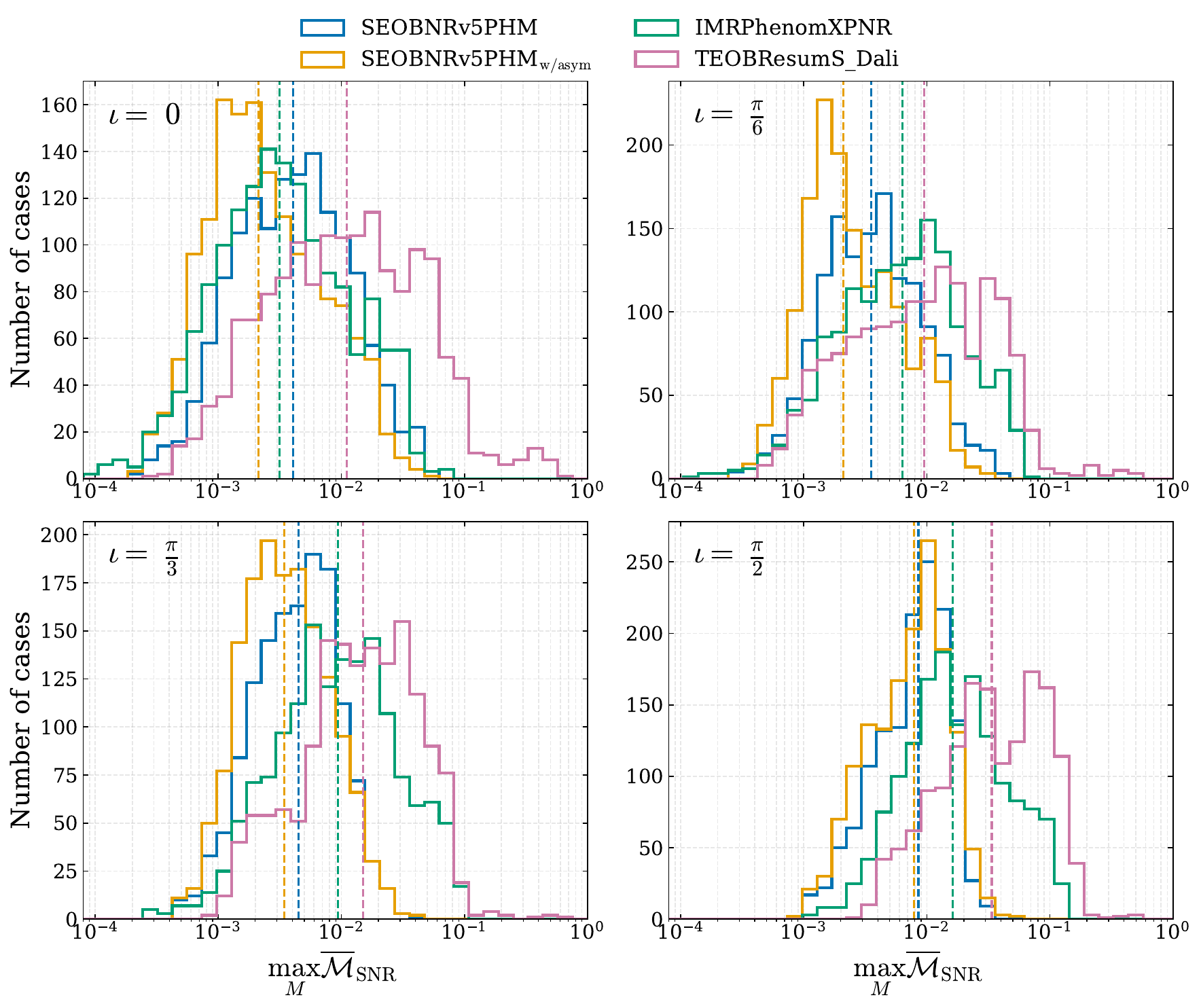}
    \caption{Histogram of the maximum SNR-weighted averaged unfaithfulness with the NR SXS dataset across a total mass range $[20,300]\,M_{\odot}$, for four source inclinations $\iota_s = \{0,\pi/6,\pi/3,\pi/2\}$. Vertical dashed lines mark the median of each distribution.}
    \label{fig:hist_max_mm_comparison}
\end{figure*}

\begin{table*}[ht]
    \centering
    \renewcommand{\arraystretch}{1.2}
    \setlength{\tabcolsep}{10pt}
    \begin{tabular}{c|cccc}
        \hline
        $\iota_s$ & \texttt{SEOBNRv5PHM} & \textbf{\SEOBASYM{}} & \texttt{IMRPhenomXPNR} & \texttt{TEOBResumS\_Dali} \\
        \hline
        $0$       & 0.00405 (47.6\%) & 0.00212 & 0.00316 (32.7\%) & 0.01103 (80.8\%) \\
        $\pi/6$   & 0.00353 (40.3\%) & 0.00211 & 0.00632 (66.6\%) & 0.00952 (77.8\%) \\
        $\pi/3$   & 0.00449 (23.9\%) & 0.00342 & 0.00935 (63.4\%) & 0.01507 (77.3\%) \\
        $\pi/2$   & 0.00855 (8.0\%)  & 0.00786 & 0.01632 (51.8\%) & 0.03374 (76.7\%) \\
        \hline
    \end{tabular}
    \caption{Median SNR-weighted averaged unfaithfulness for different waveform models and source inclinations. Parentheses indicate the relative difference of \SEOBASYM{} with respect to the other models, with positive values meaning a decrease of the median.}
    \label{tab:median_unfaithfulness}
\end{table*}

\begin{figure*}[ht!]
    \centering
    \includegraphics[width=\textwidth]{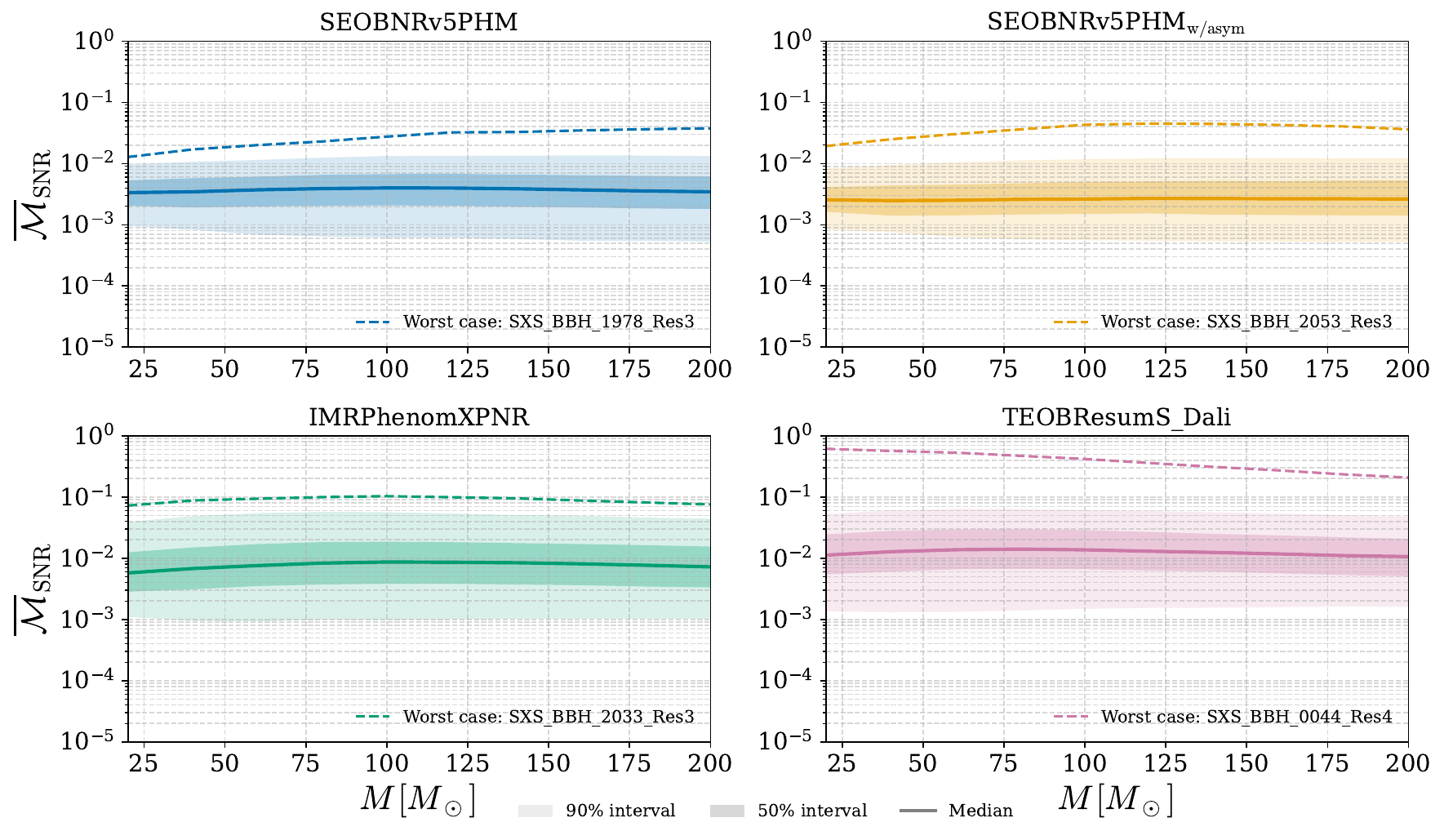}
    \caption{
        SNR-weighted averaged unfaithfulness with the SXS NR dataset as a function of the total mass for inclination $\iota_s=\pi/3$.
        For each model, the shaded regions denote the 5--95\% (light) and 25--75\% (dark) quantile envelopes of the unfaithfulness across all SXS systems, while the solid line shows the median. The dashed curve marks the worst-case system for each model.
        Colors correspond to the waveform models: Blue: \texttt{SEOBNRv5PHM}, Orange: \SEOBASYM, Green: \texttt{IMRPhenomXPNR}, Pink: \texttt{TEOBResumS\_Dali}.
    }
    \label{fig:spaghetti_asym_xprn}
\end{figure*}

\begin{table*}[ht!]
    \centering
    \renewcommand{\arraystretch}{1.2}
    \setlength{\tabcolsep}{10pt}
    \begin{tabular}{c|cccc}
        \hline
        $\iota_s$ & \texttt{SEOBNRv5PHM} & \textbf{\SEOBASYM{}} & \texttt{IMRPhenomXPNR} & \texttt{TEOBResumS} \\
        \hline
        $0$       & 1.43\% (14.19\%) & 0.24\% (7.03\%) & 1.50\% (15.02\%) & 21.53\% (46.84\%) \\
        $\pi/6$   & 0.39\% (10.04\%) & 0.01\% (4.87\%) & 6.38\% (28.59\%) & 16.53\% (43.64\%) \\
        $\pi/3$   & 0.06\% (9.08\%)  & 0.11\% (6.82\%) & 11.77\% (40.95\%) & 20.41\% (57.57\%) \\
        $\pi/2$   & 0.03\% (17.23\%) & 0.32\% (16.61\%) & 21.39\% (56.85\%) & 43.18\% (82.28\%) \\
        \hline
    \end{tabular}
    \caption{Fraction of cases with SNR-weighted unfaithfulness exceeding 3\% (in parentheses: exceeding 1\%), for each waveform model and inclination.}
    \label{tab:unfaithfulness_percentages}
\end{table*}

We define the \emph{faithfulness} between a signal waveform $h_s$ and a template waveform $h_t$ as
\begin{equation}
\mathcal{F}(M_s, \iota_s, \varphi_{\mathrm{ref}\,s}, \kappa_s) = \max_{t_c, \varphi_{\mathrm{ref}\,t}, \kappa_t}
\left. \frac{\langle h_s, h_t \rangle}
{\sqrt{\langle h_s, h_s \rangle \langle h_t, h_t \rangle}} \right|_{\substack{\iota_t = \iota_s \\ \bm{\lambda}_t = \bm{\lambda}_s}}.
\label{eq:faithfulness}
\end{equation}
The maximization over $\varphi_{\mathrm{ref}\,t}$ and $t_c$ is performed numerically, while the maximization over $\kappa_t$ is done analytically~\cite{Capano:2013raa}. To align the intrinsic parameters, we rotate the in-plane spin components of the template waveform by a rigid angle $\delta \in [0, 2\pi]$ at the reference frequency~\cite{Pratten:2020ceb,Gamba:2021gap}:
\begin{equation}
\begin{split}
\chi_{ix}^{(t)} &= \chi_{ix}^{(s)} \cos \delta - \chi_{iy}^{(s)} \sin \delta, \\
\chi_{iy}^{(t)} &= \chi_{ix}^{(s)} \sin \delta + \chi_{iy}^{(s)} \cos \delta, \quad i=1,2.
\end{split}
\label{eq:spin_rotation}
\end{equation}
This rotation mitigates the ambiguity in aligning the reference time across waveform models.

To assess model performance, we marginalize over the azimuthal phase $\varphi_{\mathrm{ref}\,s}$ and effective polarization $\kappa_s$, but keep the inclination $\iota_s$ fixed. The resulting azimuthal- and polarization-averaged faithfulness is
\begin{equation}
\overline{\mathcal{F}}(M_s, \iota_s) = \frac{1}{8\pi^2} \int_0^{2\pi} \!\mathrm{d}\varphi_{\mathrm{ref}\,s} \int_0^{2\pi} \!\mathrm{d}\kappa_s\, \mathcal{F}.
\label{eq:avg_faith}
\end{equation}
In practice, this integral is evaluated over a grid of four values each for $\varphi_{\mathrm{ref}\,s}$ and $\kappa_s$.

We also define the SNR-weighted faithfulness~\cite{Ossokine:2020kjp}:
\begin{equation}
\overline{\mathcal{F}}_{\mathrm{SNR}}(M_s, \iota_s) = \left[
\frac{
\int \mathrm{d}\Omega\, \mathcal{F}^3\, \mathrm{SNR}^3
}{
\int \mathrm{d}\Omega\, \mathrm{SNR}^3
}
\right]^{1/3},
\label{eq:SNRweight}
\end{equation}
where $\mathrm{d}\Omega = \mathrm{d}\varphi_{\mathrm{ref}\,s} \mathrm{d}\kappa_s$ and the SNR is given by
\begin{equation}
\mathrm{SNR}(\iota_s, \varphi_{\mathrm{ref}\,s}, \kappa_s) = \sqrt{\langle h_s, h_s \rangle}.
\label{eq:snr}
\end{equation}
Finally, the unfaithfulness or mismatch is defined as
\begin{equation}
\overline{\mathcal{M}}_{\mathrm{SNR}} = 1 - \overline{\mathcal{F}}_{\mathrm{SNR}}.
\label{eq:mismatch}
\end{equation}
This is the quantity we use in the remainder of this section to compare waveform models to NR simulations.

\subsection{Unfaithfulness against the SXS dataset}

Figure~\ref{fig:hist_max_mm_comparison} presents the distribution of SNR-weighted unfaithfulness, maximized over total mass in the range $[20,200]\,M_{\odot}$, for four waveform models evaluated against the SXS NR dataset. We compute the metric at four different source inclinations, $\iota_s = \{0, \pi/6, \pi/3, \pi/2\}$, to probe inclination dependence.

\begin{figure*}[ht!]
    \centering
    \includegraphics[width=\textwidth]{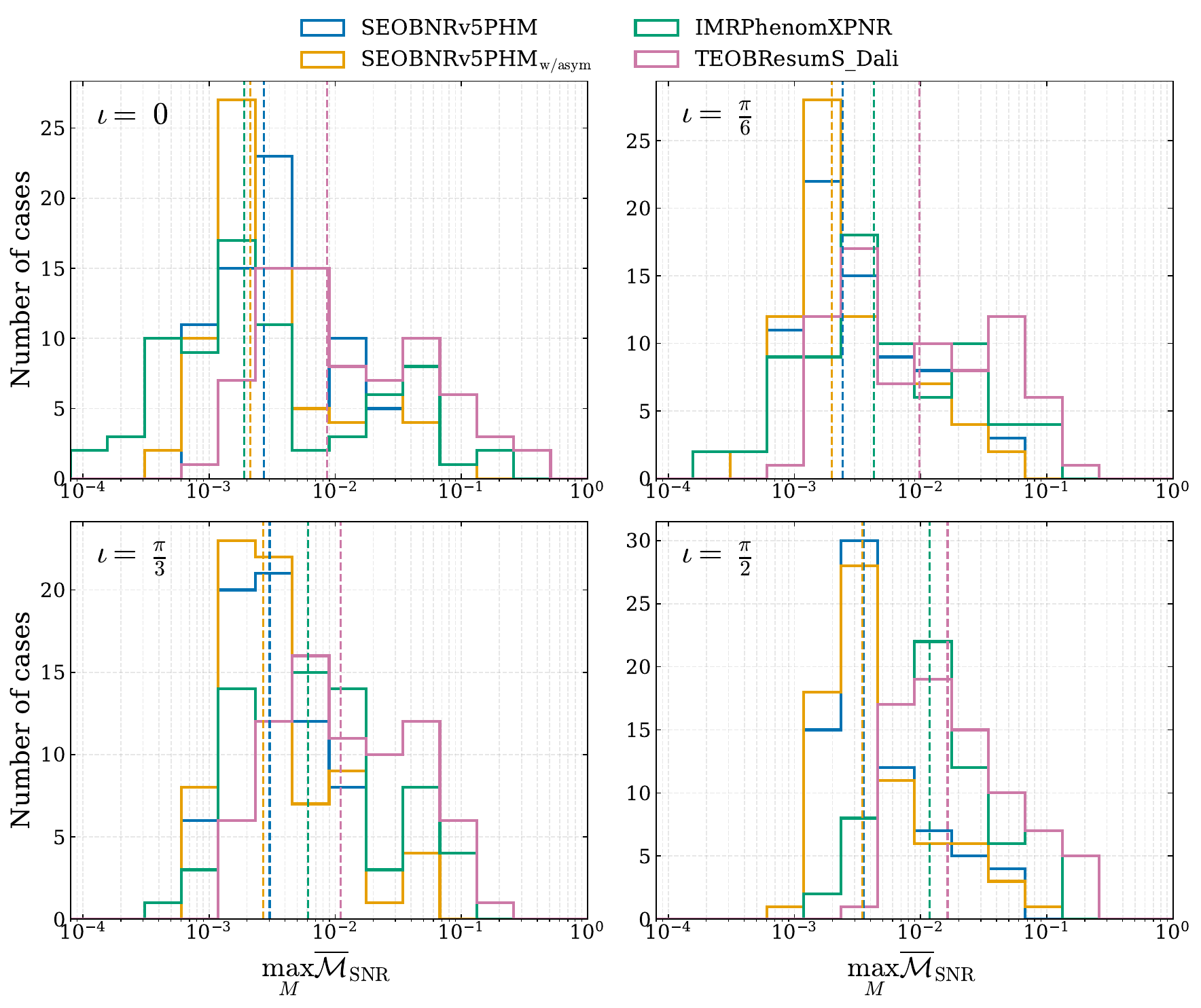}
    \caption{Histogram distribution of the maximum SNR-weighted averaged unfaithfulness of the polarizations with the NR BAM dataset of \cite{Hamilton:2023qkv} across a total mass range $[20,300]M_{\odot}$, for four different source inclination values $\iota_s=\{0,\pi/6,\pi/3,\pi/2\}$. Vertical dashed lines show the median value of the distributions.}
    \label{fig:hist_max_mm_comparison_bam}
\end{figure*}

Comparing \texttt{SEOBNRv5PHM} and \SEOBASYM{}, we observe a clear improvement from the inclusion of antisymmetric modes, particularly at low inclinations. As expected, the antisymmetric multipoles are most pronounced perpendicular to the orbital plane, which enhances the gain in face-on configurations. This shift is reflected in the distributions and their medians (reported in Table~\ref{tab:median_unfaithfulness}). Specifically, \SEOBASYM{} achieves a nearly 50\% reduction in median unfaithfulness for face-on systems. The improvement becomes more modest ($\sim$8\%) for edge-on cases. We note that the SXS NR waveforms employed here include all the simulations used in calibrating the asymmetric-mode fits of \SEOBASYM{}. The comparison against SXS is therefore not fully independent, and is meant to quantify the internal consistency of the model.

\texttt{IMRPhenomXPNR} outperforms \texttt{SEOBNRv5PHM} for face-on systems (by 22\% in the median), but performs worse at higher inclinations. When including antisymmetric modes, \SEOBASYM{} consistently outperforms \texttt{IMRPhenomXPNR} across all inclinations, improving by 32\% for face-on binaries and by 50--60\% at other orientations. Compared to \texttt{TEOBResumS\_Dali}, \SEOBASYM{} exhibits median unfaithfulness values about 80\% lower. We also inspected the small subset of cases in which IMRPhenomXPNR
achieves particularly low unfaithfulness values. These configurations
correspond to nearly equal-mass binaries with weak in-plane spin
components, for which the waveform is dominated by the leading
quadrupolar modes and spin-precession and asymmetric-mode effects are
suppressed.

Figure~\ref{fig:spaghetti_asym_xprn} shows the 90\% and 50\% quantile envelopes of the unfaithfulness as a function of total mass for each model at $\iota_s = \pi/3$. \SEOBASYM{} closely tracks \texttt{SEOBNRv5PHM}, but with a reduced median unfaithfulness (by 24\%). While \texttt{IMRPhenomXPNR} achieves lower values at low masses, it shows a broader spread and a tail extending toward 10\% unfaithfulness. \texttt{TEOBResumS\_Dali} performs worst overall, with several cases exceeding 10\%.

We also identify the simulations with the highest unfaithfulness per model. For \texttt{SEOBNRv5PHM} and \SEOBASYM{}, these are consistent across models, suggesting that these are challenging cases for all current models. In contrast, the worst case for \texttt{TEOBResumS\_Dali} is not an outlier in other models, hinting at different error sources.

Table~\ref{tab:unfaithfulness_percentages} quantifies the fraction of cases with unfaithfulness above 1\% and 3\%. \SEOBASYM{} is the only model with fewer than 1\% of cases exceeding 3\% for face-on binaries. It consistently reduces the number of high-unfaithfulness cases relative to \texttt{SEOBNRv5PHM}, except for a slight increase (below 0.12\%) at $\iota_s = \pi/3$ and $\pi/2$. \texttt{IMRPhenomXPNR} shows worse performance overall, with more than 10\% (20\%) of cases above 3\% at $\iota_s = \pi/3$ ($\pi/2$). \texttt{TEOBResumS\_Dali} exhibits the highest overall percentages, with nearly 43\% of edge-on cases above 3\% unfaithfulness.

\subsection{Unfaithfulness against BAM simulations}

We compute the SNR-weighted averaged unfaithfulness between the publicly available NR simulations from the BAM catalog release of 2023 \cite{Hamilton:2023qkv} -which includes 80 single-spin precessing simulations gridded at mass-ratios $q\in\{1,2,4,8\}$, primary spin magnitudes $a_1\in\{0.2,0.4,0.6,0.8\}$ and primary spin tilt angles uniformly distributed- and the models (\texttt{SEOBNRv5PHM}, \SEOBASYM, \texttt{IMRPhenomXPNR}, \texttt{TEOBResumS\_Dali}).

\begin{figure}[h]
    \includegraphics[width=\columnwidth]{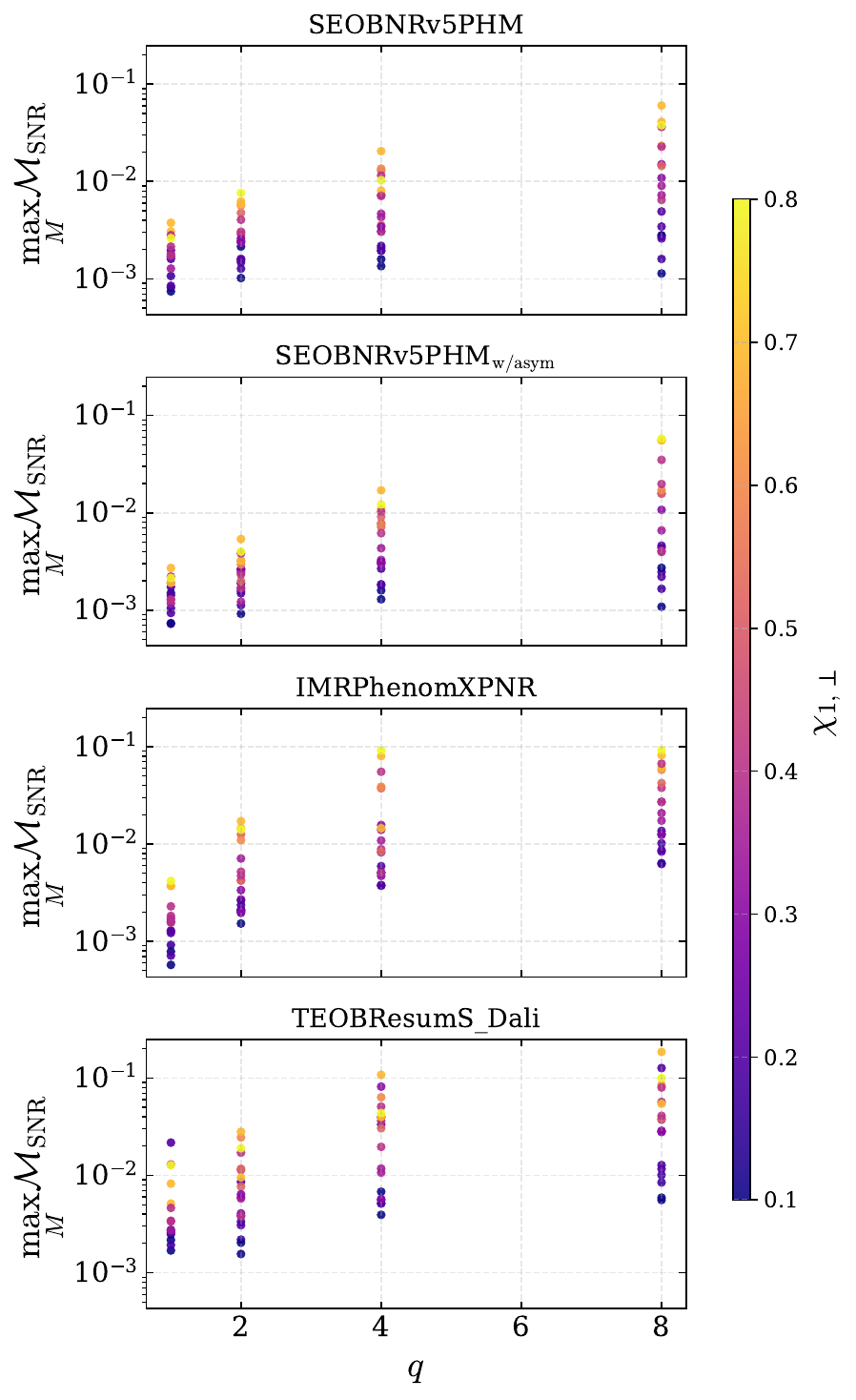}
    \caption{Maximum (across total mass) SNR-weighted averaged unfaithfulness with BAM simulations for $\iota_s=\pi/3$ as a function of mass-ratio $q$ and in-plane primary spin magnitude $\chi_{1,\perp}$.}
    \label{fig:scatter_q_mm_bam}
\end{figure}

In Fig.~\ref{fig:hist_max_mm_comparison_bam} we show the results for the SNR-weighted averaged unfaithfulness between \texttt{IMRPhenomXPHM}, \texttt{SEOBNRv5PHM}, \texttt{IMRPhenomXPNR}, \SEOBASYM and the BAM simulations. We observe an improvement of the median unfaithfulness of \SEOBASYM with respect to \texttt{SEOBNRv5PHM} at all inclinations values, although as with the SXS simulations the improvement degrades with higher inclinations. \texttt{IMRPhenomXPNR} is the most accurate model for face-on inclination, although the difference with \SEOBASYM is tiny, and we expect this result since several aspects of the \texttt{IMRPhenomXPNR} model, apart from the dominant antisymmetric multipole, have been calibrated to these set of simulations. Nevertheless, for other inclination values, both \texttt{IMRPhenom} versions have higher unfaithfulness than the \texttt{SEOBNRv5PHM} versions.

\begin{table*}[ht]
    \centering
    \renewcommand{\arraystretch}{1.2}
    \setlength{\tabcolsep}{10pt}
    \begin{tabular}{c|cccc}
        \hline
        $\iota_s$ &
        \texttt{SEOBNRv5PHM} &
        \textbf{\SEOBASYM{}} &
        \texttt{IMRPhenomXPNR} &
        \texttt{TEOBResumS\_Dali} \\
        \hline
        $0$ &
        0.00273 (21.9\%) &
        0.00213 &
        0.00190 ($-12.2$\%) &
        0.00863 (75.3\%) \\
        $\pi/6$ &
        0.00242 (18.4\%) &
        0.00197 &
        0.00425 (53.6\%) &
        0.00977 (79.8\%) \\
        $\pi/3$ &
        0.00303 (11.3\%) &
        0.00269 &
        0.00608 (55.9\%) &
        0.01100 (75.6\%) \\
        $\pi/2$ &
        0.00354 (2.6\%) &
        0.00345 &
        0.01178 (70.7\%) &
        0.01637 (78.9\%) \\
        \hline
    \end{tabular}
    \caption{
        Median SNR-weighted averaged unfaithfulness for the BAM dataset.
        Parentheses indicate the relative difference of \SEOBASYM{} with respect to the
        other models, with positive values indicating a decrease of the median.
    }
    \label{tab:bam_median_unfaithfulness}
\end{table*}

\begin{table*}[ht!]
    \centering
    \renewcommand{\arraystretch}{1.2}
    \setlength{\tabcolsep}{10pt}
    \begin{tabular}{c|cccc}
        \hline
        $\iota_s$ &
        \texttt{SEOBNRv5PHM} &
        \textbf{\SEOBASYM{}} &
        \texttt{IMRPhenomXPNR} &
        \texttt{TEOBResumS} \\
        \hline
        $0$ &
        6.22\% (21.49\%) &
        6.62\% (16.76\%) &
        11.08\% (23.11\%) &
        25.95\% (42.97\%) \\
        $\pi/6$ &
        3.38\% (15.00\%) &
        3.78\% (12.84\%) &
        12.57\% (28.38\%) &
        23.78\% (44.05\%) \\
        $\pi/3$ &
        4.73\% (15.68\%) &
        4.59\% (13.65\%) &
        15.27\% (35.95\%) &
        23.51\% (47.70\%) \\
        $\pi/2$ &
        5.27\% (16.62\%) &
        6.22\% (17.43\%) &
        18.24\% (45.27\%) &
        29.73\% (62.16\%) \\
        \hline
    \end{tabular}
    \caption{
        Fraction of cases in the BAM dataset with SNR-weighted unfaithfulness
        exceeding 3\% (in parentheses: exceeding 1\%), for each waveform model
        and inclination.
    }
    \label{tab:bam_unfaithfulness_percentages}
\end{table*}

In Fig.~\ref{fig:scatter_q_mm_bam} we show the results for $\iota_s=\pi/3$ as a function of the mass-ratio $q$ and in-plane primary spin magnitude $\chi_{1,\perp}$, observing a clear correlation of the unfaithfulness with these quantities. The accuracy of all models degrade with higher mass-ratio and higher in-plane spin magnitude, although one can observe that for closer-to-equal mass ($q=1,2$) the \SEOBASYM model performs better than the other models.

\begin{figure*}[htpb]
    \includegraphics[width=\textwidth]{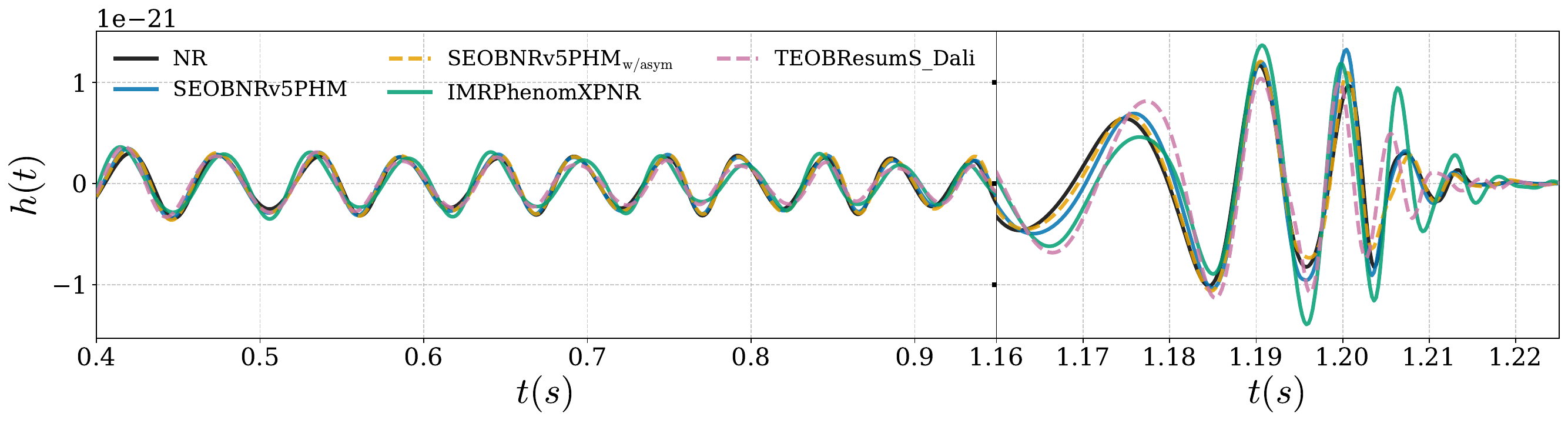}
    \caption{Time-domain comparison of \texttt{SEOBNRv5PHM}, \SEOBASYM, \texttt{IMRPhenomXPNR} and \texttt{TEOBResumS\_Dali} with the NR simulation \texttt{BAM:q4a08t90dPm1D\_T\_96\_384} from Ref.~\cite{Hamilton:2023qkv} with parameters $M=100M_{\odot}$, $q=4$, $\chi_{1,\perp}=0.8$, $\iota_s=\pi/6$, $\varphi_{\mathrm{ref}\,s}=3\pi/2$ and $\kappa_s=\pi$. Models are evaluated with the optimal values obtained for $\varphi_{\mathrm{ref}\,t}$, $\kappa_t$ and $\delta$ from the unfaithfulness optimization.}
    \label{fig:nr_waveform_comparison}
\end{figure*}

The unfaithfulness with the BAM dataset exhibits the same qualitative behaviour as the SXS catalogue, although with some quantitative differences. In terms of the median unfaithfulness (Table~\ref{tab:bam_median_unfaithfulness}), the improvement of \SEOBASYM{} over \texttt{SEOBNRv5PHM} is more modest than in the SXS case, and \texttt{IMRPhenomXPNR} performs slightly better at face-on inclination, as already discussed in this section. When considering the fraction of systems above the 1\% and 3\% thresholds (Table~\ref{tab:bam_unfaithfulness_percentages}), \SEOBASYM{} reduces the high-unfaithfulness tail relative to \texttt{SEOBNRv5PHM} for all viewing angles except $\iota_s=\pi/2$, and the SEOBNR models continue to exhibit the lowest overall fractions of cases above these thresholds. The quantitative differences with respect to the SXS results can be attributed to the fact that the BAM catalogue contains a larger proportion of high-mass-ratio configurations, for which antisymmetric-mode contributions play a less pronounced role and other sources of modeling error become dominant.

To illustrate one of the cases within the challenging region of parameter space, Fig.~\ref{fig:nr_waveform_comparison} shows a comparison between the signal and template waveforms for the simulation \texttt{BAM:q4a08t90dPm1D\_T\_96\_384}, characterized by $M=100M_{\odot}$, $q=4$, $\chi_{1,\perp}=0.8$, $\iota_s=\pi/6$, $\varphi_{\mathrm{ref},s}=3\pi/2$, and $\kappa_s=\pi$. For this particular configuration, the unfaithfulness exceeds 1\% for all models except \SEOBASYM, with values of $1.2$\% for \texttt{SEOBNRv5PHM}, $0.8$\% for \SEOBASYM, $8.7$\% for \texttt{IMRPhenomXPNR}, and $5.6$\% for \texttt{TEOBResumS\_Dali}. All models show good agreement with the NR waveform during the inspiral, with \texttt{SEOBNRv5PHM} and \SEOBASYM displaying notably better phase and amplitude consistency. In the plunge-merger-ringdown phase (right panel), where higher-order modes and asymmetries become more significant, larger discrepancies emerge. \SEOBASYM shows improved agreement with the NR waveform in the merger cycles  --- where antisymmetric contributions are most relevant --- although \texttt{SEOBNRv5PHM} seems to agree better in the ringdown cycles. The other models exhibit larger deviations in both amplitude and phase, which is reflected in the unfaithfulness, despite \texttt{IMRPhenomXPNR} being calibrated to a set of simulations that contains this case.

\begin{figure}[ht!]
    \centering
        \includegraphics[width=\columnwidth]{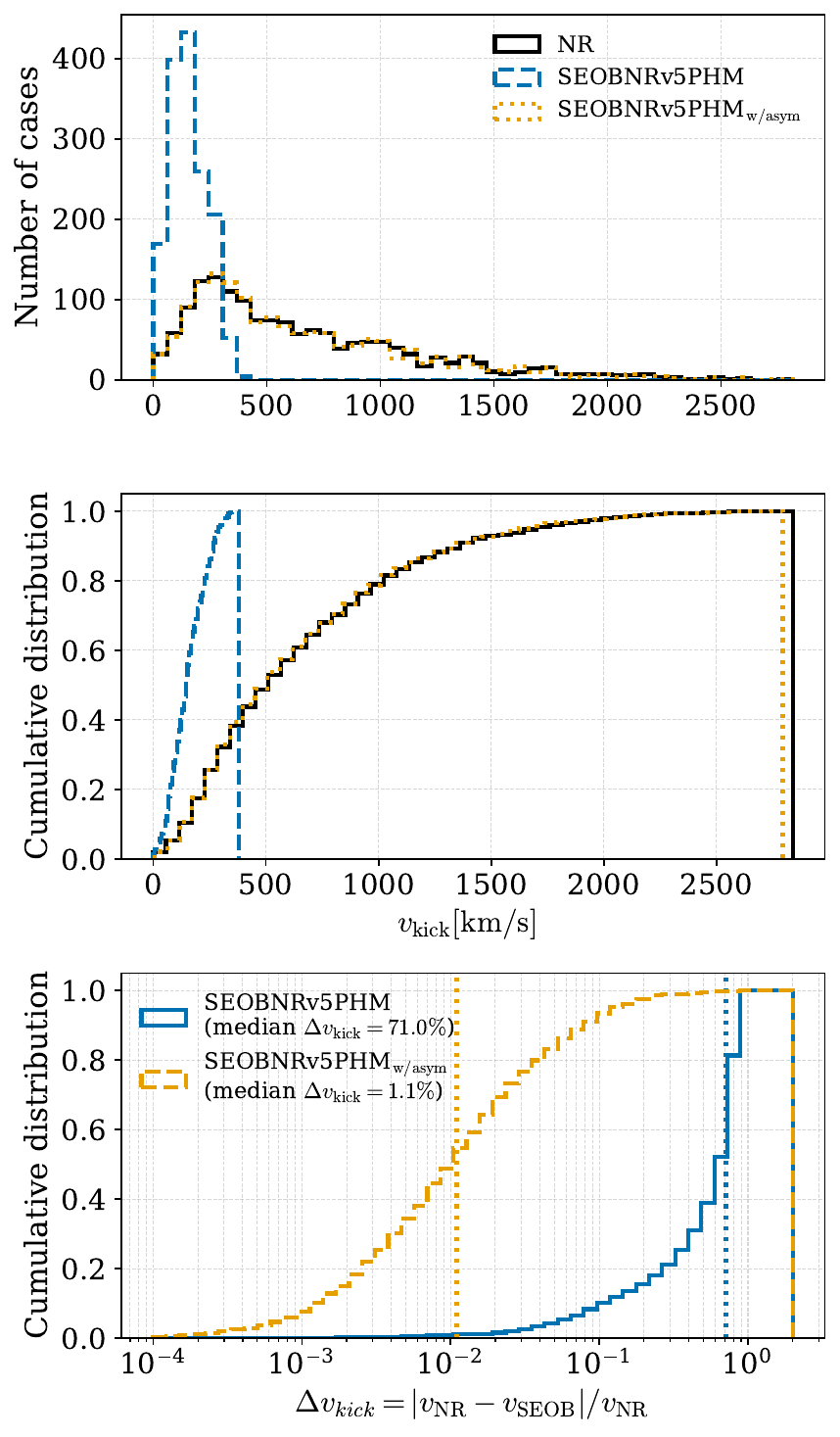}
    
    \caption{Kick velocity of the remnant black hole due to gravitational recoil comparison between the NR SXS dataset (black), \texttt{SEOBNRv5PHM} (blue) and \SEOBASYM (orange). \textit{Upper panel}: Distribution of the kick velocity. \textit{Middle panel}: Cumulative distribution of the kick velocity. \textit{Bottom panel}: Cumulative distribution of the relative error between models and NR.}
    \label{fig:recoil}
\end{figure}

\section{Gravitational recoil}
\label{subsec:recoil}

One of the main motivations for incorporating antisymmetric modes into waveform models is to improve predictions of the GW recoil, or ``kick,'' imparted to the remnant black hole. In this section, we evaluate the accuracy improvement of kick velocity predictions of \SEOBASYM with respect to \texttt{SEOBNRv5PHM}, comparing to results from NR simulations.

The recoil velocity can be obtained from the conservation of linear momentum:
\begin{equation}
v_{\rm rem} = \frac{1}{c}|\bm{P}_{\rm rem}(t_f)| = \frac{1}{c}|\bm{P}_{\rm rad}(t_f)|,
\end{equation}
where $t_f$ is a time at which the GW emission is negligible, and $\bm{P}_{\rm rad}$ is the total radiated linear momentum up to $t_f$. This momentum can be computed by integrating the linear momentum flux at null infinity, expressible in terms of the time derivatives of the SWSH modes of the strain.

The flux components are given by~\cite{Ruiz:2007yx}
\begin{align}
\dot{P}_x + i\dot{P}_y &= \lim_{r\to\infty} \frac{r^2}{8\pi} \sum_{\ell,m} \dot{h}_{\ell m} \Big( a_{\ell m} \dot{h}^*_{\ell,m+1} \nonumber\\
&+ b_{\ell,-m} \dot{h}^*_{\ell-1,m+1} - b_{\ell+1,m+1} \dot{h}^*_{\ell+1,m+1} \Big),
\\
\dot{P}_z &= \lim_{r\to\infty} \frac{r^2}{16\pi} \sum_{\ell,m} \dot{h}_{\ell m} \Big( c_{\ell m} \dot{h}^*_{\ell m} \nonumber\\
&+ d_{\ell m} \dot{h}^*_{\ell-1,m} + d_{\ell+1,m} \dot{h}^*_{\ell+1,m} \Big),
\end{align}
with the coupling coefficients
\begin{subequations}
\begin{align}
a_{\ell m} &= \frac{\sqrt{(\ell - m)(\ell + m + 1)}}{\ell(\ell + 1)}, \\
b_{\ell m} &= \frac{1}{2\ell} \sqrt{\frac{(\ell - 2)(\ell + 2)(\ell + m)(\ell + m - 1)}{(2\ell - 1)(2\ell + 1)}}, \\
c_{\ell m} &= \frac{2m}{\ell(\ell + 1)}, \\
d_{\ell m} &= \frac{1}{\ell} \sqrt{\frac{(\ell - 2)(\ell + 2)(\ell - m)(\ell + m)}{(2\ell - 1)(2\ell + 1)}}.
\end{align}
\end{subequations}

When these expressions are evaluated in the co-precessing frame, the $x$ and $y$ components of the flux correspond to momentum emitted within the orbital plane, while the $z$ component captures momentum emitted perpendicular to the plane. For spin-precessing binaries, this out-of-plane component often dominates the recoil. Notably, in the absence of equatorial asymmetries, this contribution vanishes, leading to a systematic underestimation of the recoil in waveform models that do not include these contributions.

We evaluate the radiated momentum using the \texttt{scri} package~\cite{Boyle:2013nka, Boyle:2015nqa}, both for NR simulations and waveform models. To reduce uncertainties associated with the reference spin orientation, we perform an optimization over the direction of the total spin vector. Figure~\ref{fig:recoil} presents the resulting distributions of recoil velocities.

The top panel of Fig.~\ref{fig:recoil} shows the histogram of recoil velocities recovered from NR simulations, \texttt{SEOBNRv5PHM}, and \SEOBASYM. The model without asymmetries (\texttt{SEOBNRv5PHM}) systematically underestimates the distribution. In contrast, \SEOBASYM closely reproduces the NR distribution, including the maximum velocity predicted in the dataset.

The middle panel shows the cumulative distribution of recoil velocities, emphasizing that the maximum kick magnitude in the NR dataset is accurately captured by \SEOBASYM. Finally, the bottom panel displays the cumulative distribution of the relative error with respect to the NR value. While \SEOBASYM achieves low median errors, approximately 10\% of cases exceed 10\% relative error. A likely explanation is the omission of the antisymmetric $(3, 2)$ mode, recently identified as a relevant contributor to subdominant kick components~\cite{Mielke:2024kya}.

\section{NR-injection recovery}\label{subsec:nrinj}

We study the effect of the inclusion of the antisymmetric modes on parameter estimation by performing Bayesian inference on synthetic signals generated from NR simulations. We select three cases from the SXS Collaboration Catalog: 

\begin{itemize}
    \item \texttt{SXS:BBH:0963}: an equal-mass binary with high, equal-magnitude, oppositely aligned in-plane spins ($\chi_1 = \chi_2 = 0.8$, $\theta_1 = \theta_2 = \pi/2$, $\phi_{12} = \pi$). This configuration exhibits minimal orbital-plane precession but significant excitation of antisymmetric modes, leading to one of the largest remnant kick velocities in our dataset ($v_{\rm rem} \sim 2800\,\mathrm{km/s}$).
    \item \texttt{SXS:BBH:1622}: a mass ratio $1/q = 0.383$ binary with a large in-plane primary spin ($\chi_1 = 0.72$, $\theta_1 = 1.1$).
    \item \texttt{SXS:BBH:2070}: a more asymmetric binary ($1/q = 0.25$) with a large in-plane primary spin ($\chi_1 = 0.8$, $\theta_1 \approx \pi/2$).
\end{itemize}
These systems allow us to explore the influence of antisymmetric modes across varying mass ratios and spin geometries. 

\begin{figure*}[ht!]
    \includegraphics[width=0.32\textwidth]{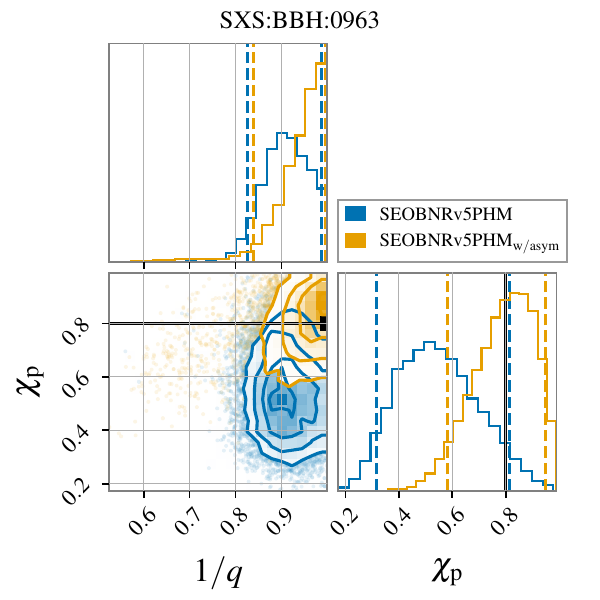}
    \includegraphics[width=0.32\textwidth]{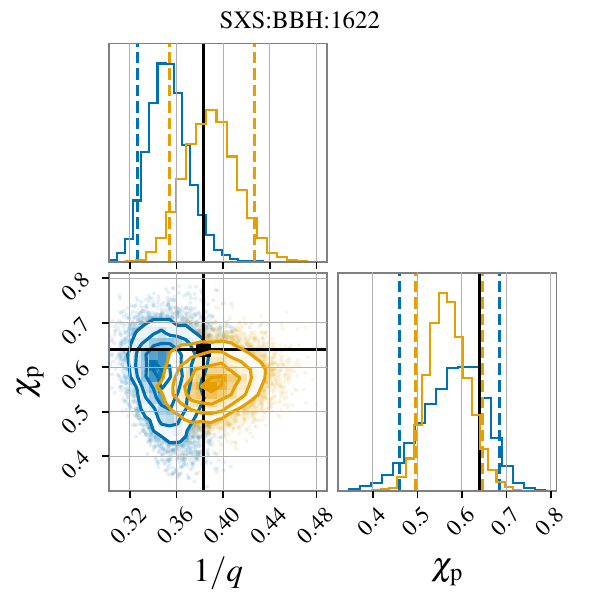}
    \includegraphics[width=0.32\textwidth]{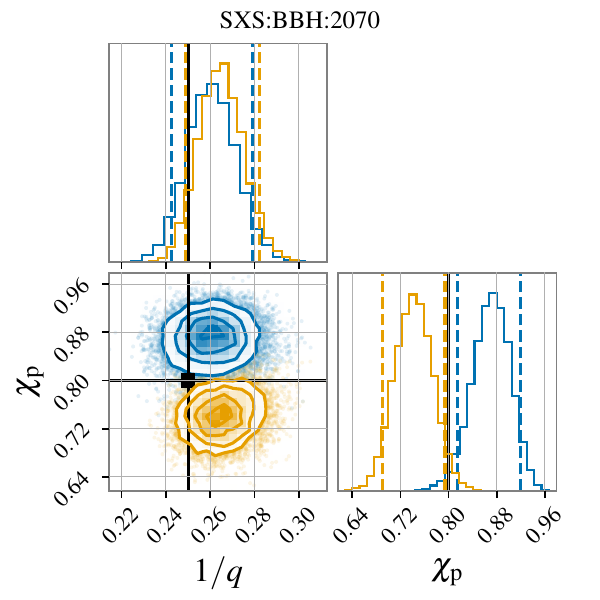}
    \includegraphics[width=0.32\textwidth]{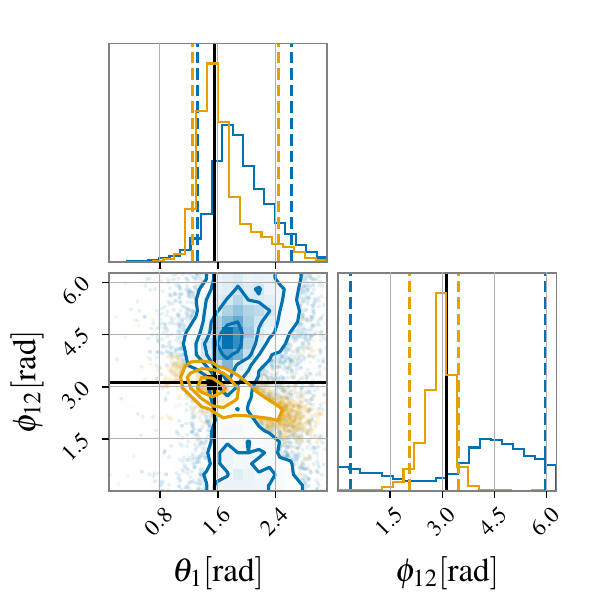}
    \includegraphics[width=0.32\textwidth]{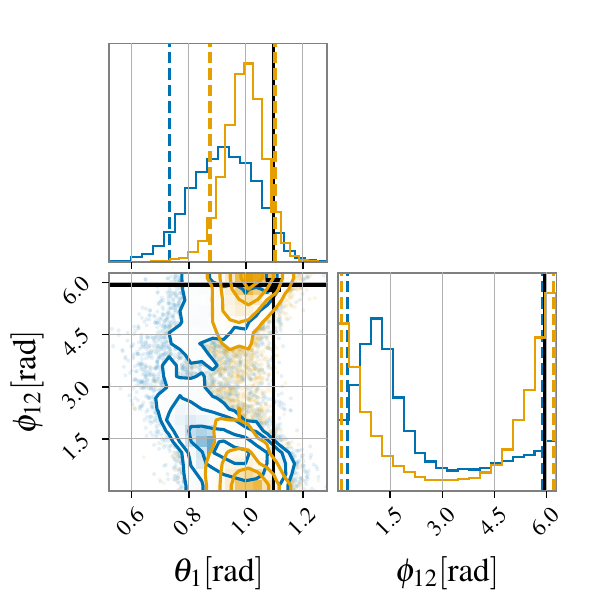}
    \includegraphics[width=0.32\textwidth]{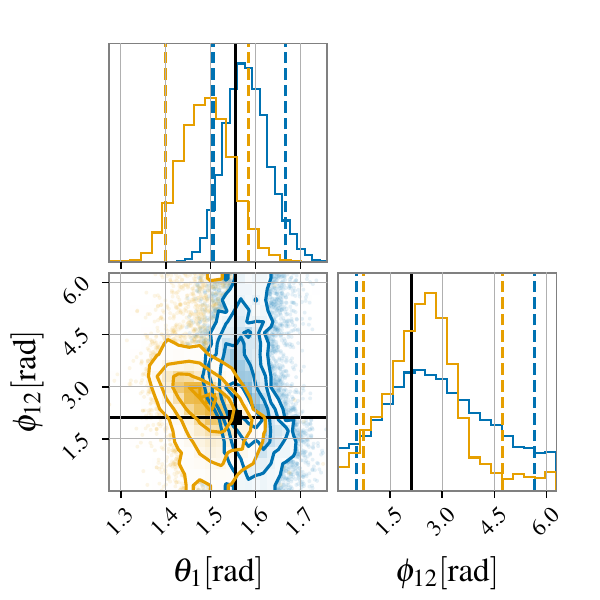}
    \caption{Recovered marginal posteriors for mass ratio $1/q$ and effective precessing spin $\chi_{\rm p}$ (top row), and primary spin tilt angle $\theta_1$ and relative in-plane spin angle $\phi_{12}$ (bottom row). Solid black lines show injected values; dashed lines indicate 90\% credible intervals.}
    \label{fig:nr_inj}
\end{figure*}

We perform injections into zero noise, which is equivalent to averaging over many noise realizations, to isolate waveform modeling biases from statistical noise effects. We simulate the signals using a detector-frame total mass of $M = 90\,M_{\odot}$, ensuring that the merger and ringdown phases fall within the most sensitive frequency band of the detectors. The inclination angle of the binary with respect to the line of sight is fixed at $\iota = \pi/3$ rad. We set the reference phase to $\varphi_{\rm ref} = 0$ rad and the polarization angle to $\psi = 2.87$ rad. The sky location corresponds to a right ascension of $0.83$ rad and a declination of $-0.3$ rad, at a geocentric GPS time of 1376089759.8 s. The luminosity distance is chosen as $d_{\rm L} = 509$ Mpc, such that the first injection (\texttt{SXS:BBH:0963}) yields a network SNR of 60 in the HLV detector network, assuming the LIGO and Virgo design sensitivity noise curves~\cite{Barsotti:2018hvm}. Full injection parameters, including spin configurations, are provided in Table~\ref{tab:nr-injections}.

\begin{table*}[htbp]
    \centering
    \renewcommand{\arraystretch}{1.2}
    \setlength{\tabcolsep}{8pt}
    \resizebox{\textwidth}{!}{%
      \begin{tabular}{c||c|c|c}
      \hline
      Parameter & \texttt{SXS:BBH:0963} & \texttt{SXS:BBH:1622} & \texttt{SXS:BBH:2070} \\
      \hline\hline
      \multirow{2}{*}{$M[M_{\odot}]$} & \bf{90} & \bf{90} & \bf{90} \\
       & (${88.91}_{-2.14}^{+2.28}$) / [${91.38}_{-2.14}^{+2.22}$] & (${92.38}_{-2.04}^{+2.18}$) / [${90.70}_{-2.13}^{+2.34}$] & (${91.19}_{-2.67}^{+2.77}$) / [${88.92}_{-1.90}^{+1.87}$] \\ 
      \multirow{2}{*}{$1/q$} & \bf{1} & \bf{0.383} & \bf{0.25} \\
       & (${0.91}_{-0.05}^{+0.05}$) / [${0.95}_{-0.06}^{+0.03}$] & (${0.35}_{-0.02}^{+0.02}$) / [${0.39}_{-0.02}^{+0.02}$] & (${0.26}_{-0.01}^{+0.01}$) / [${0.26}_{-0.01}^{+0.01}$] \\ 
      \multirow{2}{*}{$\chi_{1}$} & \bf{0.8} & \bf{0.72} & \bf{0.8} \\
       & (${0.44}_{-0.24}^{+0.22}$) / [${0.80}_{-0.27}^{+0.12}$] & (${0.73}_{-0.04}^{+0.03}$) / [${0.68}_{-0.04}^{+0.04}$] & (${0.87}_{-0.03}^{+0.03}$) / [${0.75}_{-0.03}^{+0.03}$] \\ 
      \multirow{2}{*}{$\chi_{2}$} & \bf{0.8} & \bf{0.499} & \bf{0.8} \\
       & (${0.44}_{-0.27}^{+0.28}$) / [${0.59}_{-0.15}^{+0.24}$] & (${0.66}_{-0.25}^{+0.18}$) / [${0.32}_{-0.16}^{+0.21}$] & (${0.72}_{-0.32}^{+0.19}$) / [${0.47}_{-0.28}^{+0.28}$] \\ 
      \multirow{2}{*}{$\theta_{1} [\mathrm{rad}]$} & \bf{1.56} & \bf{1.1} & \bf{1.55} \\
       & (${1.87}_{-0.30}^{+0.44}$) / [${1.58}_{-0.21}^{+0.39}$] & (${0.92}_{-0.12}^{+0.11}$) / [${1.00}_{-0.07}^{+0.06}$] & (${1.58}_{-0.05}^{+0.05}$) / [${1.49}_{-0.06}^{+0.06}$] \\ 
      \multirow{2}{*}{$\theta_{2} [\mathrm{rad}]$} & \bf{1.59} & \bf{2.04} & \bf{1.47} \\
       & (${1.38}_{-0.49}^{+0.35}$) / [${1.39}_{-0.34}^{+0.32}$] & (${2.51}_{-0.37}^{+0.31}$) / [${2.05}_{-0.47}^{+0.50}$] & (${0.65}_{-0.32}^{+0.40}$) / [${1.86}_{-0.47}^{+0.53}$] \\ 
      \multirow{2}{*}{$\phi_{12} [\mathrm{rad}]$} & \bf{3.12} & \bf{5.93} & \bf{2.11} \\
       & (${4.21}_{-2.91}^{+1.22}$) / [${2.95}_{-0.45}^{+0.28}$] & (${1.46}_{-0.80}^{+3.25}$) / [${4.39}_{-4.02}^{+1.60}$] & (${2.79}_{-1.43}^{+1.81}$) / [${2.48}_{-1.06}^{+0.94}$] \\ 
      \multirow{2}{*}{$\phi_{\rm JL} [\mathrm{rad}]$} & \bf{3.45} & \bf{3.31} & \bf{5.39} \\
       & (${4.82}_{-0.20}^{+0.22}$) / [${4.88}_{-0.33}^{+0.38}$] & (${3.71}_{-0.22}^{+0.20}$) / [${3.63}_{-0.15}^{+0.17}$] & (${5.26}_{-0.11}^{+0.11}$) / [${5.24}_{-0.11}^{+0.11}$] \\ 
       \multirow{2}{*}{$\iota [\mathrm{rad}]$} & \bf{1.05} & \bf{1.05} & \bf{1.05} \\
       & (${1.26}_{-0.08}^{+0.07}$) / [${0.95}_{-0.16}^{+0.14}$] & (${1.18}_{-0.09}^{+0.08}$) / [${1.14}_{-0.07}^{+0.08}$] & (${1.08}_{-0.04}^{+0.04}$) / [${0.99}_{-0.05}^{+0.06}$] \\ 
      \multirow{2}{*}{$d_{L} [\mathrm{Mpc}]$} & \bf{509} & \bf{509} & \bf{509} \\
       & (${581.04}_{-68.95}^{+69.22}$) / [${721.98}_{-93.30}^{+104.88}$] & (${528.14}_{-20.13}^{+19.47}$) / [${524.14}_{-21.90}^{+21.58}$] & (${525.54}_{-19.02}^{+17.97}$) / [${527.07}_{-14.10}^{+14.67}$] \\ 
      \multirow{2}{*}{$\alpha_{\rm sky} [\mathrm{rad}]$} & \bf{0.83} & \bf{0.83} & \bf{0.83} \\
       & (${0.83}_{-0.01}^{+0.01}$) / [${0.84}_{-0.01}^{+0.00}$] & (${0.83}_{-0.00}^{+0.00}$) / [${0.83}_{-0.00}^{+0.00}$] & (${0.83}_{-0.00}^{+0.00}$) / [${0.83}_{-0.00}^{+0.00}$] \\ 
      \multirow{2}{*}{$\delta_{\rm sky} [\mathrm{rad}]$} & \bf{-0.3} & \bf{-0.3} & \bf{-0.3} \\
       & (${-0.30}_{-0.01}^{+0.01}$) / [${-0.31}_{-0.01}^{+0.01}$] & (${-0.30}_{-0.01}^{+0.01}$) / [${-0.30}_{-0.01}^{+0.01}$] & (${-0.30}_{-0.01}^{+0.01}$) / [${-0.30}_{-0.01}^{+0.01}$] \\ 
      \multirow{2}{*}{$\Psi [\mathrm{rad}]$} & \bf{2.87} & \bf{2.87} & \bf{2.87} \\
       & (${3.01}_{-2.93}^{+0.09}$) / [${2.92}_{-0.26}^{+0.12}$] & (${2.99}_{-0.10}^{+0.09}$) / [${3.01}_{-2.91}^{+0.09}$] & (${2.74}_{-0.09}^{+0.09}$) / [${2.67}_{-0.09}^{+0.10}$] \\ 
      \multirow{2}{*}{$\varphi_{\rm ref} [\mathrm{rad}]$} & \bf{0} & \bf{0} & \bf{0} \\
       & (${5.46}_{-0.46}^{+0.38}$) / [${3.13}_{-0.58}^{+0.73}$] & (${5.73}_{-5.57}^{+0.41}$) / [${0.69}_{-0.53}^{+5.43}$] & (${0.79}_{-0.41}^{+0.47}$) / [${0.71}_{-0.54}^{+5.39}$] \\  & & & \\ 
      $\log_{10}\rm{BF}$ & ($748.8\pm 0.1$)[$753.2\pm 0.1$] & ($1318.3\pm 0.1$)[$1319.6\pm 0.1$] & ($1010.5\pm 0.1$)[$1013.1\pm 0.1$] \\ \hline
      \end{tabular}%
    } 
    \caption{Injected values, recovered medians with 90\% credible intervals and $\log_{10}$ Bayes factors. The reported $\log_{10}\mathrm{BF}$ values denote the Bayes factor comparing the spin-precessing hypothesis to the non-precessing one. Rows show, for each parameter: the injected value (in boldface), recovery with \texttt{SEOBNRv5PHM} (round brackets), and recovery with \SEOBASYM (square brackets). Last row shows the $\log_{10}$ Bayes factors (with associated uncertainty) for the \texttt{SEOBNRv5PHM} (round brackets) and \SEOBASYM (square brackets) analysis.}
    \label{tab:nr-injections}
\end{table*}

Each injection is recovered using both \texttt{SEOBNRv5PHM} without antisymmetric modes and our extended \SEOBASYM{} model with them included, to evaluate their impact on parameter recovery. We perform Bayesian parameter estimation using \texttt{parallel Bilby}~\cite{Smith:2019ucc}, a highly parallel implementation of the \texttt{Bilby} inference framework~\cite{Ashton:2018jfp,Romero-Shaw:2020owr}\footnote{All parameter-estimation runs were performed with the standard frequency-domain Whittle likelihood as implemented in \texttt{Bilby}, using a Tukey window to taper the data. At the time of our analysis, the default implementation included the correction factor accounting for the fractional power loss of the window (the ``$\beta$'' factor). Recent work \cite{Talbot:2025vth} has shown that including this factor leads to a mildly over-constrained likelihood, whereas omitting it yields unbiased posteriors and Bayes factors for signals with network SNR $\lesssim \mathcal{O}(100)$, provided the window does not affect the part of the segment containing the signal. In our analyses, the impact of the windowing issue is expected to be small and, moreover, it affects all waveform models in the same way. We therefore do not expect our conclusions regarding the relative performance of \SEOBASYM{} and \texttt{SEOBNRv5PHM}, nor the evidence for spin precession in GW200129, to be significantly biased.}. The sampler is configured with $n_{\mathrm{live}} = 1000$ live points and an acceptance threshold of $n_{\mathrm{accept}} = 60$, following standard recommendations. The sampling parameters not explicitly specified are kept at their default values. We adopt a prior uniform in component masses, with the following bounds in inverse mass ratio, $1/q \in [0.05, 1]$, and in chirp mass, $\mathcal{M} \in [15, 45]\,M_\odot$. The spin magnitudes $\chi_i$ are sampled uniformly in $[0, 0.99]$, while spin orientations are drawn isotropically over the unit sphere. Remaining prior choices follow the conventions detailed in Appendix~C of Ref.~\cite{LIGOScientific:2018mvr}.

In Table~\ref{tab:nr-injections}, we report the injected values and the recovered medians with 90\% credible intervals for the three NR injections analyzed with \texttt{SEOBNRv5PHM} and \SEOBASYM, as well as the $\log_{10}$ Bayes factors ($\log_{10}\rm{BF}$) quantifying the
evidence in favour of the spin-precessing hypothesis relative to the
non-precessing one. For the first injection (\texttt{SXS:BBH:0963}), where antisymmetric contributions are expected to be most significant, we find a noticeable reduction in biases for several spin-related parameters --- particularly the primary spin magnitude $\chi_1$, the tilt angle $\theta_1$, and the in-plane spin angle $\phi_{12}$. We also observe a slight improvement in the recovery of the mass ratio. However, this improvement in intrinsic spin parameters comes at the cost of larger biases in the luminosity distance, which we discuss in more detail later.

For the second injection (\texttt{SXS:BBH:1622}), which involves more unequal masses, the inclusion of antisymmetric modes improves the recovery of the total mass and mass ratio. The tilt angles of both spins are recovered more accurately, and the bias in $\phi_{12}$ is reduced. The primary spin magnitude shows a mild shift in the median, though the injected value remains within the 90\% credible interval. For the secondary spin magnitude, the posterior moves from a mild overestimation (for \texttt{SEOBNRv5PHM}) to a mild underestimation (for \SEOBASYM).
    
The third injection (\texttt{SXS:BBH:2070}) shows more modest differences between the two model variants. There is a mild improvement in the recovery of $a_1$, which becomes more consistent with the true value for \SEOBASYM, and a clearer constraint on $\phi_{12}$. The recovery of the secondary spin tilt angle also improves.
    
Figure~\ref{fig:nr_inj} shows the posteriors for mass ratio, effective precessing spin $\chi_{\rm p}$, the primary spin tilt angle, and the in-plane spin angle for the three injections. As anticipated, the most significant improvements appear in the first case, where antisymmetric multipoles play a larger role. Across all three cases, we find that using \SEOBASYM leads to generally better constraints on spin orientations, which could be crucial for identifying spin-precessing formation channels.
    
    \begin{figure}[ht!]
    \centering
    \includegraphics[width=\columnwidth]{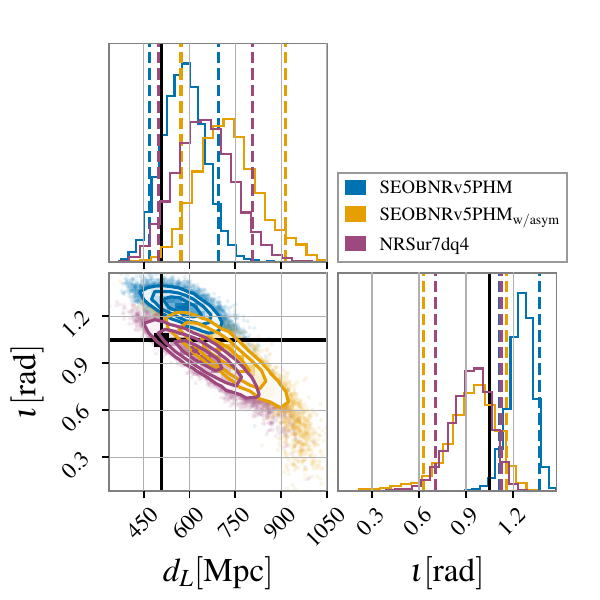}
    \caption{Recovered marginal posterior distributions for luminosity distance and inclination angle for the NR injection \texttt{SXS:BBH:0963}. Solid black lines indicate the injected values, and dashed lines show 90\% credible intervals.}
    \label{fig:sxs_0963_distance_incl}
    \end{figure}

Overall, the $\log_{10}$ Bayes factors are higher for all the injections analyzed with \SEOBASYM, with a significant improvement for the first injection (from $748.8$ with \texttt{SEOBNRv5PHM} to $753.2$ with \SEOBASYM), and milder improvements for the other two injections ($1318.3\rightarrow1319.6$ for \texttt{SXS:BBH:1622}, and $1010.5\rightarrow1013.1$ for \texttt{SXS:BBH:2070}). This confirms that the inclusion of the antisymmetric modes in \SEOBASYM leads to a better fit to the NR waveforms, with the most substantial improvement for the case where antisymmetric contributions are expected to be most significant, \texttt{SXS:BBH:0963}.

To better understand the increased biases in luminosity distance observed in \texttt{SXS:BBH:0963} when including antisymmetric modes, we perform a comparative analysis using the \texttt{NRSur7dq4} model, which also includes these effects. Figure~\ref{fig:sxs_0963_distance_incl} shows the 2D posterior distributions for luminosity distance and inclination. We find that the recovery using \SEOBASYM aligns more closely with the results from \texttt{NRSur7dq4} than \texttt{SEOBNRv5PHM}. In particular, we observe a reduction in the systematic bias for the inclination parameter in \SEOBASYM and \texttt{NRSur7dq4}. Both models including antisymmetric modes show similar bias trends in luminosity distance, although the \texttt{SEOBNRv5PHM} recovery appears to slightly overestimate it. The 2D posteriors from \SEOBASYM and \texttt{NRSur7dq4} are closer to the injected value than those from \texttt{SEOBNRv5PHM}, suggesting a projection effect when interpreting the luminosity distance posterior, potentially amplified by non-uniform priors for luminosity distance and inclination. These results suggest that the inclusion of antisymmetric modes can influence the recovery of extrinsic parameters such as distance and inclination.


\section{Re-analysis of GW200129}\label{sec:gw200129}

GW200129 is a notably strong GW signal detected by the LVK collaboration with a network SNR of 26.5. The event was observed in the two Advanced LIGO detectors and the Advanced Virgo detector, and corresponds to the merger of two black holes with component masses approximately $34.5\,M_\odot$ and $28.9\,M_\odot$. It was first reported in the GWTC-3 catalog~\cite{KAGRA:2021vkt}. The initial parameter estimation results showed a disparity in spin-precession measurements: one waveform model recovered significant spin-precession, while the other did not.

Subsequent studies re-analyzed this event using more accurate models. Ref.~\cite{Hannam:2021pit} used the \texttt{NRSur7dq4} waveform model and found compelling evidence for spin-precession, estimating a spin-precessing SNR \cite{Green:2020ptm} of 4 and a Bayes factor of 30:1 favoring the precessing hypothesis. Using the same waveform model and a NR surrogate model for remnant quantities, Ref.~\cite{Varma:2022pld} constrained the kick velocity of the remnant black hole to exceed $698~\mathrm{km/s}$ at 90\% credibility, with a median of $1542~\mathrm{km/s}$. However, Ref.~\cite{Payne:2022spz} showed that the evidence for spin-precession is sensitive to the glitch subtraction method applied to the data, with some techniques significantly reducing the support for spin-precession in the signal. Most recently, Ref.~\cite{Kolitsidou:2024vub} investigated the role of antisymmetric modes in spin-precessing signals, finding that removing these components from \texttt{NRSur7dq4} degraded the recovery of spin-precession for this signal, additionally impacting the recovery of other parameters such as the mass ratio.

In this work, we re-analyze GW200129 using \SEOBASYM, the extension of the \texttt{SEOBNRv5PHM} model that incorporates antisymmetric spin-precessing modes presented in this work. Our goals are twofold: (i) to independently verify the findings of Ref.~\cite{Kolitsidou:2024vub} using an independent waveform model, and (ii) to assess the robustness of these findings under different glitch subtraction techniques, following the methodology of Ref.~\cite{Payne:2022spz}.

\subsection{Analysis overview}

We follow the Bayesian inference setup described in the previous section. Parameter estimation is performed using \texttt{bilby}~\cite{Ashton:2018jfp} with its parallel implementation \texttt{parallel-bilby}~\cite{Smith:2019ucc}, across five versions of the detector data surrounding the event. These include the public data released through GWOSC~\cite{Vallisneri:2014vxa}, one dataset processed with the \texttt{gwsubtract} algorithm~\cite{Davis:2018yrz,Davis:2022ird}, and three datasets processed with the \texttt{BayesWave} glitch-subtraction algorithm~\cite{Cornish:2014kda,Hourihane:2022doe}, corresponding to three different draws from the glitch posterior labeled as \texttt{BW-A}, \texttt{BW-B}, and \texttt{BW-C}.

For each of the five datasets, we conduct four independent analyses: using (i) \texttt{NRSur7dq4}, (ii) \texttt{SEOBNRv5PHM}, (iii) \SEOBASYM, and (iv) the aligned-spin model \texttt{SEOBNRv5HM}. The difference between \SEOBASYM and \texttt{SEOBNRv5PHM} lies exclusively in the inclusion of antisymmetric contributions to the $\ell=m$ co-precessing modes. In all cases, we analyze 8 seconds of strain data centered around the trigger time $t_c = 1264316110.4$ (GPS), sampled at $2048$ Hz, using a minimum frequency of 20 Hz and a template starting frequency of 15 Hz. We employ PSDs estimated from the data using \texttt{BayesWave}.

We adopt uniform priors in component masses, with bounds in chirp mass and mass ratio: $\mathcal{M}_c \in [14.48, 48.94]\,M_\odot$ and $q \in [0.05, 1]$, consistent with previous analyses. Spin priors are isotropic for the spin-precessing models, while the aligned-spin analyses employ the aligned-spin projections of the isotropic spin prior as prior for $\chi_{1,l}, \chi_{2,l}$. The remaining priors follow those employed in GWTC-3~\cite{KAGRA:2021vkt}. Sampling is performed with \texttt{dynesty}~\cite{Speagle:2019ivv}, using the \texttt{acceptance-walk} method with $n_{\mathrm{accept}} = 60$ and 1000 live points.

\subsection{Impact of antisymmetric modes on parameter recovery}

\begin{figure*}[ht!]
    \centering
    \includegraphics[width=\textwidth]{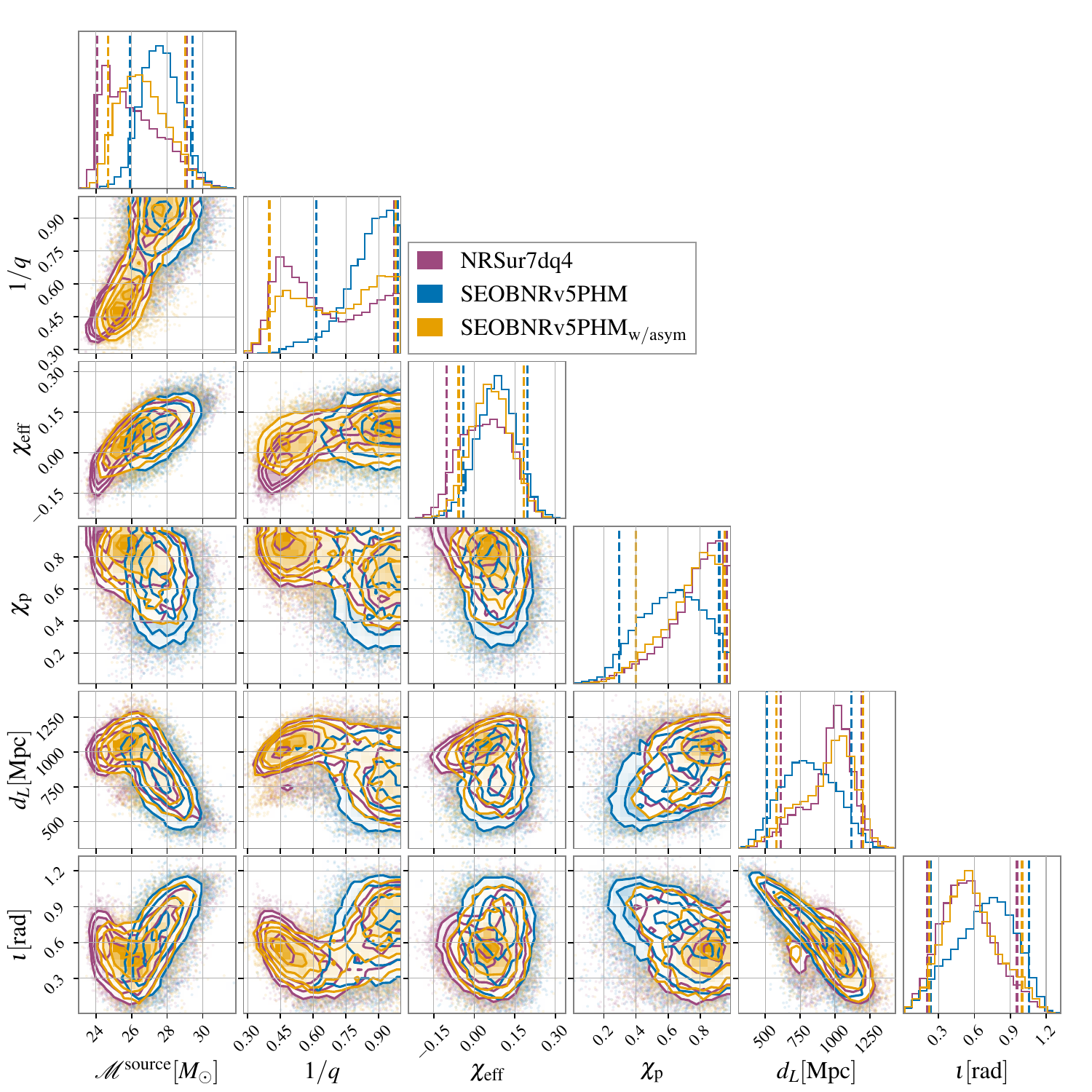}
    \caption{Comparison of the spin-precessing analyses of GW200129 using the public GWOSC data. We show the 1D and 2D marginalized posteriors for selected parameters. Dashed lines denote 90\% credible intervals.}
    \label{fig:gw200129_6params}
\end{figure*}

Figure~\ref{fig:gw200129_6params} compares the posterior distributions obtained from the public GWOSC data. We find that \SEOBASYM improves agreement with \texttt{NRSur7dq4} relative to \texttt{SEOBNRv5PHM}, confirming the relevance of antisymmetric modes for this event. In particular, both \SEOBASYM and \texttt{NRSur7dq4} favor higher values of the effective spin-precession parameter $\chi_{\rm p}$ and more asymmetric mass ratios (with a peak near $q \sim 0.6$), whereas \texttt{SEOBNRv5PHM} prefers more symmetric binaries. Additionally, \SEOBASYM shifts the inferred distributions of luminosity distance and inclination, suggesting that the added angular structure helps reduce degeneracies in those parameters.

\begin{figure*}[ht!]
    \centering
    \begin{minipage}{0.47\textwidth}
        \centering
        \includegraphics[width=\linewidth]{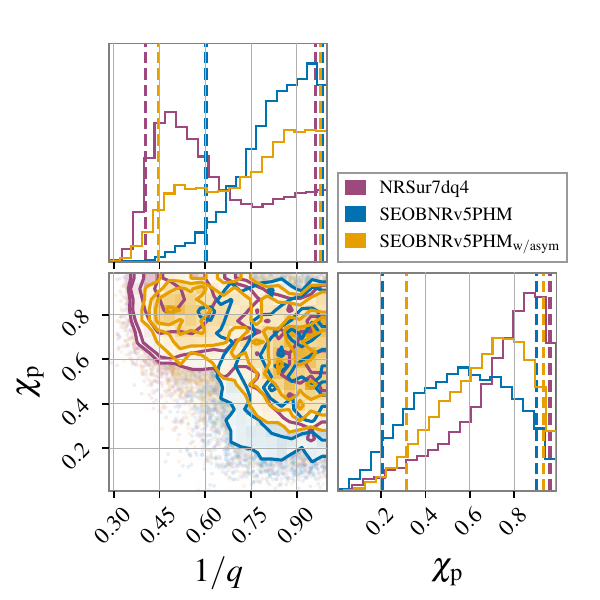}
        \subcaption[]{\texttt{GWSubstract}}
        \label{fig:gw200129_gwsub}
    \end{minipage}
    \hfill
    \begin{minipage}{0.47\textwidth}
        \centering
        \includegraphics[width=\linewidth]{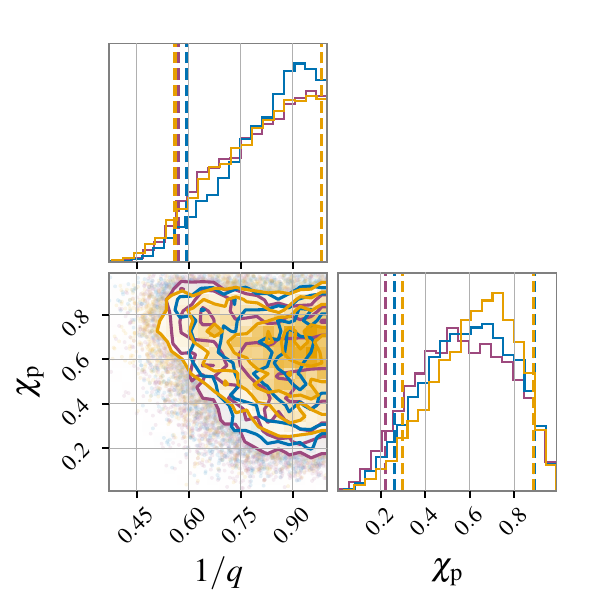}
        \subcaption[]{\texttt{BayesWave-A}}
        \label{fig:gw200129_bwa}
    \end{minipage}
    
    \vspace{0.3cm}
    \begin{minipage}{0.47\textwidth}
        \centering
        \includegraphics[width=\linewidth]{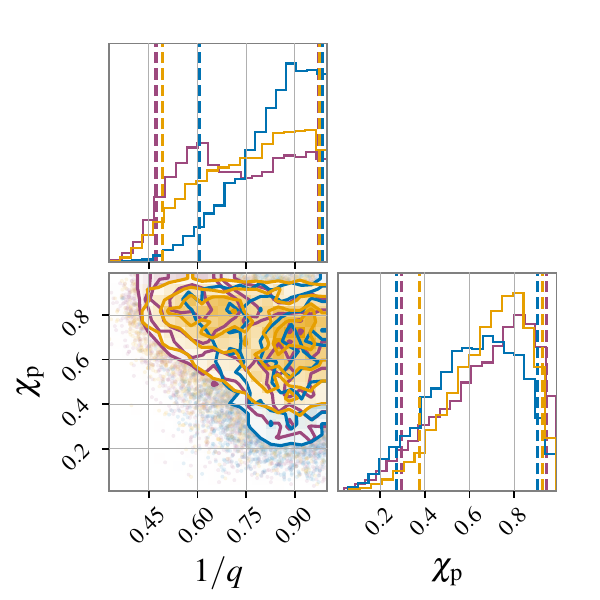}
        \subcaption[]{\texttt{BayesWave-B}}
        \label{fig:gw200129_bwb}
    \end{minipage}
    \hfill
    \begin{minipage}{0.47\textwidth}
        \centering
        \includegraphics[width=\linewidth]{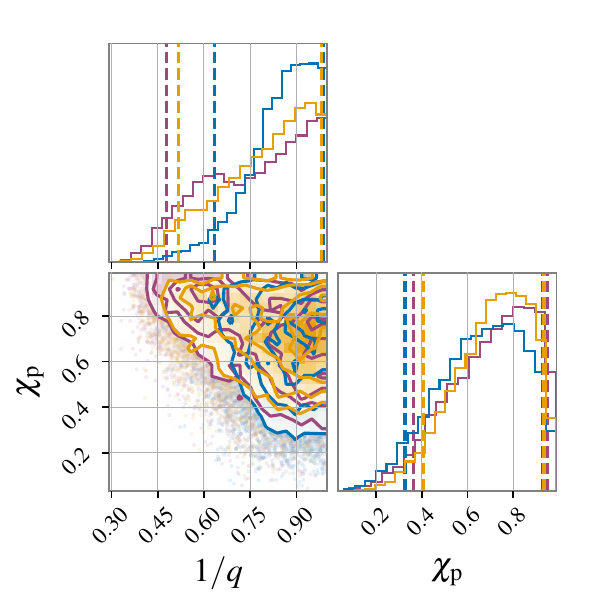}
        \subcaption[]{\texttt{BayesWave-C}}
        \label{fig:gw200129_bwc}
    \end{minipage}
    
    \caption{Comparison of mass ratio and effective spin-precession parameter $\chi_{\rm p}$ across different glitch subtraction methods for GW200129. Each panel shows posteriors from \texttt{NRSur7dq4}, \texttt{SEOBNRv5PHM}, and \SEOBASYM.}
    \label{fig:gw200129_2params}
\end{figure*}

Figure~\ref{fig:gw200129_2params} highlights the recovery of $\chi_{\rm p}$ and $q$ for the four glitch-subtracted datasets. The most significant differences appear in the \texttt{gwsubtract} case, where \texttt{SEOBNRv5PHM} underestimates spin-precession and mass asymmetry compared to both \SEOBASYM and \texttt{NRSur7dq4}. Inclusion of antisymmetric modes in \SEOBASYM partially reduces the systematics with \texttt{NRSur7dq4}. The \texttt{BayesWave} datasets exhibit more consistent behavior across models, with \SEOBASYM closely matching \texttt{NRSur7dq4}, particularly for \texttt{BW-B} and \texttt{BW-C}. For \texttt{BW-A}, the results from \SEOBASYM remain closer to those of \texttt{SEOBNRv5PHM}, indicating residual sensitivity to the glitch treatment.

\subsection{Bayesian model selection for spin-precession}

\begin{table}[h!]
\centering
\begin{tabular}{lcccc}
\toprule
& \texttt{gwsubtract} & \texttt{BW-A} & \texttt{BW-B} & \texttt{BW-C} \\
\midrule
\texttt{SEOBNRv5PHM}       & 6.3 & 3.7 & 6.7 & 23.9 \\
\SEOBASYM & 15.9 & 13.1 & 30.5 & 69.8 \\
\bottomrule
\end{tabular}
\caption{Bayes factors comparing the spin-precessing hypothesis to the non-precessing one. Each value represents the ratio of the evidence for the corresponding spin-precessing model to that of the aligned-spin model \texttt{SEOBNRv5HM}, for each glitch-subtracted dataset. 
}
\label{tab:bf_precession}
\end{table}

Table~\ref{tab:bf_precession} presents the Bayes factors for the spin-precessing versus aligned-spin hypotheses. In all cases, the inclusion of antisymmetric modes in \SEOBASYM strengthens support for the spin-precessing hypothesis, often by factors of 2--3 relative to \texttt{SEOBNRv5PHM}. The strongest support appears for the \texttt{BW-C} dataset, with a Bayes factor of nearly 70, while \texttt{BW-A} yields the weakest but still non-negligible support (Bayes factor $\sim 13$).

These results confirm and extend the findings of Ref.~\cite{Kolitsidou:2024vub}, demonstrating that the inclusion of antisymmetric modes in \SEOBASYM plays a key role in the interpretation of GW200129 as a spin-precessing binary. While \texttt{SEOBNRv5PHM} alone also supports this hypothesis, the evidence becomes significantly stronger when the antisymmetric contributions are included. Nevertheless, recent works employing eccentric aligned-spin waveform models \cite{Gupte:2024jfe,Planas:2025jny} show that the eccentric hypothesis is highly supported, with Bayes factor for eccentric aligned-spin versus quasi-circular spin-precessing orders of magnitude greater. Future reanalysis of this event employing eccentric spin-precessing waveform models will be key to assess the actual amount of spin-precession exhibited by this signal.

\section{Conclusions}\label{sec:conclusions}

In this work, we have developed a new waveform model for the spin-induced equatorial-asymmetric effects included in the GW modes with $\ell = m \leq 4$, formulated within the \texttt{SEOBNR} framework and incorporated into \SEOBASYM, an extension of the \texttt{SEOBNRv5PHM} model for quasi-circular, spin-precessing BBH systems. The model construction combines analytical PN results with calibration to numerical data, including NR simulations for spin-precessing systems and plunging geodesic test-body results for non-equatorial orbits.

We have evaluated the accuracy of \SEOBASYM{} by computing the SNR-weighted unfaithfulness between the model and two NR datasets: the SXS dataset described in Sec.~\ref{sec:datasets} and the BAM dataset from Ref.~\cite{Hamilton:2023qkv}. The inclusion of equatorial asymmetric effects in the modes significantly improves the agreement with NR across all inclination angles, with the most pronounced gains occurring at low inclinations, where asymmetric effects in the $(2,2)$ mode become more relevant. For face-on binaries, \SEOBASYM{} achieves nearly a 50\% reduction in the median unfaithfulness relative to \texttt{SEOBNRv5PHM}, and is the only model with fewer than 1\% of configurations exceeding 3\% unfaithfulness. Compared to \texttt{IMRPhenomXPNR}, \SEOBASYM{} performs comparably for face-on sources, and significantly better at higher inclinations, reducing median unfaithfulness by up to 60\%. The model also shows a reduced spread in unfaithfulness and fewer high-unfaithfulness outliers compared to all other models evaluated.

Against the BAM simulations, which were not used in the calibration of \SEOBASYM{}, we find consistent improvements in median unfaithfulness across all inclinations. While \texttt{IMRPhenomXPNR} slightly outperforms \SEOBASYM{} for face-on sources --- likely due to calibration to this specific dataset --- \SEOBASYM{} delivers more accurate results at all other inclinations. In particular, for moderate inclinations ($\iota = \pi/3$), \SEOBASYM{} shows improved performance at low mass ratios and in-plane spin magnitudes, where it outperforms all other models. Overall, these results confirm that the inclusion of antisymmetric modes in \SEOBASYM{} leads to a systematic and robust improvement in waveform accuracy for precessing BBH systems.

Furthermore, the model yields more accurate predictions for the GW recoil (kick) velocity of the remnant, reproducing the qualitative shape of the NR kick distribution, including the maximum estimated kick velocity from the dataset. It significantly reduces the median relative error compared to the symmetric model, although there is a percentage of cases disagreeing more than 10\% with the predicted NR value, potentially related with the lack of asymmetric effects in the $(3,2)$ mode.

We have validated the model in Bayesian inference scenarios using three synthetic injections in zero-noise data. These injections were based on NR simulations of systems with strong spin-precession and high remnant recoil. We find that including antisymmetric modes in \SEOBASYM improves the recovery of several key parameters, notably the primary spin orientation $\theta_1$, the in-plane spin angle difference $\phi_{12}$, and mass-related parameters. However, one injection exhibited a systematic bias in luminosity distance. A comparative analysis using \texttt{NRSur7dq4} revealed a similar trend, suggesting that the bias arises from the intrinsic properties of the signal, not the particular modeling of the antisymmetric modes.

We have also re-analyzed the astrophysical event GW200129 using multiple glitch-subtracted versions of the strain data. Across all datasets, we find that \SEOBASYM systematically reduces the discrepancy with the results obtained using the \texttt{NRSur7dq4} surrogate model. In particular, the inclusion of antisymmetric modes increases the recovered effective precession parameter $\chi_{\rm p}$ and shifts the mass-ratio posterior towards more asymmetric configurations, consistent with earlier findings~\cite{Kolitsidou:2024vub}. Bayesian model selection shows a clear enhancement in the support for the spin-precessing hypothesis when using \SEOBASYM, increasing the Bayes factor by approximately a factor of 3 compared to \texttt{SEOBNRv5PHM} alone, and reaching a maximum of about $\sim$70:1 in favor of spin-precession for one dataset.

Despite these improvements, further refinement is likely required for modeling higher-SNR signals or capturing subdominant antisymmetric effects. In particular, the physical origin and modeling of mode-mixing in the $(3,3)$ antisymmetric mode remain to be clarified, and discrepancies between PN predictions and NR data for antisymmetric $\ell \ne m$ modes persist. Including the antisymmetric $(3,2)$ mode may help reduce residual differences in the recoil velocity for some systems. Finally, because antisymmetric modes depend sensitively on the spin-precession frequency, improved modeling of the precessing dynamics --- currently uncalibrated to NR --- could yield further gains in both the accuracy of antisymmetric modes and the overall model performance.

The improvements described in \SEOBASYM have been released in the \texttt{pyseobnr} python package \cite{Mihaylov:2023bkc}, and they can be employed for any version greater than \texttt{v0.3.3}. To activate the inclusion of antisymmetric contributions, the following option has to be passed in the waveform arguments while calling the \texttt{SEOBNRv5PHM} approximant: \texttt{enable\_antisymmetric\_modes=True}. To ensure that the subdominant contributions are enabled, the following option has to be specified: \texttt{antisymmetric\_modes\_hm=True}.

\section*{Acknowledgments}
We are especially grateful to Serguei Ossokine for initiating this project and for valuable early discussions that helped shape its direction. We would like to thank Scott A. Hughes, Anuj Apte, Gaurav Khanna and Halston Lim for providing the Teukolsky waveforms used in this project. It is our pleasure to thank Eleanor Hamilton, Marta Colleoni, Shrobana Ghosh, Lucy Thomas and Yumeng Xu for perfoming the LVK Collaboration internal review of the implementation of the antisymmetric modes into \texttt{SEOBNRv5PHM}. We would like to warmly thank Shrobana Ghosh, Jannik Mielke, Antoni Ramos-Buades, Arnab Dhani, Nihar Gupte and Yifan Wang for useful discussions. 

The computational work for this manuscript was carried out on the computer cluster Hypatia at the Max Planck Institute for Gravitational Physics in Potsdam. 

A. Buonanno's research is supported in part by the European Research Council (ERC) Horizon Synergy Grant “Making Sense of the Unexpected in the Gravitational-Wave Sky” grant agreement no. GWSky–101167314.

This research has made use of data or software obtained from the Gravitational Wave Open Science Center (gwosc.org), a service of LIGO Laboratory, the LIGO Scientific Collaboration, the Virgo Collaboration, and KAGRA. LIGO Laboratory and Advanced LIGO are funded by the United States National Science Foundation (NSF) as well as the Science and Technology Facilities Council (STFC) of the United Kingdom, the Max-Planck-Society (MPS), and the State of Niedersachsen/Germany for support of the construction of Advanced LIGO and construction and operation of the GEO600 detector. Additional support for Advanced LIGO was provided by the Australian Research Council. Virgo is funded, through the European Gravitational Observatory (EGO), by the French Centre National de Recherche Scientifique (CNRS), the Italian Istituto Nazionale di Fisica Nucleare (INFN) and the Dutch Nikhef, with contributions by institutions from Belgium, Germany, Greece, Hungary, Ireland, Japan, Monaco, Poland, Portugal, Spain. KAGRA is supported by Ministry of Education, Culture, Sports, Science and Technology (MEXT), Japan Society for the Promotion of Science (JSPS) in Japan; National Research Foundation (NRF) and Ministry of Science and ICT (MSIT) in Korea; Academia Sinica (AS) and National Science and Technology Council (NSTC) in Taiwan.

\appendix

\section{Unfaithfulness against the \texttt{NRSur7dq4} model}
\label{sec:nrsur}

To further validate the performance of our model, we assess the SNR-weighted unfaithfulness against the \texttt{NRSur7dq4} surrogate model \cite{Varma:2019csw} across a large set of randomly generated spin-precessing configurations. This offers a complementary benchmark from the NR unfaithfulness study of Sec.~\ref{sec:unfaith}, since now the unfaithfulness is evaluated at points that have not been employed in the calibration of \SEOBASYM nor in the calibration of \texttt{IMRPhenomXPNR}. We note, however, that \texttt{NRSur7dq4} is itself trained on the SXS catalog, and is therefore not fully independent of the data used to calibrate the antisymmetric modes in \SEOBASYM{}.

\begin{figure}[h]
    \includegraphics[width=\columnwidth]{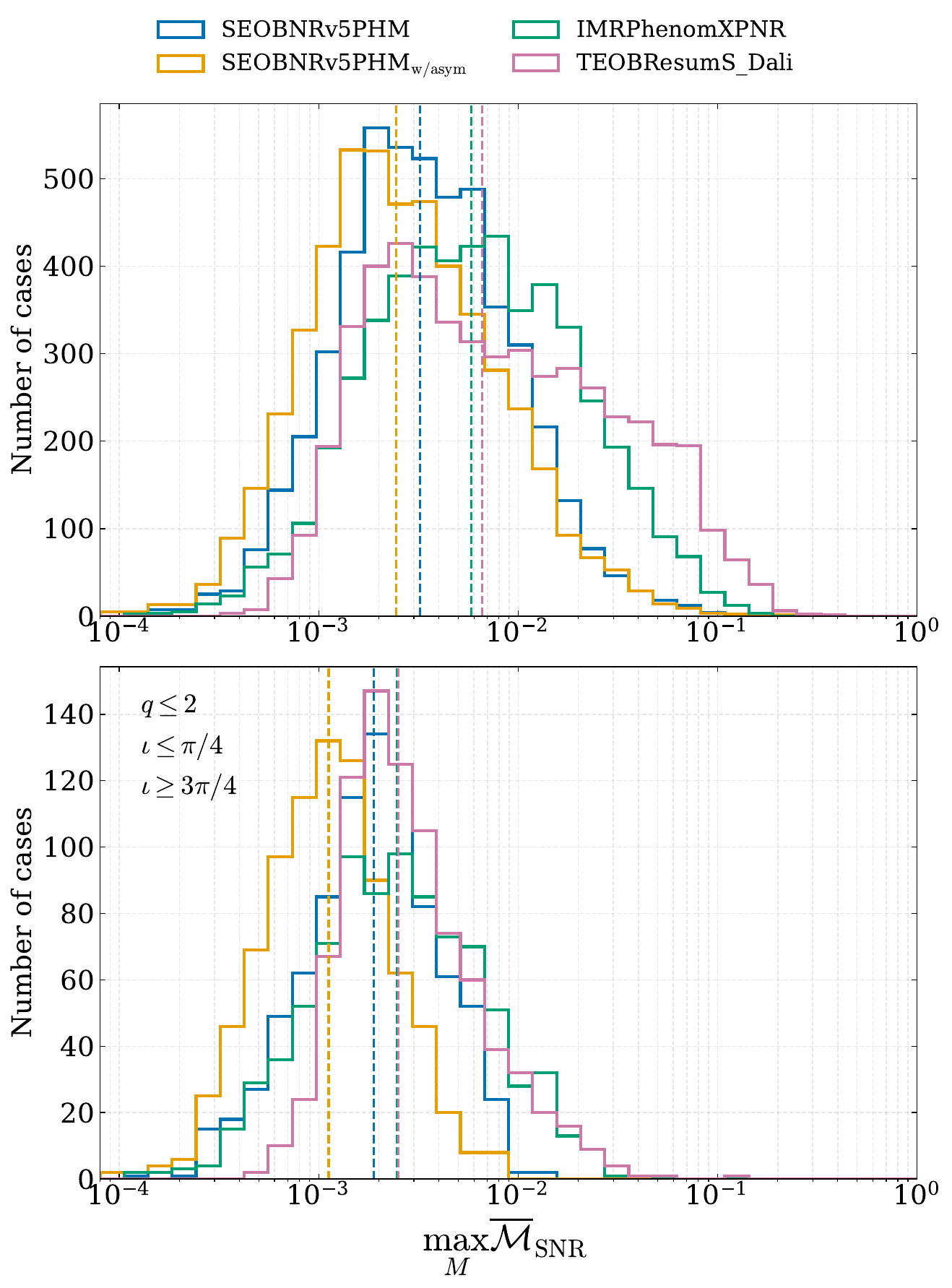}
    \caption{Distribution of maximum (across total mass) SNR-weighted unfaithfulness with \texttt{NRSur7dq4} for 5000 random configurations (top panel) and a subset of 1250 configurations with $q\leq 2$ and inclinations close to face-on and face-off (bottom panel). Vertical lines denote median values of the distributions}
    \label{fig:hist_nrsur_all_and_lowincl}
\end{figure}

We simulate 5000 BBH configurations using \texttt{NRSur7dq4} with the following parameter distributions:
\begin{itemize}
    \item Mass ratio \( q \in [1,4] \), uniformly distributed,
    \item In-plane spin magnitudes \( \chi_{1,2,\perp} \in [0,0.8] \), uniformly distributed,
    \item Aligned spin components \( \chi_{1,2,\ell} \in [\chi_{1,2,\perp}, 0.8] \),
    \item Uniform distributions in inclination \( \iota \), reference phase, polarization angle, and in-plane spin orientations.
\end{itemize}

\begin{figure}[h]
    \includegraphics[width=\columnwidth]{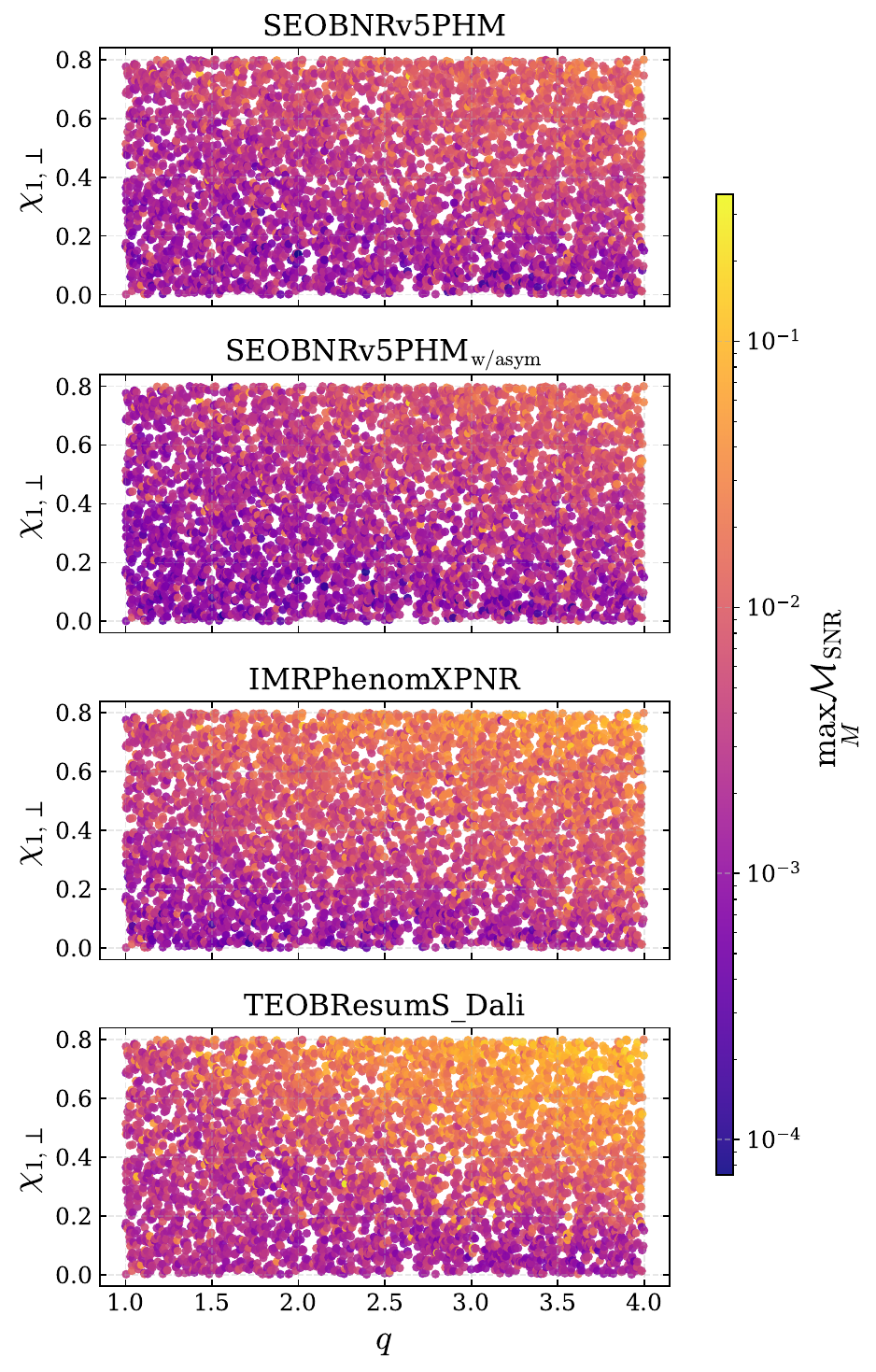}
    \caption{Maximum (across total mass) SNR-weighted unfaithfulness with \texttt{NRSur7dq4} for 5000 random configurations as a function of mass-ratio $q$ and in-plane primary spin magnitude $\chi_{1,\perp}$.}
    \label{fig:scatter_q_chi1perp}
\end{figure}

The unfaithfulness is evaluated without averaging over extrinsic parameters (since we already randomize over effective polarization and reference phase), across a total mass range of \( [20, 200]\,M_\odot \). Figure~\ref{fig:hist_nrsur_all_and_lowincl} displays the distribution of the maximum unfaithfulness over total mass, and Fig.~\ref{fig:scatter_q_chi1perp} shows the dependence on mass ratio and primary in-plane spin.

We observe that \SEOBASYM{} achieves the lowest median unfaithfulness (0.0024), outperforming \texttt{SEOBNRv5PHM} (0.0032), \texttt{IMRPhenomXPNR} (0.0058), and \texttt{TEOBResumS\_Dali} (0.0066). The tail of the distribution is comparable for \texttt{SEOBNRv5PHM} and \SEOBASYM{}, reaching in both cases 10\% unfaithfulness, and pointing to accuracy limitations shared by \texttt{SEOBNRv5PHM} and \SEOBASYM and, therefore, independent from the asymmetric content.

For a subset of 1250 near equal-mass and near face-on/off configurations (\( q \leq 2 \), \( \iota < \pi/4 \) or \( \iota > 3\pi/4 \)) (results shown in the bottom panel of Fig.~\ref{fig:hist_nrsur_all_and_lowincl}), \SEOBASYM{} reaches a median unfaithfulness of 0.0011, nearly halving that of \texttt{SEOBNRv5PHM} (0.0019), consistent with the results against NR at low inclination discussed in Sec.~\ref{sec:unfaith}. This confirms the relevance of including the antisymmetric modes especially in the region where they are expected to contribute more (near face-on and face-off systems). Additionally, the median unfaithfulness and the tail of the distribution is reduced for all models in this subset, pointing to a common accuracy limitation in all models for more mass-asymmetric systems and higher inclinations, suggesting that the treatment of subdominant harmonics in the models is not optimal.

In Fig.~\ref{fig:scatter_q_chi1perp} we show the dependence of the maximum (across total mass) SNR-weighted unfaithfulness on the mass-ratio $q$ and in-plane primary spin magnitude $\chi_{1,\perp}$ for all the models. We can observe that the accuracy of all models degrades with more asymmetric-mass systems and higher spin-precession, as it is commonly observed in waveform modeling publications \cite{Pratten:2020ceb,Ossokine:2020kjp,Estelles:2021gvs,Gamba:2021ydi,Ramos-Buades:2023ehm}. We observe similar dependence for \texttt{SEOBNRv5PHM} and \SEOBASYM, consistent with the similarity of the tail of the distribution in the median unfaithfulness shown in Fig. \ref{fig:hist_nrsur_all_and_lowincl}. \texttt{IMRPhenomXPHM} and \texttt{TEOBResumS\_Dali} show an increased number of cases reaching higher unfaithfulness in this region, which is consistent with the more pronounced tail observed in in Fig.~\ref{fig:hist_nrsur_all_and_lowincl}.

\begin{figure*}[htbp!]
    \centering
    \begin{minipage}{\textwidth}
        \centering
        \includegraphics[width=\linewidth]{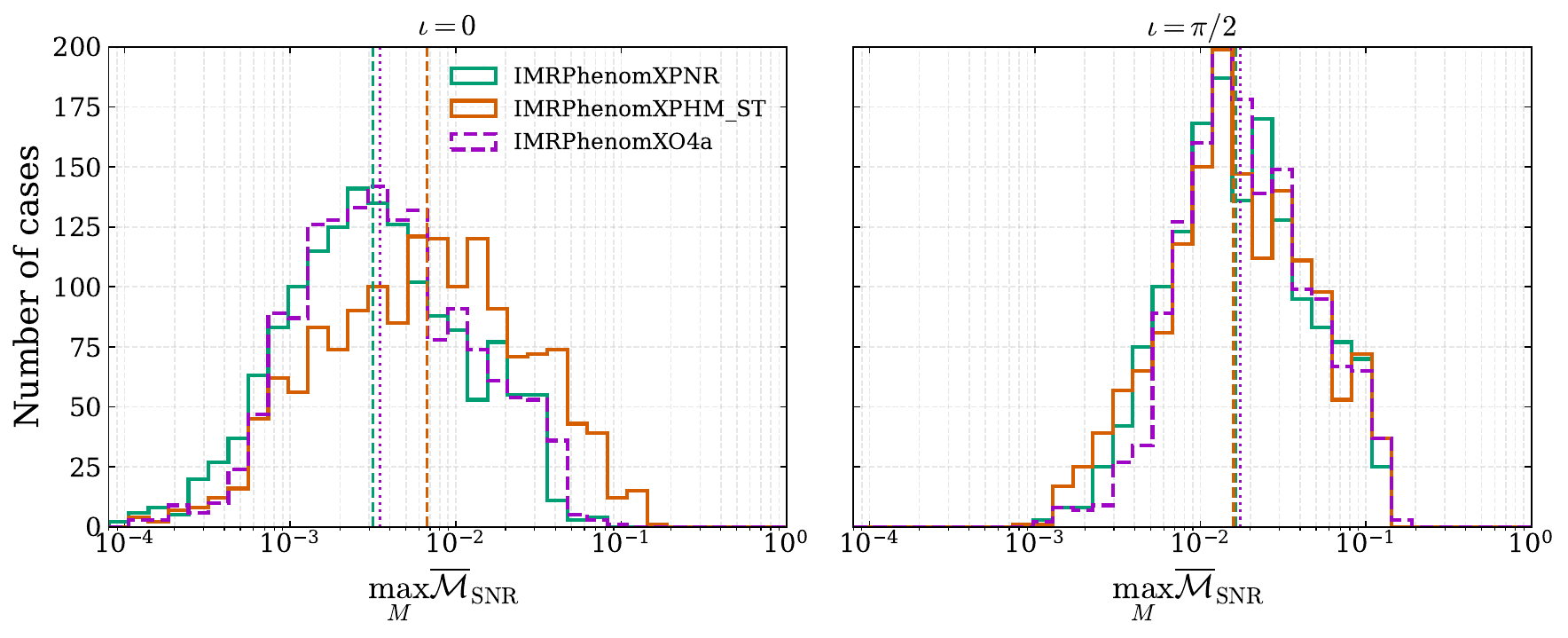}
        \subcaption[]{Unfaithfulness against SXS dataset}
        \label{fig:unf_xphm_sxs}
    \end{minipage}
    
    \vspace{0.3cm}
    \begin{minipage}{\textwidth}
        \centering
        \includegraphics[width=\linewidth]{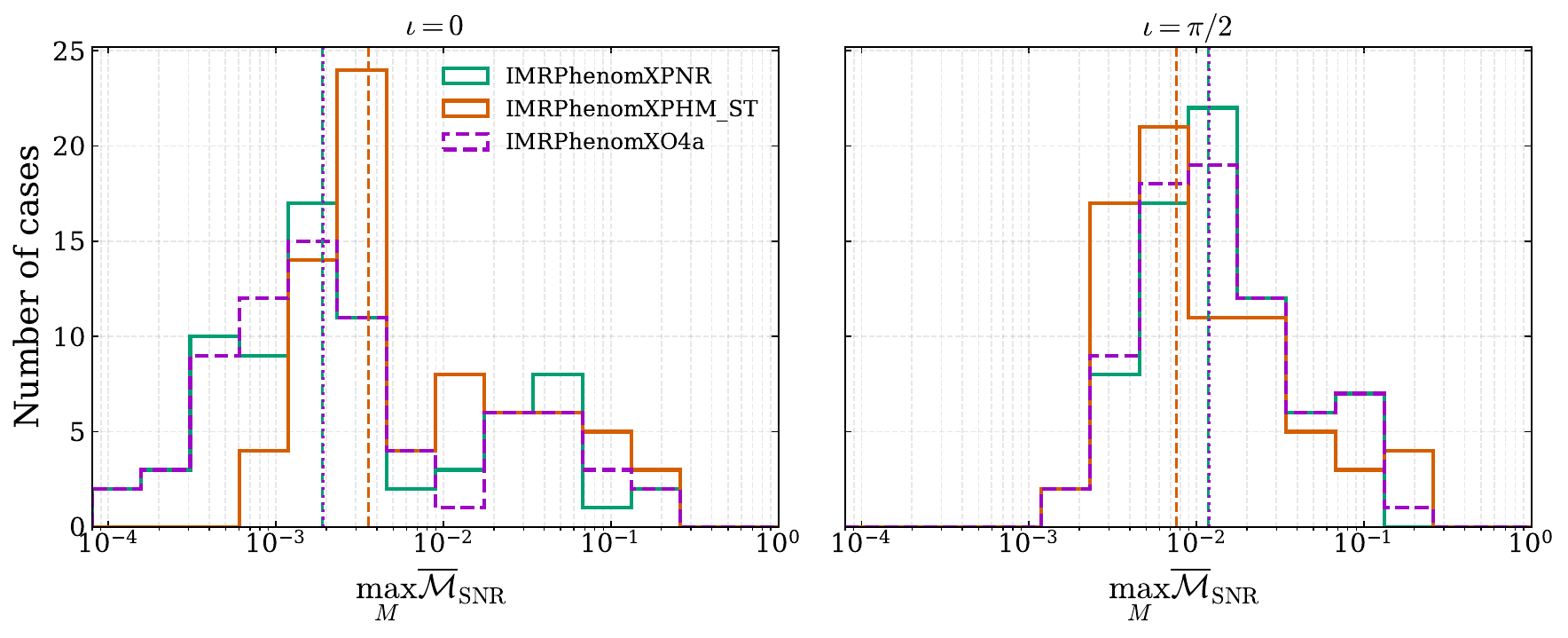}
        \subcaption[]{Unfaithfulness against BAM dataset}
        \label{fig:unf_xphm_bam}
    \end{minipage}
    
    \caption{Histogram of the maximum SNR-weighted averaged unfaithfulness against the NR datasets across a total mass range $[20,300]\,M_{\odot}$, for inclinations $\iota_s = \{0,\pi/2\}$. Top panel: comparison against SXS dataset. Bottom panel: comparison against BAM dataset. Vertical dashed lines mark the median of each distribution.}
    \label{fig:unf_xphm}
\end{figure*}

\section{Unfaithfulness against NR with \texttt{IMRPhenom} models}
\label{sec:xphm}

We compare the accuracy of three Fourier-domain phenomenological models: \texttt{IMRPhenomXPHM\_ST} \cite{Colleoni:2024knd}, \texttt{IMRPhenomXO4a} \cite{Hamilton:2021pkf,Thompson:2023ase}, and the recent \texttt{IMRPhenomXPNR} model \cite{xpnr_inprep}. The \texttt{IMRPhenomXO4a} model incorporates several improvements over \texttt{IMRPhenomXPHM} \cite{Pratten:2020ceb,Garcia-Quiros:2020qpx,Pratten:2020fqn}, including calibration to NR simulations in the merger-ringdown regime \cite{Hamilton:2021pkf}, and the inclusion of dominant-mode antisymmetries in the co-precessing frame \cite{Ghosh:2023mhc}. The \texttt{IMRPhenomXPHM\_ST} model includes a more accurate treatment of spin dynamics during the inspiral, and \texttt{IMRPhenomXPNR} combines the improvements of both models. 

We compute the maximum SNR-weighted unfaithfulness against NR waveforms from both the SXS and BAM datasets, for two representative source inclinations: face-on (\( \iota = 0 \)) and edge-on (\( \iota = \pi/2 \)). Figure~\ref{fig:unf_xphm} shows the unfaithfulness distributions for the three models. In general, we observe very similar results for \texttt{IMRPhenomXO4a} and \texttt{IMRPhenomXPNR}. For face-on systems, we observe a substantial reduction in median unfaithfulness in both models compared to \texttt{IMRPhenomXPHM\_ST}, across both datasets. For the SXS dataset, the median is reduced from 0.0081 (\texttt{XPHM\_ST}) to 0.0035 (\texttt{XO4a}) and 0.0032 (\texttt{XPNR}), a 60\% improvement. For the BAM dataset, the median is reduced from 0.0035 to 0.0019 (\texttt{XO4a} and \texttt{XPNR}), yielding a 46\% improvement. 

At edge-on orientation (\( \iota = \pi/2 \)), the difference is less pronounced, consistent with the calibration in \texttt{IMRPhenomXO4a} and \texttt{IMRPhenomXPNR} being performed for the dominant mode (and the inclusion of the dominant asymmetry) and therefore being more relevant for face-on systems. In the SXS dataset, \texttt{IMRPhenomXPNR} still outperforms \texttt{IMRPhenomXPHM\_ST} with a median unfaithfulness of 0.0163 vs 0.0217, slightly improving the results of \texttt{IMRPhenomXO4a} (0.0174), while in the BAM dataset, \texttt{IMRPhenomXPNR} exhibits a slightly higher median unfaithfulness than \texttt{IMRPhenomXPHM\_ST} (0.0118 vs 0.0085), with \texttt{IMRPhenomXO4a} showing very similar results. 

Interestingly, the three models perform better overall with the BAM dataset than with SXS, for both inclinations. Since the BAM dataset only contains single-spin precessing simulations, this may reflect a better accuracy in the description of the spin dynamics for these models in the single-spin case.

\begin{figure*}[ht!]
    \centering
    \includegraphics[width=\textwidth]{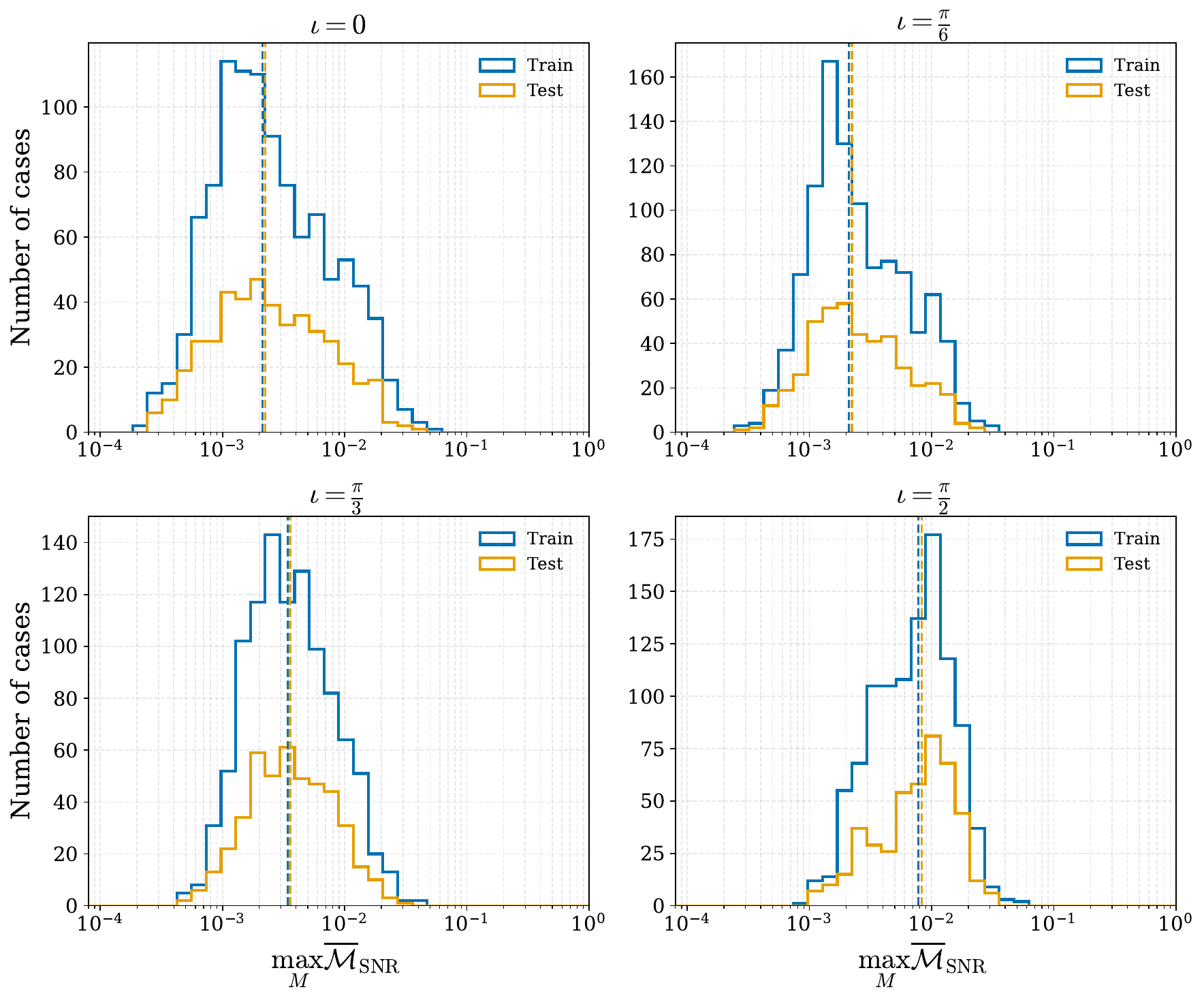}
    \caption{
    Comparison of the maximum SNR-weighted unfaithfulness for the 
    \SEOBASYM{} model between the training (blue) and testing (orange)
    subsets of the SXS catalog, for four values of the inclination.  
    Dashed vertical lines mark the respective median values.  
    The close agreement between the training and testing distributions 
    across all inclinations demonstrates that the antisymmetric-mode 
    fits generalize well and do not exhibit signs of overfitting.
    }
    \label{fig:train_test_hist}
\end{figure*}

\section{Unfaithfulness against training and testing SXS subsets.}
\label{app:train_test}

We assess explicitly the generalization properties of the NR calibration included of the antisymmetric modes included in \SEOBASYM{} by splitting the unfaithfulness results discussed in Sec.~\ref{sec:unfaith} between the training and testing SXS datasets employed in the calibration. As described in Sec.~\ref{sec:omp}, the split is performed independently for the SXS and Teukolsky datasets, using a 70/30 division with a fixed random seed.

Figure~\ref{fig:train_test_hist} shows, for each inclination, the distribution of the maximum SNR-weighted unfaithfulness across the total-mass range for the \SEOBASYM{} model, evaluated separately on the training and testing subsets of the SXS dataset. The two distributions overlap closely for all inclinations, and their median values differ only at the level of a few $\times 10^{-4}$.  This demonstrates that the fits do not overfit the training set and that their performance generalizes well to unseen NR waveforms.


\bibliographystyle{apsrev4-2}
\bibliography{bibliography}

\end{document}